\documentclass[floatfix,aps,amsmath,prd,twocolumn,eqsecnum,showpacs,nofootinbib]{revtex4}
\usepackage{graphicx,graphics,amsfonts,amsmath,amssymb,bm,hyperref,latexsym,natbib}

\allowdisplaybreaks

\newcommand{\be}{\begin{equation}}
\newcommand{\ee}{\end{equation}}
\newcommand{\bs}{\begin{subequations}}
\newcommand{\es}{\end{subequations}}

\begin{document}

\title{Post-Newtonian corrections to the gravitational-wave memory for quasicircular, inspiralling compact binaries}
\author{Marc Favata}
\email{favata@kitp.ucsb.edu}
\affiliation{Kavli Institute for Theoretical Physics, University of California, Santa Barbara, California 93106-4030, USA}
\received{30 November 2008}
\begin{abstract}
The Christodoulou memory is a nonlinear contribution to the gravitational-wave field that is sourced by the gravitational-wave stress-energy tensor.  For quasicircular, inspiralling binaries, the Christodoulou memory produces a growing, nonoscillatory change in the gravitational-wave ``plus'' polarization, resulting in the permanent displacement of a pair of freely-falling test masses after the wave has passed. In addition to its nonoscillatory behavior, the Christodoulou memory is interesting because even though it originates from 2.5 post-Newtonian (PN) order multipole interactions, it affects the waveform at leading (Newtonian/quadrupole) order. The memory is also potentially detectable in binary black-hole mergers. While the oscillatory pieces of the gravitational-wave polarizations for quasicircular, inspiralling compact binaries have been computed to 3PN order, the memory contribution to the polarizations has only been calculated to leading order (the next-to-leading order 0.5PN term has previously been shown to vanish).  Here the calculation of the memory for quasicircular, inspiralling binaries is extended to 3PN order. While the angular dependence of the memory remains qualitatively unchanged, the PN correction terms tend to reduce the memory's magnitude. Explicit expressions are given for the memory contributions to the plus polarization and the spin-weighted spherical-harmonic modes of the metric and curvature perturbations.
Combined with the recent results of Blanchet et al.~[Class.~Quantum Grav.~{\bf 25}, 165003 (2008)], this completes the waveform polarizations to 3PN order. This paper also discusses: (i) the difficulties in extracting the memory from numerical relativity simulations, (ii) other nonoscillatory effects that enter the waveform polarizations at high PN orders, and (iii) issues concerning the observability of the memory in gravitational-wave detectors.
\end{abstract}
\maketitle

\section{\label{sec:intro}Introduction}
The primary purpose of this paper is to compute the post-Newtonian (PN) corrections to the nonlinear memory piece of the gravitational-wave (GW) polarizations for quasicircular\footnote{\emph{quasicircular} means that the binary orbit is circular up 2.5PN order radiation-reaction effects that cause the binary to slowly inspiral.}, inspiralling binaries. We begin by reviewing the gravitational-wave memory (Sec.~\ref{sec:memintro}) and then motivate its further study (Sec.~\ref{sec:motivation}). A summary of results is given in Sec.~\ref{sec:summary}.
\subsection{\label{sec:memintro}What is gravitational-wave memory?}
Gravitational-wave memory refers to the permanent displacement of an ``ideal'' GW detector after the GW has passed.  An ideal detector is one which is only sensitive to gravitational forces\textemdash e.g., a ring of freely-falling test-masses\textemdash and is isolated from local tidal interactions. After the passage of a GW \emph{without} memory, the detector returns to its initial state of internal displacement (its state long before the passage of the wave). After the passage of a GW \emph{with} memory, the initial and final displacement states differ.

There are two types of GW memory: linear and nonlinear. The linear memory has been known since the 1970s \cite{zeldovich-polnarev,braginskii-grishchuk,braginskii-thorne} and arises from gravitational sources that produce a net change in the time derivatives of one or more of their \emph{source-multipole moments}.
For example, at leading order in a PN expansion the linear memory causes a net change in the GW field given by
\be
\label{eq:hTTmem}
\Delta h_{jk}^{\rm TT} = \frac{2}{R} \Delta(\ddot{\mathcal I}_{jk}^{\rm TT}) ,
\ee
where ${\mathcal I}_{ij}$ is the source mass-quadrupole moment, $R$ is the distance to the source, TT means to take the transverse-traceless projection [Eq.~\eqref{eq:TT}], and the $\Delta$ refers to the difference between late and early times. Geometric units ($G=c=1$) are used here and throughout.

A simple example of a source with linear memory is a binary on a hyperbolic orbit (gravitational two-body scattering) \cite{turner-unbound,turner-will,kovacs-thorne-IV}. To see this consider the TT piece of the second-time derivative of the mass-quadrupole moment at Newtonian order and in the center-of-mass frame:
\be
\label{eq:ddIij}
\ddot{\mathcal I}_{jk}^{\rm TT} = 2 \mu \left[ \dot{x}_j \dot{x}_k - \frac{M}{r^3} x_j x_k \right]^{\rm TT} ,
\ee
where $\mu=m_1 m_2/M$ is the reduced mass, $M=m_1+m_2$ is the total mass, $x_j$ is the relative orbital separation vector with magnitude $r$, $\dot{x}_j$ is the relative orbital velocity, and we have replaced second-time derivatives with the equation of motion $\ddot{x}_j = -M x_j/r^3$. If we take $t=0$ to be the time of closest approach, then at very early ($t\rightarrow -\infty$) and very late ($t\rightarrow +\infty$) times the relative velocities will be finite ($v_{\infty} = \sqrt{2E/\mu}$ where $E$ is the orbital energy), while the second term in Eq.~\eqref{eq:ddIij} falls off like $1/r$ and is negligible. The memory for a hyperbolic binary thus arises from the difference in the direction of the velocity vectors at late and early times:
\be
\Delta h_{jk}^{\rm TT} = \frac{4 \mu}{R} \Delta [\dot{x}_j \dot{x}_k]^{\rm TT} .
\ee

Other systems with linear memory are those whose components change from being bound to unbound (or vice versa). These include binaries whose components are captured, disrupted, or undergo mass loss. Gravitational-waves with linear memory have been studied in the context of supernova explosions and their resulting neutron star kicks \cite{burrows-hayesPRL96,buonanno-stochasticBGmemory,muller-GWcorecollapse,kotake-sato-SNmemory,davies-king-SNkick-memory,cuesta-pulsarkickmem,kotake-etal-sasiApJL09}, asymmetric mass loss due to neutrino emission \cite{turner-neutrinomemory,epstein-neutrinomemory,loveridge-pulsarkickmemory,burrows-hayesPRL96,kotake-etal-sasiApJL09} (see Ref.~\cite{ott-corecollapsereview} for a recent review), or gamma-ray-burst jets \cite{sago-GRBmemory,hiramatsu-kotake-GRBmemory,segalis-ori-GRBmemory}. Linear memory can also arise from GW recoil in binary black-hole mergers \cite{favata-lisa7confproc}. Sources whose components remain bound generally do not display linear memory (but see Sec.~\ref{sec:linmem} for a caveat).

A general formula for the linear memory produced by a system of $N$ bodies with changing masses $M_A$ or velocities ${\bm v}_A$ is given by Thorne \cite{kipmemory} (also Ref.~\cite{braginskii-thorne}):
\be
\label{eq:hijlinmem}
\Delta h_{jk}^{\rm TT} = \Delta \sum_{A=1}^N \frac{4 M_A}{R\sqrt{1-v_A^2}} \left[ \frac{v_A^j v_A^k}{1-{\bm v}_A \cdot {\bm N}} \right]^{\rm TT} \,,
\ee
where the masses are unbound in their initial or final states (or both), the $\Delta$ means to take the difference between the final and initial values of the summation, and ${\bm N}$ is a unit vector that points from the source to the observer. For example, the masses $M_A$ and velocities ${\bm v}_A$ might refer to two or more particles on a scattering orbit (as discussed above), or they might refer to the various pieces of a star that becomes unbound. This formula is essentially the standard Li\'{e}nard-Wiechert solution of the space-space part of the linearized Einstein equations (with a source term given by the stress-energy tensor of $N$ noninteracting point particles).

The nonlinear memory was discovered independently by Payne \cite{payne-zfl},  Blanchet and Damour \cite{blanchet-thesis,blanchet-damour-hereditary}, and Christodoulou \cite{christodoulou-mem}. It is often referred to as the ``Christodoulou memory.''  The nonlinear memory\footnote{Throughout this paper the terms ``nonlinear memory'' and ``Christodoulou memory'' are used interchangeably. When not otherwise specified, ``memory'' refers to the ``Christodoulou memory''.} arises from a change in the \emph{radiative-multipole moments} that is sourced by the energy flux of the radiated gravitational waves. One can heuristically understand the origin of the nonlinear memory as follows: Consider the Einstein field equations in harmonic gauge \cite{kiprmp}:
\bs
\label{eq:EFEeqs}
\be
\label{eq:EFE}
\Box \bar{h}^{\alpha \beta} = -16 \pi (-g) ( T^{\alpha \beta} + t_{\rm LL}^{\alpha \beta} ) -  {\bar{h}^{\alpha \mu}}_{\;\;\; , \nu} {\bar{h}^{\beta \nu}}_{\;\;\; , \mu} + \bar{h}^{\mu \nu} {\bar{h}^{\alpha \beta}}_{\;\;\; , \mu \nu} ,
\ee
\be
\label{eq:harmonicgauge}
{{\bar h}^{\alpha \beta}}_{\;\;\; , \beta}= 0,
\ee
\es
where
\be
\label{hbardef}
\bar{h}^{\alpha \beta} \equiv \eta^{\alpha \beta} - \sqrt{-g} g^{\alpha \beta}
\ee
is the gravitational-field tensor\footnote{\label{ftnt:sign-h}Different notations for $\bar{h}^{\alpha \beta}$ are used in the literature. For example, Blanchet \cite{blanchetLRR} and Will and Wiseman (WW) \cite{will-wiseman-2pn} both use the symbol $h^{\alpha \beta}$, but their conventions differ by a sign: $\bar{h}^{\alpha \beta} = h^{\alpha \beta}_{\rm WW} = - h^{\alpha \beta}_{\rm B}$. However, all agree on the sign convention for the waveform $h_{jk}^{\rm TT}$.}, $g$ is the determinant of the metric $g_{\alpha \beta}$, $T^{\alpha \beta}$ is the matter stress-energy tensor, $t_{\rm LL}^{\alpha \beta}$ is the Landau-Lifshitz pseudotensor, $\Box \equiv -\partial^2_t + \nabla^2$ is the flat-space wave operator, a comma denotes a partial derivative (${}_{,\mu}\equiv \partial_{\mu}$), and $\nabla^2$ is the flat-space Laplacian. While there are many nonlinear terms that enter the right-hand side of Eq.~\eqref{eq:EFE}, there is a particular piece of $t_{\rm LL}^{jk}$ [see the last term of Eq.~(2.7) in Ref.~\cite{will-wiseman-2pn}] that is equal to the GW stress-energy tensor \cite{isaacson,mtw}:
\be
T^{\rm gw}_{jk} = \frac{1}{32\pi} \left\langle h^{\rm TT}_{ab, j} h^{\rm TT}_{ab, k} \right\rangle \approx T_{00}^{\rm gw} n_j n_k = \frac{1}{R^2} \frac{dE^{\rm gw}}{dt d\Omega} n_j n_k ,
\ee
where $\frac{dE^{\rm gw}}{dt d\Omega}$ is the GW energy flux, $n_j$ is a unit radial vector, the angle-brackets mean to average over several wavelengths,  $h_{ab}^{\rm TT} = \bar{h}_{ab}^{\rm TT}$, and we have used the plane-wave approximation $h^{\rm TT}_{ab} \approx F_{ab}(t-R)/R +O(R^{-2})$.   When applying the standard Green's function to the right-hand side of Eq.~\eqref{eq:EFE}, this piece---proportional to the energy flux of the emitted GWs---yields the following correction term to the GW field \cite{wiseman-will-memory}:
\be
\label{eq:hTTnonlinmem}
\delta h^{\rm TT}_{jk} = \frac{4}{R} \int_{-\infty}^{T_R} dt'\, \left[ \int \frac{dE^{\rm gw}}{dt' d\Omega'} \frac{n'_j n'_k}{(1-{\bm n}' \cdot {\bm N})} d\Omega' \right]^{\rm TT},
\ee
where $T_R$ is the retarded time. This equation shows that part of the distant GW field is sourced by the loss of GW energy.  Note that while the magnitude of the nonlinear memory approximately scales with the total radiated GW energy $\Delta E^{\rm gw}$, the angular-dependent unit vectors in the integrand of Eq.~\eqref{eq:hTTnonlinmem} imply that the nonlinear memory is \emph{not} directly proportional to $\Delta E^{\rm gw}$. The memory should not be mistaken as a change in the monopolar piece of the $1/R$ expansion of the metric. Rather it is a change in the quadrupolar (and higher-order) pieces of the $1/R$ part of the TT projection of the asymptotic spatial metric (see Sec.~\ref{sec:hered} below).

Since the nonlinear memory occurs in any system that radiates GWs, systems that are usually considered to have vanishing linear memory (such as bound binaries) have a nonvanishing nonlinear memory. Thorne \cite{kipmemory} has shown that the nonlinear memory [Eq.~\eqref{eq:hTTnonlinmem}] can be described by his formula for the linear memory [Eq.~\eqref{eq:hijlinmem}] if the unbound objects in the system are taken to be the individual gravitons with energies $E_A=M_A/(1-v_A^2)^{1/2}$ and velocities $v^j_A = c \, n_A'^j$. The correspondence between these two equations also holds for null sources that contribute only to the linear memory. For example, the linear memory from a massless neutrino (or any other null particle) can be described either as a discrete sum over the memory from each individual particle [Eq.~\eqref{eq:hijlinmem}], or by replacing the GW energy flux in Eq.~\eqref{eq:hTTnonlinmem} with the energy flux of neutrinos \cite{epstein-neutrinomemory}.

The leading-order PN expansion of Eq.~\eqref{eq:hTTnonlinmem} is proportional to [see Eqs.~\eqref{eq:hijTT}, \eqref{eq:Uij}, and \eqref{eq:canonical}]
\be
\label{eq:hTTmempropto}
\delta h^{\rm TT}_{jk} \propto \frac{1}{R} \left[ \int_{-\infty}^{T_R} {\mathcal I}_{aj}^{(3)}(\tau) {\mathcal I}_{ka}^{(3)}(\tau) d\tau  \right]^{\rm TT},
\ee
where $ {\mathcal I}_{ij}^{(3)}$ is the third-time derivative of the source-quadrupole moment. If we specialize to quasicircular, inspiralling binaries with total mass $M=m_1+m_2$, reduced mass ratio $\eta=m_1 m_2/M^2$, and angular orbital frequency $\omega(t)$, the integrand of Eq.~\eqref{eq:hTTmempropto} contains oscillatory terms proportional to $\eta^2 (M\omega)^{10/3} e^{\pm 4i\omega t}$ and nonoscillatory terms proportional to  $\eta^2 (M\omega)^{10/3}$. When performing the time integration, the oscillatory terms are effectively multiplied by the \emph{orbital timescale} $T_{\omega}\propto 1/\omega$:
\bs
\label{eq:memintscaling}
\begin{align}
\label{eq:memintscaling1}
\delta h_{jk}^{\textrm{TT, osc}} &\propto \frac{1}{R} \int_{-\infty}^{T_R} dt' \, \eta^2 (M\omega)^{10/3} e^{\pm 4i\omega t'} \nonumber \\
&\propto \frac{\eta^2}{R} (M\omega)^{10/3} e^{\pm 4i\omega t} T_{\omega} \propto \frac{\eta^2 M}{R} (M\omega)^{7/3} e^{\pm 4i\omega t},
\end{align}
while the nonoscillatory terms are effectively multiplied by the \emph{radiation-reaction timescale} $T_{\rm rr} \propto (M/\eta) (M\omega)^{-8/3}$:
\begin{align}
\label{eq:memintscaling2}
\delta h_{jk}^{\textrm{TT, non-osc}} &\propto \frac{1}{R} \int_{-\infty}^{T_R} dt' \, \eta^2 (M\omega)^{10/3} \propto \frac{\eta^2}{R} (M\omega)^{10/3} T_{\rm rr} \nonumber
\\
&\propto \frac{\eta M}{R} (M\omega)^{2/3}.
\end{align}
\es
This shows that while the oscillatory pieces of $\delta h^{\rm TT}_{jk}$ are a 2.5PN correction to the waveform amplitudes, the nonoscillatory piece enters at the same order as the familiar quadrupole-order piece of the waveform:\footnote{This also follows from the approximate scaling of the nonlinear memory with the total radiated energy, $\delta h_{jk}^{\rm TT} \sim \Delta E^{\rm gw}/R$. Since the energy radiated during the inspiral is equal to the change in the orbital energy, $\Delta E^{\rm gw} \approx (\eta M/2)(M/r)$, the nonlinear memory has the same scaling as the quadrupolar waveform $h_{jk}^{\rm TT,(0)} \propto (\eta M/R)(M/r)$, where $r$ is the orbital separation.}
\be
\label{eq:quadapprox}
h_{jk}^{\rm TT,(0)} = \frac{2}{R} \ddot{\mathcal I}_{jk}^{\rm TT} \propto \frac{\eta M}{R} (M \omega)^{2/3} .
\ee
\subsection{\label{sec:motivation}Motivation}
The Christodoulou memory is a unique and interesting manifestation of the nonlinearity of general relativity. Analytic computations of gravitational radiation involve many types of nonlinearities, the origins of which are often obscured by the complex, iterative algorithms involved in solving the Einstein field equations. The Christodoulou memory, however, has a clear physical interpretation: It arises from the loss of GW energy from the system and the effect of this loss (through the GW stress-energy tensor) on the system's radiative mass-multipole moments. For comparison, GW tails are another interesting nonlinear, general-relativistic effect (see Sec.~3.4 of Ref.~\cite{blanchet-marck-lasota} for references). They arise from the last term in Eq.~\eqref{eq:EFE}, which modifies the flat-spacetime wave operator on the left-hand side and causes backscattering of the gravitational radiation as it propagates through the curved spacetime around the binary \cite{wiseman-tails}.  Tail effects enter the waveform at 1.5PN order.
The Christodoulou memory, on the other hand, arises from nonlinear interactions at 2.5PN order, but affects the gravitational waveform at leading (0PN) order. Both tails and the Christodoulou memory are \emph{hereditary}---their contribution to the GW field at any given retarded time depends on the entire past history of the source [see e.g., Eq.~\eqref{eq:Uij}]. However, the Christodoulou memory is especially sensitive to the motion of the source in the distant past (see Sec.~\ref{sec:NRmemory}), while tails are primarily sensitive to the recent past (see also Sec.~4 of Ref.~\cite{arun25PNamp}). Tails and most other nonlinear effects that enter the waveform cause oscillatory corrections to both the $+$ and $\times$ polarizations. In contrast, the Christodoulou memory causes a nonoscillatory shift in the amplitude of the $+$ polarization only\footnote{This statement is only true for standard choices of the polarization triad (see Sec.~\ref{sec:polmodes}). Since a rotation of this triad by an angle $\Psi$ about the propagation direction transforms the GW polarizations via $h_+ - i h_{\times} \rightarrow (h_+ - i h_{\times}) e^{2i \Psi}$, a purely $+$ polarized wave can become mixed-polarized.}. This amplitude shift starts small at early times (when the binary is widely separated) and slowly grows during the inspiral. As the binary components merge the memory rapidly grows and then saturates to a final value during the ringdown phase. The details of how the memory reaches its saturation value are explored in Refs.~\cite{favata-memory-saturation,favata-lisa7confproc}. Unlike tails, the nonoscillatory pieces of the memory do not affect the orbital phase of a quasicircular binary up to 3.5PN order (the highest order to which the phase has been computed), but they could modify the phase at higher PN orders\footnote{For example (and using notation defined later), the GW luminosity ${\mathcal L}=-\dot{E}\propto \sum_{lm} |\dot{h}_{lm}|^2$ will have a term proportional to $|\dot{h}_{20}|^2$ which will affect the luminosity and hence the phase at relative 5PN order (see Sec.~\ref{sec:selecrules} below for scalings). However, one should note that the \emph{oscillatory} terms that arise from the memory integral [cf.~Eq.~\eqref{eq:memintscaling1} and associated discussion] affect the phasing beginning at relative 2.5PN order.}.

Over the past three decades the PN corrections to the oscillatory pieces of the GW polarizations have been computed \cite{wagoner-will,blanchet-iyer-will-wiseman-CQG-2PNwaveform,arun25PNamp,kidder-blanchet-iyer-25PNwaveform}, most recently at the 3PN order \cite{blanchet3pnwaveform}. These computations were motivated by the development---and finally the operation---of a global network of ground-based GW interferometric detectors \cite{ligoweb,geoweb,virgoweb,tamaweb}.  These detectors are primarily sensitive to the oscillatory components of the GW signal, as most of the signal power lies in the lowest-order oscillatory modes of the radiation (for nearly circular orbits). More recently, successes in numerical relativity (NR) \cite{pretorius-PRL2005,pretorius-CQG2006,pretorius-BBHreview,baker-etal-PRL2006,baker-etal-PRD2006,campanelli-etal-PRL2006,campanelli-etal-PRD2006,herrmann-etal-CQG2007,sperhake-PRD2007,scheel-etal-PRD2006,koppitz-etal-PRL2007,pollney-etal-spinorbitrecoil,scheel-merger,samurai09,ninja09} have necessitated the need for accurate PN waveforms to compare with the results of binary black-hole (BH) merger simulations \cite{buonanno-cook-pretorius,boyle-etal-PRD2007,mroue-kidder-saul-PRD2008,boyle-etal-Efluxcomparison,baker-etal-PRL2007-NRPN,baker-etal-PRD2007-NRPN,hinder-etal-eccentricPN-NR,campanelli-etal-spinning-NR-PN-compare,hannam-etal-PN-NR-meet,berti-etal-multipolarnonspinning,berti-etal-mulitpolarspinning,hannam-gopa-NRPN-spin,damour-nagar-jena,damour-nagar-AEI,damour-nagar-caltechcornell,damour-nagar-PRD09,gopa-jena-eccentricPNNR,pan-buonanno-baker-etal-NRPN,buonanno-pan-baker-etal-nonspinningEOB,buonanno-caltechEOB09}.

Computations of the nonlinear memory's contribution to the waveform have not progressed as far.
Wiseman and Will \cite{wiseman-will-memory} first calculated the nonlinear memory's leading-order effect on the waveform polarizations for a quasicircular, inspiralling binary. Their result was confirmed\footnote{The factor of $2$ discrepancy in the memory waveform computed in Ref.~\cite{wiseman-will-memory} is either a typo or due to a different choice of normalization for the polarization tensors in Eq.~\eqref{eq:epluscross} below. Note that the relative amplitudes shown in Fig.~1 of Ref.~\cite{wiseman-will-memory} have the correct values at large separation.} in Refs.~\cite{kennefick-memory,arun25PNamp,blanchet3pnwaveform}. The 0.5PN corrections to the memory were calculated in Ref.~\cite{blanchet3pnwaveform} and found to vanish.  The primary purpose of this paper is to compute the corrections to the leading-order formula for the nonlinear memory to 3PN order.

Why are these corrections needed? First, while GW interferometers are mostly sensitive to the oscillatory parts of the GW, the memory piece of the signal could be detectable for certain sources. During a signal's observation time, the nonlinear memory causes a growing change in the signal's amplitude. This change contributes power at low frequencies. For many sources this power is swamped by the detector's low-frequency noise. But for sources with large signal-to-noise ratios---especially supermassive black-hole binaries in the low-frequency LISA \cite{lisaweb} band---the memory can be detectable. For example, the memory from the merger of two $10^6 M_{\odot}$ black holes should be detectable by LISA out to a redshift of $z\approx 2$ (see Fig.~3 of Ref.~\cite{favata-memory-saturation}).  The detectability of the nonlinear memory was previously considered by Thorne \cite{kipmemory} and Kennefick \cite{kennefick-memory} and is discussed further in Sec.~\ref{sec:memdetec} below and in Refs.~\cite{favata-memory-saturation,favata-lisa7confproc}. Central to the issue of detecting the memory is knowing its magnitude, and computing higher-PN corrections to the memory will help to determine this more accurately.

Second, the comparison of PN waveforms with numerical relativity simulations could also benefit from more accurate expressions for the memory. While numerical relativity accounts for all of the nonlinear effects of general relativity, it is difficult to compute the memory accurately in these simulations: The modes of the waveform that have memory are very small and depend sensitively on the initial separation of the binary (see Sec.~\ref{sec:NRmemory} for details). The PN corrections to the memory computed here could supply initial conditions for the waveform modes in the numerical simulations.
As has been done with the oscillatory pieces of the waveform \cite{buonanno-cook-pretorius,boyle-etal-PRD2007,mroue-kidder-saul-PRD2008,boyle-etal-Efluxcomparison,baker-etal-PRL2007-NRPN,baker-etal-PRD2007-NRPN,hinder-etal-eccentricPN-NR,campanelli-etal-spinning-NR-PN-compare,hannam-etal-PN-NR-meet,berti-etal-multipolarnonspinning,berti-etal-mulitpolarspinning,hannam-gopa-NRPN-spin,damour-nagar-jena,damour-nagar-AEI,damour-nagar-caltechcornell}), it will also be insightful to compare the memory calculated in future simulations with the PN expressions computed here.

Lastly, it is aesthetically pleasing to have all of the pieces---both oscillatory and nonoscillatory---of the waveform amplitudes computed consistently to the same post-Newtonian order. Combined with the results of Ref.~\cite{blanchet3pnwaveform}, this work completes the waveform polarization amplitudes to 3PN order.
\subsection{\label{sec:summary}Summary}
The bulk of this paper is devoted to calculating the nonlinear memory's contribution to the $+$ waveform polarization and its PN corrections. Readers uninterested in the details should skip to the results presented in Sec.~\ref{sec:results}. Readers uninterested in lengthy formulas can skip directly to Figs.~\ref{fig:H-theta} and \ref{fig:x-hlm}, the discussion (Sec.~\ref{sec:discussion}), and conclusions (Sec.~\ref{sec:conclusion}). An outline of the paper and a summary of its main results are presented here:

The necessary PN wave-generation formalism is briefly reviewed in Sec.~\ref{sec:pnformalism}. Section \ref{sec:polmodes} gives expressions for the waveform polarizations $h_{+,\times}$ in terms of the radiative-multipole moments. Expressions for the spin-weighted spherical-harmonic mode decomposition of the waveform in terms of the ``scalar'' radiative-multipole moments are also introduced.
Post-Newtonian calculations of the radiative-multipole moments are usually performed entirely in terms of symmetric-trace-free (STF) tensors, and are then (if needed) decomposed into spherical-harmonic modes at the end of the calculation. However, the computation of the memory is significantly easier if one works almost entirely from the start in terms of ``scalar'' quantities that are the coefficients of tensor quantities decomposed on the basis of scalar or spin-weighted spherical harmonics.
Sections \ref{sec:MPM} and \ref{sec:canonicalmoments} review the multipolar-post-Minkowskian (MPM) formalism which relates the radiative-multipole moments ${\mathcal U}_L$ and ${\mathcal V}_L$ to the canonical moments ${\mathcal M}_L$ and ${\mathcal S}_L$ and the source moments ${\mathcal I}_L$ and ${\mathcal J}_L$.
Section \ref{sec:hered} focuses on the hereditary contributions to the radiative multipoles: tails and memory. The results of Blanchet and Damour \cite{blanchet-damour-hereditary}, which serve as the starting point of the memory calculation, are reviewed. The main result of this section is Eq.~\eqref{eq:Ulmmem}, which expresses the memory piece of the radiative mass-multipole moments in terms of time and angular integrals over the GW energy flux from the source.

The main part of the memory calculation---explicitly evaluating the nonlinear memory contributions to the radiative-mass multipoles for quasicircular binaries---is detailed in Sec.~\ref{sec:evalUmem}. The computation of the angular integrals is discussed in Sec.~\ref{sec:angularint}. Section \ref{sec:selecrules} discusses which $(l,m)$ modes need to be calculated to determine the memory contribution to the waveform to 3PN order. In Sec.~\ref{sec:timeintegrals} we discuss how the time integral over the past history of the source is computed. In particular we need to use a model for the adiabatic inspiral that is accurate to at least 3PN order. This model is contained in Eq.~\eqref{eq:xdot35pn}, which gives the time evolution of the PN parameter $x\equiv (M\omega)^{2/3}$ to 3.5PN order. This formula (which easily follows from the results of Refs.~\cite{blanchet35PNphase,blanchet35PNphaseerratum}) expresses the well-known result for the 3.5PN frequency evolution in a form that is useful for computations involving the PN parameter $x$.

The main results of this work are listed in Sec.~\ref{sec:results}: the 3PN memory contributions to the $+$ polarization $h_{+}$ of the gravitational waveform and its spin-weighted spherical-harmonic modes $h_{lm}$. Those expressions can be directly combined with the oscillatory pieces of the waveform given to 3PN order in Ref.~\cite{blanchet3pnwaveform}. Figure \ref{fig:H-theta} illustrates the memory's angular dependence and the relative sizes of the various PN corrections. The PN corrections have little effect on the memory's angular dependence, but tend to decrease the overall magnitude of the memory.

Aside from the Christodoulou memory, there are additional nonoscillatory (DC or ``direct current'') time-varying terms present in the waveform. Section \ref{sec:crosmem} discusses a new type of \emph{nonlinear, nonhereditary} DC term discovered by Arun et al.~\cite{arun25PNamp}. Unlike the Christodoulou memory, this nonlinear, nonhereditary DC term affects the $\times$ polarization and first enters the waveform at 2.5PN order. Section \ref{sec:linmem} discusses an additional class of \emph{linear, nonhereditary} DC terms that enter the waveform at 5PN and higher orders. These terms arise from the effects of radiation-reaction or nongravitational forces on the source-multipole moments.

Section \ref{sec:NRmemory} discusses the challenges in extracting the nonlinear memory from numerical simulations of binary black holes. It will be difficult for current simulations to accurately extract the memory waveform. One of the reasons for this is illustrated in Fig.~\ref{fig:x-hlm}. Numerical relativity simulations can most accurately compute the $(l,m)=(2,2)$ mode of the waveform; but the nonlinear memory is present only in the $m=0$ modes (for binaries orbiting in the $x$-$y$ plane). For simulations that can directly compute the metric-perturbation modes $h_{lm}$, the largest memory mode $h_{20}$ is an order-of-magnitude smaller than the $h_{22}$ mode. For the majority of simulations that compute the spin-weighted spherical-harmonic modes of the $\Psi_4$ Weyl scalar, $\psi_{lm}=\ddot{h}_{lm}$, the situation is significantly worse: The largest memory mode $\psi_{20}$ is nearly \emph{three orders of magnitude} smaller than the dominant $\psi_{22}$ mode in the late inspiral. There are also difficulties in determining the two integration constants needed when computing the $h_{lm}$ modes from the $\psi_{lm}$ modes. The memory's strong dependence on the past history of the source also introduces significant errors unless the simulations start with very large binary separations. The results of this paper could help to alleviate some of these difficulties by providing initial conditions for the $m=0$ modes.

Section \ref{sec:memdetec} discusses the detection of the memory. While the memory from a signal in the distant past is unobservable, it is possible to detect the buildup of the memory from a passing GW. Previous work by Thorne \cite{kipmemory} and Kennefick \cite{kennefick-memory} has estimated the nonlinear memory's signal-to-noise ratio in laser interferometers. However, the signal-to-noise ratio is sensitive to the details of how the memory rises to its saturation value, which previous works have not properly modeled.

Conclusions and suggestions for further work are presented in Sec.~\ref{sec:conclusion}.
Some results relegated to the appendices are potentially useful in other applications: Appendix \ref{app:angularint} gives an explicit formula for the angular integral of the product of three spin-weighted spherical harmonics.  Appendix \ref{app:psilm} gives an explicit prescription for calculating the Weyl scalar modes $\psi_{lm}$ from the known expressions for $h_{lm}$ presented here and in Ref.~\cite{blanchet3pnwaveform}.
\subsection{\label{sec:notation}Notation}
This paper borrows notations and conventions from Kidder \cite{kidder08}, Thorne \cite{kiprmp}, and Blanchet, et al.~\cite{blanchet3pnwaveform}.
We generally set $G=c=1$, except in certain equations where we wish to make explicit the post-Newtonian or post-Minkowskian order of the various terms. We use the notation $O(n)$ to denote post-Newtonian (PN) correction terms of order $O(c^{-n})$.
Spacetime indices are denoted with Greek letters, spatial indices with Latin letters.  In Sec.~\ref{sec:MPM} we use Blanchet's \cite{blanchetLRR} notation for the gravitational field, $h^{\alpha \beta} = - \bar{h}^{\alpha \beta}$ (see footnote \ref{ftnt:sign-h}). Multi-index notation is denoted with a capital subscript: $A_L = A_{i_1 i_2 \ldots i_l}$. A multi-index $L$ on a vector denotes a product of $l$ vectors: $x_L = x_{i_1} x_{i_2} \ldots x_{i_l}$. Repeated spatial indices and multi-indices are summed regardless of their relative positions. Symmetric-trace-free (STF) spatial tensors are denoted with capital script letters  (as in ${\mathcal U}_L$). The corresponding ``scalar'' versions of these moments [their coefficients on the basis of the STF spherical-harmonic tensors ${\mathcal Y}_L^{lm}$ (defined below)] are denoted with nonscript capital letters (as in $U_{lm}$). Spherical-harmonic indices $(l,m)$ are raised or lowered arbitrarily (i.e.,~$U_{lm} = U^{lm})$.  Symmetrization and STF projection are denoted by enclosing the relevant indices by $()$ or $<>$, respectively. Indices that are left out of the symmetrization or STF projection are displayed with an underbar: $A_{<i}B_{\underline{a}j>}$. Time derivatives are denoted by an overdot or by superscript parenthesis: ${\mathcal M}_L^{(p)}= d^p{\mathcal M}_L/{dT^p}$. To avoid confusion with the azimuthal harmonic index $m$, we denote the sum of the binary's component masses as $M\equiv m_1+m_2$. This is distinct from the total binary (ADM) mass or mass-monopole moment, which we denote ${\mathcal M}$.
Euler's constant is denoted by $\gamma_{E} = 0.577\,21\ldots$. The symbol $\phi$ denotes an azimuthal angle while $\varphi$ denotes an orbital phase.
\section{\label{sec:pnformalism}Post-Newtonian wave generation formalism}
\subsection{\label{sec:polmodes}Waveform polarizations and mode decomposition}
Following Kidder \cite{kidder08} we begin by introducing an asymptotically-flat radiative coordinate system described by spherical coordinates $(T, R, \Theta, \Phi)$ and their orthonormal basis vectors $(\vec{e}_T,\vec{e}_R,\vec{e}_{\Theta}, \vec{e}_{\Phi})$. The coordinate origin is the center-of-mass of the source. The retarded time in radiative coordinates is $T_R=T-R$.

The gravitational waveforms are usually expressed as the transverse-traceless (TT) piece of the asymptotic metric perturbation decomposed into a sum over radiative mass- and current-multipole moments:
\begin{multline}
\label{eq:hijTT}
h^{TT}_{ij} = \frac{4G}{c^2 R} \Pi_{ijkl} \sum_{l = 2}^\infty
\frac{1}{c^l l!} \left[ {\mathcal U}_{kl L-2}(T_R) N_{L-2}
\right. \\ \left. +
\frac{2 l}{c (l + 1)} \epsilon_{pq(k}
{\mathcal V}_{l)pL-2}(T_R)  N_{qL-2} \right] + O\left(\frac{1}{R^2}\right).
\end{multline}
Here $\Pi_{ijkl}$ is the TT projection operator
\begin{equation}
\label{eq:TT}
\Pi_{ijkl} = P_{ik} P_{jl} - \frac{1}{2} P_{ij} P_{kl},
\end{equation}
where $P_{ij} = \delta_{ij} - N_i N_j$. The radiative mass- and current-multipole moments ${\mathcal U}_L(T_R)$ and ${\mathcal V}_L(T_R)$ are STF-tensors, and, as discussed below, are related to the source-multipole moments which are defined in the near zone as integrals over the source. Equation \eqref{eq:hijTT} represents the most general outgoing-wave, transverse trace-free solution of the vacuum-wave equation $\Box h^{\rm TT}_{ij}=0$ \cite{kiprmp}.

The plus and cross GW polarizations are related to the TT piece of the asymptotic metric via $h_{+,\times}=h_{ij}^{\rm TT} e^{+,\times}_{ij}$, where the TT-polarization tensors are defined in terms of a chosen orthonormal triad $(\vec{N}, \vec{P}, \vec{Q})$:
\bs
\label{eq:epluscross}
\begin{align}
e_{ij}^{+} &= \frac{1}{2} (P_i P_j - Q_i Q_j) , \\
e_{ij}^{\times} &= \frac{1}{2} (P_i Q_j + P_j Q_i) .
\end{align}
\es
The unit vector $\vec{N}$ points from the source to the observer. While in some calculations it is convenient to choose this direction to lie along the $z$ axis $\vec{e}_Z$, here it points in a general direction specified by the spherical polar angles $(\Theta, \Phi)$. We choose the remaining vectors in accordance with Kidder's \cite{kidder08} notation: $\vec{P}=\vec{e}_{\Theta}$ and $\vec{Q}=\vec{e}_{\Phi}$.\footnote{This choice differs from that chosen by Blanchet and collaborators \cite{blanchet3pnwaveform,arun25PNamp}. The Kidder \cite{kidder08} and Blanchet \cite{blanchet3pnwaveform} polarizations are related by $h_{+,\times}^{\rm B} = - h_{+,\times}^{\rm K}(\Theta=\iota,\Phi=\pi/2)$. There is also a difference in the overall sign of $h_{lm}$ between the two conventions. However the definitions of the radiative-multipole moments and $h_{ij}^{\rm TT}$ are the same in both conventions.} Note that our choice of normalization implies
\be
\label{eq:epolnorm}
e_{ij}^{A} e_{ij}^{B} = \frac{\delta^{AB}}{2} , \; \text{for } A, B = + \text{ or } \times , \;\;\;\; \text{and}
\ee
\be
\label{eq:hTTijepol}
h_{ij}^{\rm TT} = 2( h_+ e^{+}_{ij} + h_{\times} e^{\times}_{ij} ) .
\ee

While waveform calculations usually involve computing the STF radiative multipoles ${\mathcal U}_L$ and ${\mathcal V}_L$ and then using Eqs.~\eqref{eq:hijTT}--\eqref{eq:epluscross} to compute the $+$ and $\times$ polarizations, it can be more convenient to compute the polarizations by performing a mode decomposition of the combination $h_+ - i h_{\times}$ via
\be
\label{eq:hdecompose}
h_{+} - i h_{\times} = \sum_{l=2}^{\infty} \sum_{m=-l}^{l} h^{lm} {}_{-2}Y^{lm}(\Theta,\Phi) \;,
\ee
where
\be
\label{eq:hlm}
h^{lm} = \frac{G}{\sqrt{2} R c^{l+2}} \left[ U^{lm}(T_R) - \frac{i}{c} V^{lm}(T_R) \right] .
\ee
Here the ``scalar'' mass and current multipoles are related to their STF counterparts by [Thorne \cite{kiprmp}, Eq.~(4.7)]
\bs
\label{eq:UVlmdef}
\begin{align}
\label{eq:Ulmdef}
U^{lm} &= A_l \, {\mathcal U}_L {\mathcal Y}^{lm \, \ast}_L , \\
\label{eq:Vlmdef}
V^{lm} &= B_l \, {\mathcal V}_L {\mathcal Y}^{lm \, \ast}_L , \qquad \text{where}
\end{align}
\begin{align}
\label{eq:Al}
A_l &=\frac{16 \pi}{(2l+1)!!} \sqrt{\frac{(l+1)(l+2)}{2l(l-1)}} , \\
\label{eq:Bl}
B_l &=-\frac{32 \pi l}{(2l+1)!!} \sqrt{\frac{(l+2)}{2l(l+1)(l-1)}} ,
\end{align}
\es
and $\ast$ denotes complex conjugation.

The complex conjugates of these moments satisfy [Eq.~(4.5) of Ref.~\cite{kiprmp}]:
\be
\label{eq:UVcc}
U^{lm \ast}= (-1)^m U^{l\,-m} \,, \;\;\; V^{lm \ast} = (-1)^m V^{l\,-m}.
\ee
The inverse relations of Eqs.~\eqref{eq:UVlmdef} are given by
\bs
\label{eq:UVstfdef}
\begin{align}
\label{eq:Ustfdef}
{\mathcal U}_L &= \frac{l!}{4} \sqrt{\frac{2 l (l-1)}{(l+1)(l+2)}} \sum_{m=-l}^{l} U^{lm} {\mathcal Y}^{lm}_L , \\
\label{eq:Vstfdef}
{\mathcal V}_L &= -\frac{(l+1)!}{8l} \sqrt{\frac{2 l (l-1)}{(l+1)(l+2)}} \sum_{m=-l}^{l} V^{lm} {\mathcal Y}^{lm}_L .
\end{align}
\es
These relationships also hold for the other mass-type and current-type multipole moments that are used elsewhere in this paper. The ${\mathcal Y}^{lm}_L$ are the STF spherical harmonics and are related to the familiar scalar spherical harmonics by
\be
\label{eq:Ylm}
Y^{lm} = {\mathcal Y}^{lm}_L n_L = {\mathcal Y}^{lm}_L n_{<L>} ,
\ee
where $n_i$ is a general unit radial vector. They are given explicitly by Eq.~(2.12) of Thorne \cite{kiprmp} or Eq.~(39) of Iyer \cite{iyermultipolenotes}. We also note that for nonspinning binaries the $h^{lm}$ satisfy [Eq.~(78) of Ref.~\cite{kidder08}]
\be
h^{l -m} = (-1)^l h^{lm\ast}.
\ee

The spin-weighted spherical harmonics are defined in terms of the
Wigner $d$ functions by
\begin{equation}
\label{eq:Yslm}
 {}_{-s}Y^{lm}(\Theta,\Phi) = (-1)^s \sqrt{\frac{2l + 1}{4 \pi}}
d^l_{ms}(\Theta) e^{i m \Phi}.
\end{equation}
Here
\begin{multline}
\label{eq:dlms}
 d^l_{ms}(\Theta) = \sqrt{(l + m)! (l - m)! (l + s)! (l - s)!}
 \\ \times
\sum_{k = k_i}^{k_f}
 \frac{(-1)^k (\sin{\frac{\Theta}{2}})^{2 k + s - m}
(\cos{\frac{\Theta}{2}})^{2 l + m -s - 2 k}}{ k! (l + m - k)!
(l - s - k)! (s - m + k)!},
\end{multline}
where $k_i$ = max$(0,m-s)$ and $k_f$ = min$(l + m,l -s)$. The complex conjugates of the spin-weighted spherical harmonics satisfy
\be
\label{eq:Yslmcc}
{}_{s}Y^{lm \ast} = (-1)^{s+m} {}_{-s}Y^{l -m} .
\ee

Working with the ``scalar'' multipole moments $U^{lm}$ and $V^{lm}$ can be more convenient because it allows us to work with $2l+1$ scalars instead of the $2l+1$ independent components of an $l$-index STF tensor. Using the ``scalar'' moments also allows angular integrals over the products of unit vectors to be expressed in terms of products of spin-weighted spherical harmonics, which are more easily evaluated using computer algebra programs.
\subsection{\label{sec:MPM}The multipolar post-Minkowskian formalism I: relating the radiative and canonical moments}
One of the purposes of a gravitational wave-generation formalism is to relate the \emph{radiative}-multipole moments ${\mathcal U}_L$ and ${\mathcal V}_L$ that appear in the wavezone expansion of $h^{\rm TT}_{ij}$ [Eq.~\eqref{eq:hijTT}] to some other family of multipole moments---the \emph{source}-multipole moments---which are defined in terms of integrals over the stress-energy pseudotensor of the matter and gravitational fields of the source.  One procedure for relating these families of multipole moments is the multipolar-post-Minkowskian (MPM) iteration scheme developed by Blanchet, Damour, Iyer, and collaborators. This method is briefly summarized here and reviewed in detail by Blanchet \cite{blanchetLRR} (see also Ref.~\cite{blanchet-marck-lasota} for a dated but much shorter review). Other PN wave-generation formalisms are discussed in Refs.~\cite{will-wiseman-2pn,pati-will-DIRE1,goldberger-rothstein-PRD2006}.

The first step in the MPM procedure is a post-Minkowskian (weak field) iteration of the Einstein field equations subject to the harmonic gauge condition [Eqs.~\eqref{eq:EFEeqs}]. This involves expanding the metric deviation $h^{\alpha \beta} \equiv \sqrt{-g} g^{\alpha \beta} - \eta^{\alpha \beta} = -\bar{h}^{\alpha \beta}$ in powers of the gravitation constant $G$:
\be
\label{eq:hexpansion}
h^{\alpha \beta} = G h_1^{\alpha \beta} + G^2 h_2^{\alpha \beta} + \cdots + G^n h_n^{\alpha \beta} + \cdots
\ee
and substituting into the vacuum Einstein equations and the harmonic gauge condition, resulting in a system of wave equations,
\be
\label{eq:hnMPM}
\Box h_n^{\alpha \beta} = \Lambda_n^{\alpha \beta}[h_1, \ldots h_{n-1}]\,; \;\;\;\;\; \partial_{\beta} h_n^{\alpha \beta} = 0 \;.
\ee
Here $\Lambda_n^{\alpha \beta}$ represents the appropriate expansion of the right-hand side of Eq.~\eqref{eq:EFE} with $T^{\alpha \beta}=0$.

The next step consists of performing a multipolar expansion of the $h_n^{\alpha \beta}$---an expansion in $L/r$ where $L<r$ is the size of the source and $r$ is the field point. The coefficients of the powers of $L/r$ can be expressed in terms of a new family of multipole moments.
At linear order Thorne \cite{kiprmp} has shown that the most general solution for $h_1^{\alpha \beta}$ (valid outside the source and up to an infinitesimal gauge transformation that preserves the harmonic gauge condition) is given by a multipole expansion that depends on only two types of STF moments, ${\mathcal M}_L$ and  ${\mathcal S}_L$. These moments are referred to as the \emph{canonical} or \emph{algorithmic}  mass- and current-multipole moments. They represent an intermediate family of moments in between the radiative and source moments.

Starting with this linear solution, each higher-order solution is generated by substituting the lower-order pieces into the right-hand side of Eq.~\eqref{eq:hnMPM} and solving the resulting wave equation. This results in solutions at each order that take the form of a multipole expansion depending on the canonical moments, $h_n^{\alpha \beta} = h_n^{\alpha \beta}[{\mathcal M}_L,{\mathcal S}_L]$.

Because of the singularity at $r=0$ in the multipole expansion, the ordinary retarded Green's function operator $\Box^{-1}_{\rm ret}$ yields divergent integrals. Instead one has to use a regularization procedure that consists of multiplying $\Lambda_n^{\alpha \beta}$ by a factor $(r/r_0)^B$, where $r_0$ is an arbitrary constant length scale and $B$ is a complex number. Applying the $\Box^{-1}_{\rm ret}$ operator, and taking the finite part of the Laurent series expansion about $B\rightarrow 0$ yields the following solution to the wave equation at each order $n$ (see Ref.~\cite{blanchetLRR} for details):
\be
\label{eq:ualphabeta}
u_n^{\alpha \beta} = \mathop{\mathrm{FP}}_{B=0}\,
\Box^{-1}_{\rm ret} \left[ \left(\frac{r}{r_0}\right)^B \Lambda_n^{\alpha \beta} \right] \,.
\ee
In order to satisfy the harmonic gauge condition as well as the wave equation, an additional piece $v_n^{\alpha \beta}$, which is constructed from the divergence of $u_n^{\alpha \beta}$ and is a homogeneous solution of the wave equation, must be added to Eq.~\eqref{eq:ualphabeta} to yield the full solution to Eq.~\eqref{eq:hnMPM} at order $n$:
\be
\label{eq:hnsoln}
h_n^{\alpha \beta} = u_n^{\alpha \beta} + v_n^{\alpha \beta} \,.
\ee
This result is then transformed from harmonic coordinates $(ct,x^i)$ to radiative coordinates $(cT,X^i)$ via the transformation
\bs
\label{eq:coordtrans}
\begin{align}
T_R &= t - \frac{r}{c} - \frac{2G{\mathcal M}}{c^3} \ln\left(\frac{r}{r_0}\right) + O(G^2) , \\
X^i &= x^i + O(G^2) ,
\end{align}
\es
where $r_0$ is another arbitrary length scale which is usually taken to have the same value as the $r_0$ in Eq.~\eqref{eq:ualphabeta}, and ${\mathcal M}$ is the mass-monopole moment\footnote{For quasicircular binaries this is related to the sum of the point-particle masses $M$ by Eq.~(1) of Ref.~\cite{blanchet-tailsoftails-erratrum}:
\[ {\mathcal M} = M \left[ 1 -\frac{\eta}{2} x + \frac{\eta}{24} x^2 (9 +\eta) + O(c^{-6}) \right], \] where we have used Eq.~(6.6) of Ref.~\cite{blanchet3pnwaveform} to express the result in terms of the PN parameter $x\equiv (M\omega)^{2/3}$ for orbital angular frequency $\omega$.}.

Taking the TT piece of the result and comparing with Eq.~\eqref{eq:hijTT} allows one to read off the relations between the radiative- and canonical-multipole moments. The result of this procedure is listed (to the highest PN order yet completed) in Eqs.~(5.4)-(5.8) of Ref.~\cite{blanchet3pnwaveform}. For illustration we show here the result for the radiative mass quadrupole:
\pagebreak
\begin{widetext}
\begin{multline}
\label{eq:Uij}
{\mathcal U}_{ij}(T_R) = {\mathcal M}_{ij}^{(2)}(T_R) + \frac{2G{\mathcal M}}{c^3} \int_{-\infty}^{T_R} \left[ \ln\left(\frac{T_R - \tau}{2\tau_0}\right) + \frac{11}{12} \right] {\mathcal M}_{ij}^{(4)}(\tau) \, d\tau  -\frac{2}{7} \frac{G}{c^5} \int_{-\infty}^{T_R} {\mathcal M}_{a<i}^{(3)}(\tau) {\mathcal M}_{j>a}^{(3)}(\tau) \, d\tau \\
+ \frac{G}{c^5} \left[ \frac{1}{7} {\mathcal M}_{a<i}^{(5)} {\mathcal M}_{j>a}^{} -\frac{5}{7} {\mathcal M}_{a<i}^{(4)} {\mathcal M}_{j>a}^{(1)} - \frac{2}{7} {\mathcal M}_{a<i}^{(3)} {\mathcal M}_{j>a}^{(2)} + \frac{1}{3} \epsilon_{ab<i}{\mathcal M}_{j>a}^{(4)} {\mathcal S}_b \right] \\
+ 2 \frac{G^2 {\mathcal M}^2}{c^6} \int_{-\infty}^{T_R} \left[ \ln^2\left(\frac{T_R-\tau}{2\tau_0}\right) + \frac{57}{70} \ln\left(\frac{T_R-\tau}{2\tau_0}\right) + \frac{124\,627}{44\,100} \right] {\mathcal M}_{ij}^{(5)}(\tau) \, d\tau + O\left(\frac{1}{c^7}\right) ,
\end{multline}
\end{widetext}
where the constant $\tau_0=r_0/c$.

Let us examine the various types of terms that appear in Eq.~\eqref{eq:Uij}: First is the leading-order \emph{instantaneous} term that depends directly on the retarded configuration of the source. This term is familiar from the standard quadrupole formalism and enters at leading order in $G$ [although, as shown below, this term contains $O(G)$ corrections when the ${\mathcal M}_{ij}$ are expressed in terms of the source moments].  The remaining terms on the first and second lines are the nonlinear correction terms that enter at $O(G^2)$ in the MPM iteration scheme. The first of these is the leading-order tail term, which affects the waveform at 1.5PN and higher orders.\footnote{Note however that the $11/12$ term in the integral is actually a nonlinear instantaneous term proportional to ${\mathcal M} {\mathcal M}_{ij}^{(3)}(T_R)$. The same is also true for the analogous term in the tail-of-tails integral.} Tails arise from the scattering of GWs off of the monopole moment ${\mathcal M}$ of the source \cite{blanchet-damour-hereditary}.  The next term is the leading-order nonlinear memory, which is the primary focus of this paper. Because of the integral over the infinite-past history of the source, both the tail and memory are called \emph{hereditary} terms \cite{blanchet-damour-hereditary}. On the second line we have nonlinear instantaneous terms that affect the waveform at 2.5PN and higher orders \cite{blanchet-quadquad}. The last term is the cubically-nonlinear tail-of-tails term which affects the waveform at 3PN and higher orders \cite{blanchet-tailsoftails,blanchet-tailsoftails-erratrum}. The other radiative mass- and current-multipole moments have analogous correction terms \cite{blanchet3pnwaveform}. For the rest of this paper we will primarily concern ourselves with the terms on the first line, and, in particular, the nonlinear memory term and its corrections (see Sec.~\ref{sec:hered} below).
\subsection{\label{sec:canonicalmoments}The multipolar post-Minkowskian formalism II: relating the canonical and source moments}
The canonical moments ${\mathcal M}_L$ and ${\mathcal S}_L$ do not have simple closed-form expressions in terms of integrals over the source \cite{blanchetLRR}. Instead they serve as intermediate moments that are related to a family of six types of \emph{source}-multipole moments: $\{ {\mathcal I}_L, {\mathcal J}_L, {\mathcal W}_L, {\mathcal X}_L, {\mathcal Y}_L, {\mathcal Z}_L \}$. The mass- and current-source moments ${\mathcal I}_L$ and ${\mathcal J}_L$ tend to dominate over the four remaining moments which enter as 2.5PN corrections. Expressions for all of the source moments as explicit integrals over the stress-energy pseudotensor [$\tau^{\alpha \beta} \equiv (16\pi)^{-1} \times$ the terms on the right-hand side of Eq.~\eqref{eq:EFE}] are found in Eqs.~(85)-(90) of Ref.~\cite{blanchetLRR}.

The procedure for relating the canonical and source moments is discussed in Ref.~\cite{blanchet3pnwaveform}. To summarize, it consists of performing the same MPM iteration discussed above, except the multipole expansion of the metric deviation is in terms of the source moments instead of the canonical moments. The result is a solution $h_n^{\alpha \beta}[{\mathcal I}_L,  \ldots {\mathcal Z}_L]$ at each post-Minkowskian order that is related to the canonical moment solution $h_n^{\alpha \beta}[{\mathcal M}_L, {\mathcal S}_L]$ by a gauge transformation. This relationship between the two metrics can be translated into a relationship between the canonical and source moments:
\bs
\label{eq:canonical}
\begin{align}
{\mathcal M}_L &= {\mathcal I}_L + G \delta{\mathcal I}_L + O(G^2) , \\
{\mathcal S}_L &= {\mathcal J}_L + G \delta{\mathcal J}_L + O(G^2) ,
\end{align}
\es
where the correction terms $\delta{\mathcal I}_L$ and $\delta{\mathcal J}_L$ are functions of the six source moments. These corrections modify the leading-order mass- and current-source moments at 2.5PN and higher orders and are given (up to 3PN order) by Eqs.~(5.9)-(5.11) of Ref.~\cite{blanchet3pnwaveform}.

The final step in the gravitational wave-generation formalism consists of matching the wave-zone MPM expansion of the metric in terms of the source moments with a post-Newtonian near-zone solution of the nonvacuum Einstein equations. This matching takes place in the region outside but close to the source where both approximation schemes are valid and yields an explicit relationship between the source moments and the PN expansion of the near-zone metric. Solving the PN equations of motion for the source and substituting back into the PN-expanded stress-energy pseudotensor $\tau^{\alpha \beta}$ (which depends on the metric and matter stress-energy tensor) yields explicit expressions for the source moments in terms of variables describing the source (see Ref.~\cite{blanchetLRR} and references therein for the many nontrivial details of this procedure). When specialized to quasicircular inspiralling compact binaries, the resulting expressions for the source moments are listed up to 3PN order in Eqs.~(5.12)-(5.25) of Ref.~\cite{blanchet3pnwaveform}.
\subsection{\label{sec:hered}Hereditary contributions to the radiative-multipole moments}
Blanchet and Damour \cite{blanchet-damour-hereditary} give general expressions for the leading post-Minkowskian order hereditary contributions (tail + memory) to the radiative mass- and current-multipole moments for arbitrary $l$:
\bs
\label{eq:UVLhered}
\begin{align}
\label{eq:ULhered}
{\mathcal U}_L &= {\mathcal M}_L^{(l)} + G {\mathcal U}_L^{\rm (tail)} + G {\mathcal U}_L^{\rm (mem)} + O(G^2) + O(G/c^5), \\
\label{eq:VLhered}
{\mathcal V}_L &= {\mathcal S}_L^{(l)} + G {\mathcal V}_L^{\rm (tail)} + O(G^2) + O(G/c^5) .
\end{align}
\es
The neglected terms at $O(G^2)$ are tail-of-tails like terms and other cubically nonlinear interactions. The neglected $O(G/c^5)$ terms are quadrupole-quadrupole nonlinearities [as in the second line of Eq.~\eqref{eq:Uij}] and other instantaneous ``canonical moment $\times$ canonical moment'' type terms.

Using updated notation and incorporating the leading-order \emph{synchronous} terms (those depending directly on events on the past null cone) into the tail integrals, the tail terms are given by [see Eq.~(98) of Ref.~\cite{blanchetLRR}]:
\bs
\label{eq:UVLtails}
\begin{align}
\label{eq:ULtail}
{\mathcal U}_L^{\rm (tail)} &= \frac{2G{\mathcal M}}{c^3} \int_{-\infty}^{T_R} \left[ \ln\left(\frac{T_R - \tau}{2\tau_0}\right) + \kappa_l \right] {\mathcal M}_L^{(l+2)}(\tau) \, d\tau , \\
\label{eq:VLtail}
{\mathcal V}_L^{\rm (tail)} &= \frac{2G{\mathcal M}}{c^3} \int_{-\infty}^{T_R} \left[ \ln\left(\frac{T_R - \tau}{2\tau_0}\right) + \pi_l \right] {\mathcal S}_L^{(l+2)}(\tau) \, d\tau ,
\end{align}
\es
where the constants $\kappa_l$ and $\pi_l$ are
\begin{equation}
\label{eq:kappapi}
\kappa_l = \frac{2l^2+5l+4}{l(l+1)(l+2)} + \sum_{k=1}^{l-2} \frac{1}{k} \,, \;\;\;\;\;\;\;
\pi_l = \frac{l-1}{l(l+1)} + \sum_{k=1}^{l-1} \frac{1}{k} . 
\end{equation}
The memory term only affects the mass moments at $O(G)$ and is given by [Eq.~(2.43c) of Ref.~\cite{blanchet-damour-hereditary}]:
\be
\label{eq:ULmem}
{\mathcal U}_L^{\rm (mem)} = \frac{2c^{l-2} (2l+1)!!}{(l+1)(l+2)} \int_{-\infty}^{T_R} dt \, \int d\Omega\, \frac{dE_{\rm gw}}{dt \, d\Omega} n_{<L>} \;.
\ee
Here $\frac{dE_{\rm gw}}{dt d\Omega}$ is the GW energy flux, $t$ is a dummy variable for the radiative coordinate time; $n_i$ is a general unit vector centered at the source that points in the direction of the spherical polar angles $(\theta,\phi)$ [distinct from the direction $N_i$ from the source to the observer and the corresponding angles $(\Theta,\Phi)$ appearing in the polarization waveforms]; and the angular integral is over the angles $(\theta,\phi)$. Equation \eqref{eq:ULmem} is the primary starting point for computing the nonlinear memory.

Since Ref.~\cite{blanchet-damour-hereditary}'s derivation of Eq.~\eqref{eq:ULmem} was completed to second-post-Minkowskian order, their expression for the energy flux in Eq.~\eqref{eq:ULmem} formally depended only on the metric at first-post-Minkowskian (linearized) order [see Eq.~(2.11b) of Ref.~\cite{blanchet-damour-hereditary} and associated derivation]. However it is clear from other derivations \cite{christodoulou-mem,wiseman-will-memory} that the memory depends on the full GW energy flux. So Ref.~\cite{blanchet-damour-hereditary}'s equation for the memory [their Eq.~(2.43c)] is naturally extended to higher post-Minkowskian orders by using the full energy flux to the highest-PN-order known.

Note that we are ignoring the ``linear memory'' contribution to ${\mathcal U}_L^{\rm (mem)}$ arising from changes in the derivatives of the canonical-mass moments ${\mathcal M}_L^{(l)}$.  We also note that changes in the derivatives of the canonical-current moments ${\mathcal S}_L^{(l)}$ lead to linear memory contributions to the radiative-current moments ${\mathcal V}_L$ \cite{blanchet-damour-hereditary}. The linear memory does not directly contribute to the Christodoulou memory and vanishes for quasicircular, inspiralling binaries that remain bound in the infinite past (but see Sec.~\ref{sec:linmem} below). For bound astrophysical binaries whose components were formed, captured, exchanged, or underwent mass loss long before the GW driven regime, the linear memory is negligible.
Note also that while there is no nonlinear, hereditary memory contribution to the radiative-current multipoles, there is a nonlinear, ``nonhereditary'' DC (nonoscillatory) effect that arises from the 1.5PN correction to ${\mathcal V}_{ijk}$; see the discussion in Sec.~\ref{sec:crosmem} below.

The GW energy flux can be computed from the GW stress-energy tensor and is given by \cite{kiprmp}
\begin{equation}
\label{eq:dEdtdOmega-def}
\frac{dE_{\rm gw}}{dt d\Omega} = R^2 T^{\rm gw}_{00} = \frac{R^2}{32\pi} \langle \dot{h}^{\rm TT}_{jk} \dot{h}^{\rm TT}_{jk} \rangle
=\frac{R^2}{16\pi} \langle \dot{h}_{+}^2 + \dot{h}_{\times}^2 \rangle,
\end{equation}
where the angled brackets mean to average over several wavelengths, and we have used Eq.~\eqref{eq:hTTijepol} to arrive at the last equality. Using Eq.~\eqref{eq:hdecompose} we can write the energy flux in terms of the $h_{lm}$ modes:
\begin{widetext}
\be
\label{eq:dEdtdOmega-hlm}
\frac{dE_{\rm gw}}{dt d\Omega} = \frac{R^2}{16\pi} \sum_{l'=2}^{\infty} \sum_{l''=2}^{\infty} \sum_{m'=-l'}^{l'} \sum_{m''=-l''}^{l''} \langle \dot{h}_{l'm'} \dot{h}^{\ast}_{l''m''} \rangle {}_{-2}Y^{l' m'}(\theta,\phi) {}_{-2}Y^{l'' m'' \,\ast}(\theta,\phi) .
\ee
Alternatively we can substitute Eqs.~\eqref{eq:hijTT} or \eqref{eq:hlm} into Eqs.~\eqref{eq:dEdtdOmega-def} or \eqref{eq:dEdtdOmega-hlm} to give the energy flux in terms of the radiative-multipole moments [Eq.~(4.14) of Thorne \cite{kiprmp}\footnote{To translate from Thorne's \cite{kiprmp} notation for the radiative-multipole moments to the notation used here and in Ref.~\cite{kidder08}, the following replacements are made in Thorne's formulas: ${}^{(l)}{\mathcal I}_{A_l} \rightarrow {\mathcal U}_L$, ${}^{(l)}{\mathcal S}_{A_l} \rightarrow {\mathcal V}_L$, ${}^{(l)}I_{lm} \rightarrow U_{lm}$, and ${}^{(l)}S_{lm} \rightarrow V_{lm}$. In this paper the symbols ${\mathcal I}_L$ and $I_{lm}$ refer to the source mass-multipole moments, while ${\mathcal S}_L$ and $S_{lm}$ refer to the canonical current-multipole moments.}]:
\begin{multline}
\label{eq:dEdtdOmegaSTF}
\frac{dE_{\rm gw}}{dt \, d\Omega} = \frac{1}{4\pi} \sum_{l',l''} \left\langle \frac{1}{l'! l''!} \left[ {\mathcal U}^{(1)}_{L'} {\mathcal U}^{(1)}_{L''} n_{L'} n_{L''} - 4 {\mathcal U}^{(1)}_{a L'-1} {\mathcal U}^{(1)}_{a L''-1} n_{L'-1} n_{L''-1} +2 {\mathcal U}^{(1)}_{ab L'-2} {\mathcal U}^{(1)}_{ab L''-2} n_{L'-2} n_{L''-2} \right] \right. \\ + \frac{l' l''}{(l'+1)! (l''+1)!} \left[  4 {\mathcal V}^{(1)}_{L'} {\mathcal V}^{(1)}_{L''} n_{L'} n_{L''} - 8 {\mathcal V}^{(1)}_{a L'-1} {\mathcal V}^{(1)}_{a L''-1} n_{L'-1} n_{L''-1} \right. \\ \left. +4 {\mathcal V}^{(1)}_{ab L'-2} {\mathcal V}^{(1)}_{ab L''-2} n_{L'-2} n_{L''-2}   - 4 \epsilon_{cpa}\epsilon_{dqb} {\mathcal V}^{(1)}_{cd L'-2} {\mathcal V}^{(1)}_{ab L''-2} n_p n_q n_{L'-2} n_{L''-2}  \right] \\ \left. + \frac{8l''}{l'!(l''+1)!} \epsilon_{abp} n_p \left[ - {\mathcal U}^{(1)}_{a L'-1} {\mathcal V}^{(1)}_{b L''-1} n_{L'-1} n_{L''-1} + {\mathcal U}^{(1)}_{ac L'-2} {\mathcal V}^{(1)}_{bc L''-2} n_{L'-2} n_{L''-2} \right]  \right\rangle ,
\end{multline}
or its simpler expression in terms of the ``scalar'' moments:
\be
\label{eq:dEdtdOmega-scalarmoments}
\frac{dE_{\rm gw}}{dt \, d\Omega} = \frac{1}{32\pi} \sum_{l',l'',m',m''} \! \left\langle U^{(1)}_{l'm'} U^{(1)}_{l''m''} T^{E2,l'm'}_{jk}  T^{E2,l''m''}_{jk}
\!\!\! + V^{(1)}_{l'm'} V^{(1)}_{l''m''} T^{B2,l'm'}_{jk}  T^{B2,l''m''}_{jk} \!\!\! + 2 U^{(1)}_{l'm'} V^{(1)}_{l''m''} T^{E2,l'm'}_{jk}  T^{B2,l''m''}_{jk} \right\rangle.
\ee
\end{widetext}
In these expressions the ``pure-spin'' tensor spherical harmonics $T^{E2, lm}_{jk}$ and $T^{B2, lm}_{jk}$ are related to the spin-weighted spherical harmonics by Eqs.~(2.38) of Thorne \cite{kiprmp}, and the summation limits are as in Eq.~\eqref{eq:dEdtdOmega-hlm}.

The memory contributions to the waveform polarizations are conveniently computed by directly evaluating the $h_{lm}$ modes of Eq.~\eqref{eq:hdecompose} in terms of the memory contributions to the $U_{lm}$ multipoles. This requires computing the ``scalar'' version of the memory piece of the radiative mass-multipole moment ${\mathcal U}_L^{\rm (mem)}$. Combining Eqs.~\eqref{eq:ULmem}, \eqref{eq:Ulmdef}, and \eqref{eq:Ylm} yields
\be
\label{eq:Ulmmem}
U_{lm}^{\rm (mem)} = \frac{32\pi}{c^{2-l}} \sqrt{\frac{(l-2)!}{2(l+2)!}} \int_{-\infty}^{T_R} \!\! dt \int \! d\Omega \, \frac{dE_{\rm gw}}{dt d\Omega}(\Omega) Y_{lm}^{\ast}(\Omega) .
\ee
This is the primary equation that we need to evaluate in order to compute the nonlinear memory and its PN corrections.

For completeness, the combined contributions to the waveform modes from all of the hereditary and synchronous terms discussed in this section are given by
\begin{multline}
\label{eq:hlmtot}
h_{lm}(T_R) = \frac{G}{\sqrt{2}R c^{l+2}} \bigg\{ M_{lm}^{(l)}(T_R) - \frac{i}{c} S_{lm}^{(l)}(T_R)  
\\ 
+ \frac{2G{\mathcal M}}{c^3} \int_{-\infty}^{T_R} \ln\left(\frac{T_R-\tau}{2\tau_0}\right) \left[M_{lm}^{(l+2)}(\tau) - \frac{i}{c} S_{lm}^{(l+2)}(\tau) \right] d\tau  
\\
+ \frac{2G{\mathcal M}}{c^3} \left[\kappa_l M_{lm}^{(l+1)}(T_R) - \frac{i}{c} \pi_l S_{lm}^{(l+1)}(T_R) \right] 
\\ 
+ G U_{lm}^{\rm (mem)}(T_R) + O(G^2) + O(G/c^5) \bigg\} .
\end{multline}
\section{\label{sec:evalUmem}Evaluating the memory contribution to the radiative-mass multipoles}
\subsection{\label{sec:angularint}Computing angular integrals}
Computing the nonlinear memory's contribution to the waveform primarily consists of evaluating the radiative-mass multipoles in Eqs.~\eqref{eq:ULmem} or \eqref{eq:Ulmmem}. The most involved step is computing the angular integral that appears in these expressions. If one uses Eq.~\eqref{eq:ULmem} and works with the expansion of the energy flux in terms of STF tensors [Eq.~\eqref{eq:dEdtdOmegaSTF}] one needs to compute a sequence of angular integrals of the form
\be
\int d\Omega \, n_L,
\ee
which have well-known expressions in terms of products of Kronecker deltas [Eq.~(2.3) of Ref.~\cite{kiprmp}]. However, for large $l$ values this approach involves expressions with many indices and a large number of terms from the sums in Eq.~\eqref{eq:dEdtdOmegaSTF}.  While it is possible to use tensor algebra software to perform the manipulations, it is much easier to evaluate the angular integral in Eq.~\eqref{eq:Ulmmem} using the expression for the energy flux in Eq.~\eqref{eq:dEdtdOmega-hlm}. In this case all of the angular integrals have the form
\be
\label{eq:angular}
\int d\Omega \,  {}_{-2}Y_{l'm'}(\theta,\phi) \, {}_{-2}Y_{l''m''}^{\ast}(\theta,\phi) \, Y_{lm}^{\ast} (\theta,\phi) .
\ee
Using Eq.~\eqref{eq:Yslm} one can easily derive an exact expression for this integral (Appendix \ref{app:angularint}). The result involves a complicated sum that is tedious to evaluate by hand. Computer algebra programs can easily compute the angular integrals in Eq.~\eqref{eq:angular} by either evaluating the sum in the resulting analytic solution [Eq.~\eqref{eq:int3Y}], or by computing the various harmonics via Eq.~\eqref{eq:Yslm} and performing the integrals symbolically.

If we define the time derivative of the memory mass-multipole moment in Eq.~\eqref{eq:Ulmmem}, $U_{lm}^{{\rm (mem)}(1)} \equiv dU_{lm}^{\rm (mem)}/dT_R$, then combining Eqs.~\eqref{eq:Ulmmem}, \eqref{eq:dEdtdOmega-hlm}, \eqref{eq:int3Y}, and \eqref{eq:Yslmcc} gives
\begin{multline}
\label{eq:dUlmmem}
\!\!\!\!\! U_{lm}^{{\rm (mem)}(1)} \! = \! R^2 \! \sqrt{\frac{2(l-2)!}{(l+2)!}} \! \sum_{l'=2}^{\infty} \sum_{l''=2}^{\infty} \sum_{m'=-l'}^{l'} \sum_{m''=-l''}^{l''} \!\!\!\! (-1)^{m+m''} \\ \times \left\langle \dot{h}_{l'm'} \dot{h}^{\ast}_{l''m''} \right\rangle G^{2 -2 0}_{l' l'' l m' -m'' -m},
\end{multline}
where $G^{s_1 s_2 s_3}_{l_1 l_2 l_2 m_1 m_2 m_3}$ is an angular integral related to Eq.~\eqref{eq:angular} and is given in Appendix \ref{app:angularint}.
\subsection{\label{sec:selecrules}Selection rules and time derivatives of the memory multipole moments to 3PN order}
To compute the nonlinear memory's contribution to the waveform polarizations to 3PN order, we must expand the energy flux in Eqs.~\eqref{eq:Ulmmem} or \eqref{eq:dUlmmem} to 3PN order. For a quasicircular binary with relative orbital speed $v$, orbital separation $r$, and orbital period $T$, the $\dot{h}_{lm}$ modes for $m\neq 0$ have the leading-order PN scaling
\be
\label{eq:dothlmscaling}
\dot{h}_{lm} \sim \frac{I^{(l+1)}_{lm}}{R} \sim \frac{\eta}{R} \frac{M}{r} \left(\frac{r}{T}\right)^{l+1} \sim \frac{\eta}{R} v^{l+3} .
\ee
Since the leading $l=2$ term is of order $O(v^5)$, our knowledge of the waveform to relative 3PN order implies that we must evaluate the sums in the energy flux or Eq.~\eqref{eq:dUlmmem} up to maximum $l'$ and $l''$ values of $l'_{\rm max} = l''_{\rm max} =8$. Expressions for the $h_{lm}$ modes to 3PN order and $l\leq8$ are given in Eqs.~(9.3)-(9.4) of Ref.~\cite{blanchet3pnwaveform}. The straightforward procedure for computing the time derivatives $\dot{h}_{lm}$ is discussed in Appendix \ref{app:psilm} and uses the results of Sec.~\ref{sec:timeintegrals} below.

When evaluating $U_{lm}^{\rm (mem)}$, what values of $l$ and $m$ do we need to compute? To determine this we first note that the $\phi$ integrals in Eq.~\eqref{eq:angular},
\be
G^{2 -2 0}_{l' l'' l m' -m'' -m} \propto \int e^{i (m' - m'' -m) \phi} \,  d\phi ,
\ee
imply the selection rule $m=m'-m''$. Since the $\dot{h}_{lm}$ (for $m\neq 0$) scale like
\be
\label{eq:sourcescaling}
\dot{h}_{lm} \propto \frac{\eta}{R} x^{(l+3)/2} e^{-i m \varphi} ,
\ee
the product appearing in Eq.~\eqref{eq:dUlmmem} is proportional to
\[ \dot{h}_{l'm'} \dot{h}_{l''m''}^{\ast} \propto x^n e^{-i(m' - m'')\varphi}, \]
where $\varphi$ is the orbital phase, $x\equiv (M\omega)^{2/3}$ is the standard PN expansion parameter for circular orbits, and $n \geq 5$.  Combining with the above-mentioned selection rule, the time integral in Eq.~\eqref{eq:Ulmmem} leads to two types of integrals:
\bs
\label{eq:timeinttypes}
\begin{multline}
\label{eq:oscillint}
\int_{-\infty}^{T_R} x^n e^{-i m \varphi} \, dt = i \frac{M}{m} x^{n -3/2} e^{-i m \varphi}  \\ + (\text{higher order terms}) \, \;\; \text{for }m\neq 0, \;\;\; \text{and}
\end{multline}
\begin{multline}
\label{eq:nonoscillint}
\int_{-\infty}^{T_R} x^n \, dt = \int_{-\infty}^{T_R} \frac{x^n}{\dot{x}} \, dx =  \frac{5}{64(n-4)} \frac{M}{\eta} x^{n - 4} \\ + (\text{higher order terms}) \, \;\; \text{for }m = 0,
\end{multline}
\es
where $\dot{x} = 64\eta x^5 /(5M) [1+O(2)]$ [see also Eqs.~\eqref{eq:memintscaling} and Ref.~\cite{arun25PNamp}]. The $m\neq 0$ terms yield oscillatory contributions to the waveform polarizations that enter at higher PN orders than the nonoscillatory, $m=0$ terms. In a complete computation of the waveform, these oscillatory pieces contribute to the full-waveform beginning at 2.5PN order. Since we are interested only in the nonoscillatory memory effect, we will only focus on computing the $m=0$ terms in $U_{lm}^{\rm (mem)}$. However the procedure discussed here can also be used to compute the $m \neq 0$ oscillatory terms.

We also note here that the restriction of the memory to the $m=0$ modes is largely a consequence of choosing our coordinate system such that the binary's motion is confined to the $x$-$y$ plane. This choice allows the $h_{lm}$ modes to be proportional to $e^{-im\varphi(t)}$.  A rotation of our coordinate system (or equivalently a rotation of the orbital angular momentum) would mix the $m$ modes and lead to nonoscillatory memory terms in the $m\neq0$ modes as well.

To determine the maximum value of $l$ that is needed in $U_{l0}^{\rm (mem)}$, we first consider the angular integral in Eq.~\eqref{eq:ULmem}, where the energy flux is given by Eq.~\eqref{eq:dEdtdOmegaSTF}. The STF properties of the radiative multipoles, combined with the Kronecker deltas that result from angular integrals over the products of $n_j$, require that the maximum $l$ for which ${\mathcal U}_L^{\rm (mem)}$ or $U_{lm}^{\rm (mem)}$ will be nonzero is $l_{\rm max}=l'_{\rm max} + l''_{\rm max}$. For example, the first term in the energy flux in Eq.~\eqref{eq:dEdtdOmegaSTF} contributes to the $U_{lm}^{\rm (mem)(1)}$ via terms of the form:
\be
\label{eq:coupling1}
U_{lm}^{\rm (mem)(1)} \propto {\mathcal Y}_{L}^{lm \ast} {\mathcal U}^{(1)}_{L'} {\mathcal U}^{(1)}_{L''} \oint n_{L} n_{L'} n_{L''} d\Omega + \cdots.
\ee
The STF properties of ${\mathcal Y}_{L}^{lm}$ and ${\mathcal U}_{L}$ then require that the index coupling that maximizes $l$ be of the form
\be
U_{l_{\rm max} m}^{\rm (mem)(1)} \propto {\mathcal Y}_{L'L''}^{l_{\rm max} m \ast} {\mathcal U}^{(1)}_{L'} {\mathcal U}^{(1)}_{L''} +\cdots,
\ee
implying $l_{\rm max}=l'_{\rm max} + l''_{\rm max}$. Using the 3PN waveform as input ($l'=l''\leq 8$), this implies that the angular integrals in Eq.~\eqref{eq:dUlmmem} will vanish for $l>16$.
However we note from Eqs.~\eqref{eq:dUlmmem} and \eqref{eq:dothlmscaling} that the PN scaling of $U_{lm}^{\rm (mem)(1)}$ is
\be
U_{lm}^{\rm (mem)(1)} \sim \eta^2 v^{l' + l'' +6},
\ee
which for $l_{\rm max}=16$ ($l'=l''=8$) implies a 6PN-order correction relative to the leading-order $U_{2m}^{\rm (mem)(1)}\sim O(v^{10})$ term. Since our 3PN waveform allows us to consistently compute corrections only to relative 3PN order, this implies that the maximum $l$ up to which we can consistently compute the $U_{lm}^{\rm (mem)(1)}$ is $l_{\rm max}=10$.

Before proceeding with the computation of the $U_{lm}^{{\rm (mem)}(1)}$, we note that the right-hand side of Eq.~\eqref{eq:dUlmmem} is a function of the full $h_{lm}$ modes [Eq.~\eqref{eq:hlmtot}], which are themselves functions of the memory multipoles $U_{l0}^{\rm (mem)}$ that we are trying to compute. However, the time derivatives $U_{l0}^{\rm (mem)(1)}$ of the memory modes provide a very small contribution to the energy flux and can be neglected in comparison to the higher-order $m\neq 0$ modes. For example while the oscillatory mode product in Eq.~\eqref{eq:dUlmmem} scales like
\be
\dot{h}_{l' m'} \dot{h}_{l'' m''} \sim \frac{\eta^2}{R^2} v^{10} + \text{higher order terms},
\ee
the nonoscillatory (memory) mode product scales like
\be
\dot{h}_{l' 0} \dot{h}_{l'' 0} \sim \frac{\eta^4}{R^2} v^{20} + \text{higher order terms}.
\ee
Even at their lowest order, the $\dot{h}_{l' 0} \dot{h}_{l'' 0}$ mode pairs contribute a relative order $O(v^{10})$ (5PN) correction to the right-hand side of Eq.~\eqref{eq:dUlmmem} and can be safely ignored. The oscillatory modes effectively act as generators for the nonoscillatory (memory) modes.

\begin{widetext}
The explicit computation of the $U_{lm}^{{\rm (mem)}(1)}$ modes is performed using the computer algebra program \textsc{Maple}. First the sums appearing in Eq.~\eqref{eq:dUlmmem} are expanded, and the angular integrals are evaluated via Eq.~\eqref{eq:int3Y}. The resulting expressions consist of long sums over the various $\dot{h}_{l'm'} \dot{h}_{l''m''}^{\ast}$ terms. Next the explicit PN expansions for the $h_{lm}$ in the quasicircular case [given in Eqs.~(9.3)-(9.4) of Ref.~\cite{blanchet3pnwaveform}] are explicitly differentiated with respect to time. The details of this time differentiation are discussed in detail in Appendix \ref{app:psilm}. The essence of the calculation involves expressing the amplitude and phase of $h_{lm}$ entirely in terms of the PN parameter $x$ and then computing the time derivative via $\dot{h}_{lm}[x(t)] = \dot{x} (dh_{lm}/dx)$. The formula for $\dot{x}$ is itself derived in the next section from the 3PN GW luminosity and is directly related to the rate-of-change of the orbital frequency (which is itself a crucial quantity needed for the detection of inspiralling binaries with LIGO and other detectors). Once these derivatives are computed and substituted into the expanded sum on the right-hand side of Eq.~\eqref{eq:dUlmmem}, the result is series expanded to the PN order appropriate for the given $(l,m)$ mode.\footnote{I have also checked that applying a similar procedure to the angular integral of the energy flux in Eq.~\eqref{eq:dEdtdOmega-hlm} yields the correct formula for the GW luminosity to 3PN order [Eq.~\eqref{eq:Lgw} below].} The resulting expressions for the $U_{lm}^{({\rm mem})(1)}$ are given by:
\bs
\label{eq:dUlmPNexpand}
\begin{multline}
\label{eq:U20mem1}
U_{20}^{{\rm (mem)}(1)} = \frac{256}{21}\sqrt{\frac{3\pi}{5}} \eta^2 x^5 \left\{ 1 + x\left( -\frac{1219}{288} + \frac{\eta}{24} \right) + 4\pi x^{3/2} + x^2 \left( -\frac{793}{1782} - \frac{14\,023}{6336} \eta - \frac{4201}{1584} \eta^2 \right)
\right. \\
+ \pi x^{5/2} \left( -\frac{2435}{144} - \frac{23}{12}\eta \right)
+ x^3 \left[ \frac{174\,213\,949\,439}{1\,816\,214\,400} +\frac{16\pi^2}{3} - \frac{856}{105} ( 2\gamma_E + \ln16x )
\right. \\ \left. \left.
+ \left( - \frac{126\,714\,689}{4\,447\,872} + \frac{41}{48}\pi^2 \right) \eta + \frac{4\,168\,379}{123\,552} \eta^2 + \frac{142\,471}{46\,332} \eta^3 \right] +O(7) \right\} ,
\end{multline}
\begin{multline}
U_{40}^{{\rm (mem)}(1)} = \frac{64}{315}\sqrt{\frac{\pi}{5}} \eta^2 x^5 \left\{ 1 + x\left( -\frac{10\,133}{704} + \frac{25\,775}{528} \eta \right) + 4\pi x^{3/2} + x^2 \left( \frac{322\,533}{4576} - \frac{721\,593}{2288} \eta - \frac{237\,865}{5148} \eta^2 \right)
\right. \\
+ \pi x^{5/2} \left( -\frac{1028}{11} + \frac{11\,114}{33}\eta \right) + x^3 \left[ \frac{32\,585\,924\,257}{403\,603\,200} +\frac{16\pi^2}{3} - \frac{856}{105} ( 2 \gamma_E + \ln16x )
\right. \\ \left. \left.
 + \left( \frac{4\,669\,843}{164\,736} + \frac{41}{48}\pi^2 \right) \eta + \frac{16\,531}{52} \eta^2 - \frac{1\,145\,725}{92\,664} \eta^3 \right] +O(7) \right\} ,
\end{multline}
\begin{multline}
U_{60}^{{\rm (mem)}(1)} = - \frac{839}{693}\sqrt{\frac{\pi}{2730}} \eta^2 x^6 \left[ 1 - \frac{3612}{839} \eta +  x\left( -\frac{982\,361}{75\,510} + \frac{56\,387}{839} \eta - \frac{62\,244}{839} \eta^2 \right)  + \pi x^{3/2} \left( \frac{5540}{839} - \frac{23\,184}{839}\eta \right)
\right. \\ \left.
+ x^2 \left( \frac{302\,491\,414}{4\,492\,845} - \frac{1\,516\,457\,957}{3\,851\,010} \eta + \frac{27\,377\,867}{42\,789} \eta^2 + \frac{1\,106\,868}{14\,263}\eta^3 \right) +O(5) \right] ,
\end{multline}
\begin{multline}
U_{80}^{{\rm (mem)}(1)} = \frac{75\,601}{347\,490}\sqrt{\frac{\pi}{1190}} \eta^2 x^7 \left[ 1 - \frac{452\,070}{75\,601} \eta + \frac{733\,320}{75\,601}\eta^2  \right. \\ \left.  + x \left( - \frac{7\,655\,551}{604\,808} + \frac{369\,735\,869}{4\,309\,257} \eta - \frac{248\,030\,070}{1\,436\,419} \eta^2 + \frac{135\,873\,360}{1\,436\,419}\eta^3 \right) +O(3) \right] ,
\end{multline}
\be
U_{10\,0}^{{\rm (mem)}(1)} = - \frac{525\,221}{15\,752\,880}\sqrt{\frac{\pi}{385}} \eta^2 x^8 \left[ 1 - \frac{79\,841\,784}{9\,979\,199} \eta + \frac{198\,570\,240}{9\,979\,199}\eta^2 - \frac{172\,307\,520}{9\,979\,199} \eta^3 +O(2) \right] .
\ee
\es
\end{widetext}
The odd-$l$ moments vanish up to the required PN order:
\be
U_{30}^{{\rm (mem)}(1)} = U_{50}^{{\rm (mem)}(1)} = U_{70}^{{\rm (mem)}(1)} = U_{90}^{{\rm (mem)}(1)} = 0 \;.
\ee
Note that both $U_{20}^{{\rm (mem)}(1)}$ and $U_{40}^{{\rm (mem)}(1)}$ are of order $O(x^5)$. This is a consequence of the $l_{\rm max}=l'_{\rm max} + l''_{\rm max}$ selection rule: If the energy flux is expanded only to leading order ($l'_{\rm max}=l''_{\rm max}=2$), the selection rules allow all $l\leq4$.
\subsection{\label{sec:timeintegrals}Computing time integrals over the past history of the source}
The final step in computing the memory contribution to the radiative mass-multipole moments is the evaluation of the time integral over the entire past history of the source [cf.~Equation \eqref{eq:Ulmmem}]. While these time integrals generally come in the two types listed in Eqs.~\eqref{eq:timeinttypes}, the multipole components that contribute to the memory $[U_{l0}^{\rm (mem)}]$ involve only integrals of the second type. Arun et al.~\cite{arun25PNamp} discuss in detail an adiabatic model for the evolution of the source that consists of an inspiralling binary described by the leading-order (2.5PN) radiation-reaction formulas. Since we are here computing 3PN corrections to the memory terms considered in Ref.~\cite{arun25PNamp}, we must extend our adiabatic model of the inspiral to three PN orders beyond the leading-order model considered there.

It is important to note that this adiabatic model is an idealization representing a binary whose orbit has been decaying along quasicircular orbits from infinite separation solely via gravitational radiation reaction.  In reality a binary is formed or captured with some initial separation and eccentricity. This formation or capture, as well as perturbations to its orbital elements (e.g., via mass loss or gravitational 3-body scattering) may have caused additional linear or nonlinear memory contributions.  The adiabatic model used here ignores these issues and assumes a perfect quasicircular inspiral from infinite initial orbital separation. This is likely to be a good approximation if deviations from quasicircularity occurred in the very distant past (long before the binary is observed). Future work will test this approximation by including the effects of the binary's eccentricity (which grows in the past) \cite{favata-eccentricmemory}.

The main ``trick'' in computing the time integrals of Eqs.~\eqref{eq:dUlmPNexpand} is to simply change variables from time $t$ to the PN expansion parameter $x$:
\be
\label{eq:memoryint}
\int_{-\infty}^{T_R} [x(t)]^n \, dt = \int_{0}^{x(T_R)} \frac{x^n}{\dot{x}} \, dx ,
\ee
where we have used the fact that $x\rightarrow 0$ in the infinite past.
While computing the Newtonian-order memory requires only the leading-order contribution to $\dot{x}$, calculation of the 3PN memory requires the 3PN corrections to $\dot{x}$. This 3PN formula for $\dot{x}$ (which is often expressed in terms of $\dot{\omega}$ and $\omega$ in the literature) is the essence of our 3PN adiabatic model.

Applying the chain rule we can express $\dot{x}$ in terms of the orbital energy and the GW luminosity ${\mathcal L} = -\dot{E}$,
\be
\label{eq:xdot}
\frac{dx}{dt} = \frac{(-{\mathcal L})}{dE/dx} .
\ee
The 3.5PN orbital energy is given by Eq.~(5) of Ref.~\cite{blanchet35PNphase},
\begin{multline}
\label{eq:E3pn}
\!\!\!\!\!\! E = -\frac{\eta M x}{2}  \bigg\{ 1 + x \left(-\frac{3}{4} - \frac{\eta}{12} \right) + x^2 \left( -\frac{27}{8} + \frac{19}{8} \eta - \frac{\eta^2}{24} \right) \\
 + x^3 \left[ - \frac{675}{64} + \left( \frac{34\,445}{576} - \frac{205}{96}\pi^2 \right) \eta  \right.
\\ \left.
- \frac{155}{96} \eta^2 - \frac{35 \eta^3}{5184}  \right]  +O(8) \bigg\} ,
\end{multline}
and the 3.5PN GW luminosity is given by Eq.~(2) of Ref.~\cite{blanchet35PNphaseerratum}:
\begin{widetext}
\begin{multline}
\label{eq:Lgw}
{\mathcal L} = \frac{32}{5} \eta^2 x^5 \left\{ 1 + x \left( -\frac{1247}{336} - \frac{35}{12} \eta \right) + 4\pi x^{3/2} + x^2 \left( - \frac{44\,711}{9072} + \frac{9271}{504} \eta + \frac{65}{18} \eta^2 \right) + \pi x^{5/2} \left( - \frac{8191}{672} - \frac{583}{24} \eta \right) \right. \\ + x^3 \left[ \frac{6\,643\,739\,519}{69\,854\,400} + \frac{16}{3} \pi^2 - \frac{856}{105} ( 2\gamma_E + \ln 16x ) + \left( - \frac{134\,543}{7776} + \frac{41}{48} \pi^2 \right)\eta - \frac{94\,403}{3024}\eta^2 - \frac{775}{324}\eta^3 \right] \\ \left. + \pi x^{7/2} \left( -\frac{16\,285}{504} + \frac{214\,745}{1728} \eta + \frac{193\,385}{3024} \eta^2 \right) +O(8) \right\} .
\end{multline}
Note that the 3.5PN term in Eq.~\eqref{eq:E3pn} is zero and we have substituted the values $\lambda = -1987/3080$ and $\theta=-11\,831/9240$ for the ambiguity parameters \cite{blanchet-damour-farese-dimreg,blanchet-iyer-hadreg,blanchetdamour3PNprl}.
Substituting these expressions into Eq.~\eqref{eq:xdot} and expanding to 3.5PN order yields
\begin{multline}
\label{eq:xdot35pn}
\frac{dx}{dt} = \frac{64}{5} \frac{\eta}{M} x^5 \left\{ 1 + x \left( - \frac{743}{336} - \frac{11}{4} \eta \right) + 4 \pi x^{3/2} + x^2 \left( \frac{34\,103}{18\,144} + \frac{13\,661}{2016} \eta + \frac{59}{18} \eta^2 \right) + \pi x^{5/2} \left( - \frac{4159}{672} - \frac{189}{8} \eta \right) \right. \\ + x^3 \left[ \frac{16\,447\,322\,263}{139\,708\,800} + \frac{16}{3}\pi^2 - \frac{856}{105} ( 2\gamma_E + \ln 16x ) + \left( - \frac{56\,198\,689}{217\,728} + \frac{451}{48}\pi^2 \right) \eta + \frac{541}{896} \eta^2 - \frac{5605}{2592} \eta^3 \right] \\ \left. + \pi x^{7/2} \left( -\frac{4415}{4032} + \frac{358\,675}{6048} \eta + \frac{91\,495}{1512} \eta^2 \right) + O(8) \right\} .
\end{multline}
Using Eq.~\eqref{eq:xdot35pn} in Eq.~\eqref{eq:memoryint} and expanding the integrands to the appropriate PN order, the time integrals of Eqs.~\eqref{eq:dUlmPNexpand} are easily computed.\footnote{Note that when computing integrals and PN series expansions here and elsewhere in this paper, the natural-log terms $\ln(16 x)$ and $\ln(x/x_0)$ [cf.~Equation \eqref{eq:psiphase}] are \emph{not} treated as constants.}
\section{\label{sec:results}Results: Memory contributions to the waveform modes and polarizations}
Rather than listing the resulting expressions for $U_{l0}^{\rm (mem)}$, we instead list the memory contributions to the spin-weighted spherical-harmonic modes of the polarization waveform [Eq.~\eqref{eq:hlm}]. These quantities are simply related via
\be
\label{eq:hl0mem}
h_{l0}^{\rm (mem)} = \frac{\alpha}{\sqrt{2} R} U_{l0}^{\rm (mem)} = 8 \sqrt{\frac{\pi}{5}} \frac{\eta M x}{R} \hat{H}_{l0} ,
\ee
where we have followed the notation of Sec.~9 of Ref.~\cite{blanchet3pnwaveform}. The notational parameter $\alpha$ accounts for the two commonly used choices for the polarization triad (see discussion in Sec.~\ref{sec:polmodes}):
\be
\label{eq:alpha}
\alpha =\left\{ \begin{array}{l} (+1) \;\;\; \text{for the Kidder \cite{kidder08} convention}, \\ (-1) \;\;\; \text{for the Blanchet et al.~\cite{blanchet3pnwaveform} convention}. \end{array} \right.
\ee
The resulting polarization modes in terms of $\hat{H}_{l0}$ are:
\bs
\label{eq:Hl0-mem}

\begin{multline}
\label{eq:H20}
\hat{H}_{20} = \alpha \frac{5}{14 \sqrt{6}} \left\{ 1 + x \left( -\frac{4075}{4032} + \frac{67}{48} \eta \right) + x^2 \left( - \frac{151\,877\,213}{67\,060\,224} - \frac{123\,815}{44\,352} \eta + \frac{205}{352} \eta^2 \right) + \pi x^{5/2} \left( -\frac{253}{336}
+ \frac{253}{84}\eta \right)
 \right. \\ \left.
+  x^3 \left[ - \frac{4\,397\,711\,103\,307}{532\,580\,106\,240}  + \left( \frac{700\,464\,542\,023}{13\,948\,526\,592} - \frac{205}{96}\pi^2 \right) \eta + \frac{69\,527\,951}{166\,053\,888} \eta^2 + \frac{1\,321\,981}{5\,930\,496}\eta^3 \right] +O(7) \right\} ,
\end{multline}
\begin{multline}
\label{eq:H40}
\hat{H}_{40} = \alpha \frac{1}{504 \sqrt{2}} \left\{ 1 + x \left( -\frac{180\,101}{29\,568} + \frac{27\,227}{1056} \eta \right) + x^2 \left( \frac{2\,201\,411\,267}{158\,505\,984} - \frac{34\,829\,479}{432\,432} \eta + \frac{844\,951}{27\,456} \eta^2 \right)
\right. \\
+ \pi x^{5/2} \left( -\frac{13\,565}{1232} + \frac{13\,565}{308}\eta \right)  +  x^3 \left[ \frac{15\,240\,463\,356\,751}{781\,117\,489\,152}  + \left( - \frac{1\,029\,744\,557\,245}{27\,897\,053\,184} - \frac{205}{96}\pi^2 \right) \eta
\right. \\ \left. \left.
- \frac{4\,174\,614\,175}{36\,900\,864} \eta^2 + \frac{221\,405\,645}{11\,860\,992}\eta^3 \right] +O(7) \right\} ,
\end{multline}
\begin{multline}
\label{eq:H60}
\hat{H}_{60} = -\alpha \frac{4195}{1\,419\,264 \sqrt{273}} \, x \left[ 1 -\frac{3612}{839} \eta +  x \left( -\frac{45\,661\,561}{6\,342\,840} + \frac{101\,414}{2517} \eta - \frac{48\,118}{839} \eta^2 \right) + \pi x^{3/2} \left( \frac{1248}{839} - \frac{4992}{839} \eta \right) \right. 
\\ \left.
+ x^2 \left( \frac{3\,012\,132\,889\,099}{144\,921\,208\,320} - \frac{27\,653\,500\,031}{191\,694\,720} \eta + \frac{1\,317\,967\,427}{4\,107\,744} \eta^2 - \frac{24\,793\,657}{342\,312} \eta^3 \right) +O(5) \right] ,
\end{multline}
\begin{multline}
\label{eq:H80}
\hat{H}_{80} = \alpha \frac{75\,601}{213\,497\,856 \sqrt{119}} \, x^2 \left[ 1 -\frac{452\,070}{75\,601} \eta + \frac{733\,320}{75\,601} \eta^2 \right. \\ \left.  
+  x \left( -\frac{265\,361\,599}{33\,869\,248} + \frac{18\,177\,898\,147}{321\,757\,856} \eta - \frac{722\,521\,125}{5\,745\,676} \eta^2 + \frac{261\,283\,995}{2\,872\,838} \eta^3 \right) +O(3) \right] ,
\end{multline}
\be
\label{eq:H100}
\hat{H}_{10\,0} = - \alpha \frac{525\,221}{6\,452\,379\,648 \sqrt{154}} \, x^3 \left[ 1 -\frac{79\,841\,784}{9\,979\,199} \eta + \frac{198\,570\,240}{9\,979\,199} \eta^2 - \frac{172\,307\,520}{9\,979\,199} \eta^3 +O(2) \right] .
\ee
\es

Choosing $\alpha=-1$, these expressions can be directly combined with Eqs.~(9.4) of Ref.~\cite{blanchet3pnwaveform} to give all of the 3PN contributions to the spin-weighted spherical-harmonic modes $h_{lm}$. Note that all of the contributions to the $m=0$, even-$l$ modes arise solely from the Christodoulou memory piece of $U_{lm}$. There are no hereditary-memory contributions to the radiative current-multipole moments $V_{lm}$. However, there is a \emph{nonhereditary} DC contribution to $h_{30}$ at 2.5PN order arising from the nonlinear corrections to ${\mathcal V}_{ijk}$. This is discussed in Sec.~\ref{sec:crosmem} below.

Combining the above modes with Eq.~\eqref{eq:hdecompose}, we can explicitly compute the memory contributions to the $+$ waveform polarization. Following the notation in Ref.~\cite{blanchet3pnwaveform} we factor the waveform as
\be
\label{eq:hplusfactor}
h_{+,\times} = \frac{2 \eta M x}{R} H_{+,\times} \,\, +\,\, O\left(\frac{1}{R^2}\right) , \qquad \text{where}
\ee
\be
\label{eq:Hplus}
H_{+,\times} = \sum_{n=0}^{\infty} x^{n/2} H_{+,\times}^{(n/2)} .
\ee
The memory contributions to $H_{+}^{(n/2)}$ are:
\bs
\label{eq:Hplusmem}
\be
\label{eq:Hplus0pn}
H_{+}^{(0, {\rm mem})} = \alpha \frac{1}{96} s^2_{\Theta} (17 + c^2_{\Theta} ) ,
\ee
\be
\label{eq:Hplus05pn}
H_{+}^{(0.5, {\rm mem})} = 0  ,
\ee
\be
\label{eq:Hplus1pn}
H_{+}^{(1, {\rm mem})} = \alpha s^2_{\Theta} \left[ - \frac{354\,241}{2\,064\,384} - \frac{62\,059}{1\,032\,192} c^2_{\Theta} - \frac{4195}{688\,128} c^4_{\Theta} + \left( \frac{15\,607}{73\,728} + \frac{9373}{36\,864} c^2_{\Theta} + \frac{215}{8192} c^4_{\Theta} \right) \eta \right] ,
\ee
\be
\label{eq:Hplus15pn}
H_{+}^{(1.5, {\rm mem})} = 0 ,
\ee
\begin{multline}
\label{eq:Hplus2pn}
H_{+}^{(2, {\rm mem})} = \alpha s^2_{\Theta} \left[ - \frac{3\,968\,456\,539}{9\,364\,045\,824} + \frac{570\,408\,173}{4\,682\,022\,912} c^2_{\Theta} + \frac{122\,166\,887}{3\,121\,348\,608} c^4_{\Theta} + \frac{75\,601}{15\,925\,248} c^6_{\Theta}  + \left( - \frac{7\,169\,749}{18\,579\,456} \right. \right. \\ \left. \left. - \frac{13\,220\,477}{18\,579\,456} c^2_{\Theta} - \frac{1\,345\,405}{6\,193\,152} c^4_{\Theta}  - \frac{25\,115}{884\,736} c^6_{\Theta} \right) \eta + \left(  \frac{10\,097}{147\,456} + \frac{5179}{36\,864} c^2_{\Theta} + \frac{44\,765}{147\,456} c^4_{\Theta} + \frac{3395}{73\,728} c^6_{\Theta} \right) \eta^2 \right] ,
\end{multline}
\be
\label{eq:Hplus25pn}
H_{+}^{(2.5, {\rm mem})} = -\alpha \frac{5\pi}{21\,504} (1-4\eta) s^2_{\Theta} \left( 509 + 472 c^2_{\Theta} + 39 c^4_{\Theta} \right),
\ee
\begin{multline}
\label{eq:Hplus3pn}
H_{+}^{(3, {\rm mem})} = \alpha s^2_{\Theta} \left\{ - \frac{69\,549\,016\,726\,181}{46\,146\,017\,820\,672} +\frac{6\,094\,001\,938\,489}{23\,073\,008\,910\,336} c^2_{\Theta} - \frac{1\,416\,964\,616\,993}{15\,382\,005\,940\,224} c^4_{\Theta} - \frac{2\,455\,732\,667}{78\,479\,622\,144} c^6_{\Theta}
\right. \\
- \frac{9\,979\,199}{2\,491\,416\,576} c^8_{\Theta}   + \left[ \frac{1\,355\,497\,856\,557}{149\,824\,733\,184} - \frac{3485 \pi^2}{9216}   + \left( - \frac{3\,769\,402\,979}{4\,682\,022\,912} - \frac{205 \pi^2}{9216} \right) c^2_{\Theta} + \frac{31\,566\,573\,919}{49\,941\,577\,728} c^4_{\Theta}
\right. \\ \left.
+ \frac{788\,261\,497}{3\,567\,255\,552} c^6_{\Theta}  + \frac{302\,431}{9\,437\,184} c^8_{\Theta}  \right] \eta + \left(  \frac{5\,319\,395}{28\,311\,552} - \frac{24\,019\,355}{99\,090\,432} c^2_{\Theta} - \frac{4\,438\,085}{3\,145\,728} c^4_{\Theta} - \frac{3\,393\,935}{7\,077\,888} c^6_{\Theta} - \frac{7835}{98\,304} c^8_{\Theta} \right) \eta^2
\\ \left.
+ \left(  \frac{1\,433\,545}{63\,700\,992} + \frac{752\,315}{15\,925\,248} c^2_{\Theta} + \frac{129\,185}{2\,359\,296} c^4_{\Theta} + \frac{389\,095}{1\,179\,648} c^6_{\Theta} + \frac{9065}{131\,072} c^8_{\Theta} \right) \eta^3 \right\} ,
\end{multline}
\es
\end{widetext}
\begin{figure*}[t]
$
\begin{array}{cc}
\includegraphics[angle=0, width=0.475\textwidth]{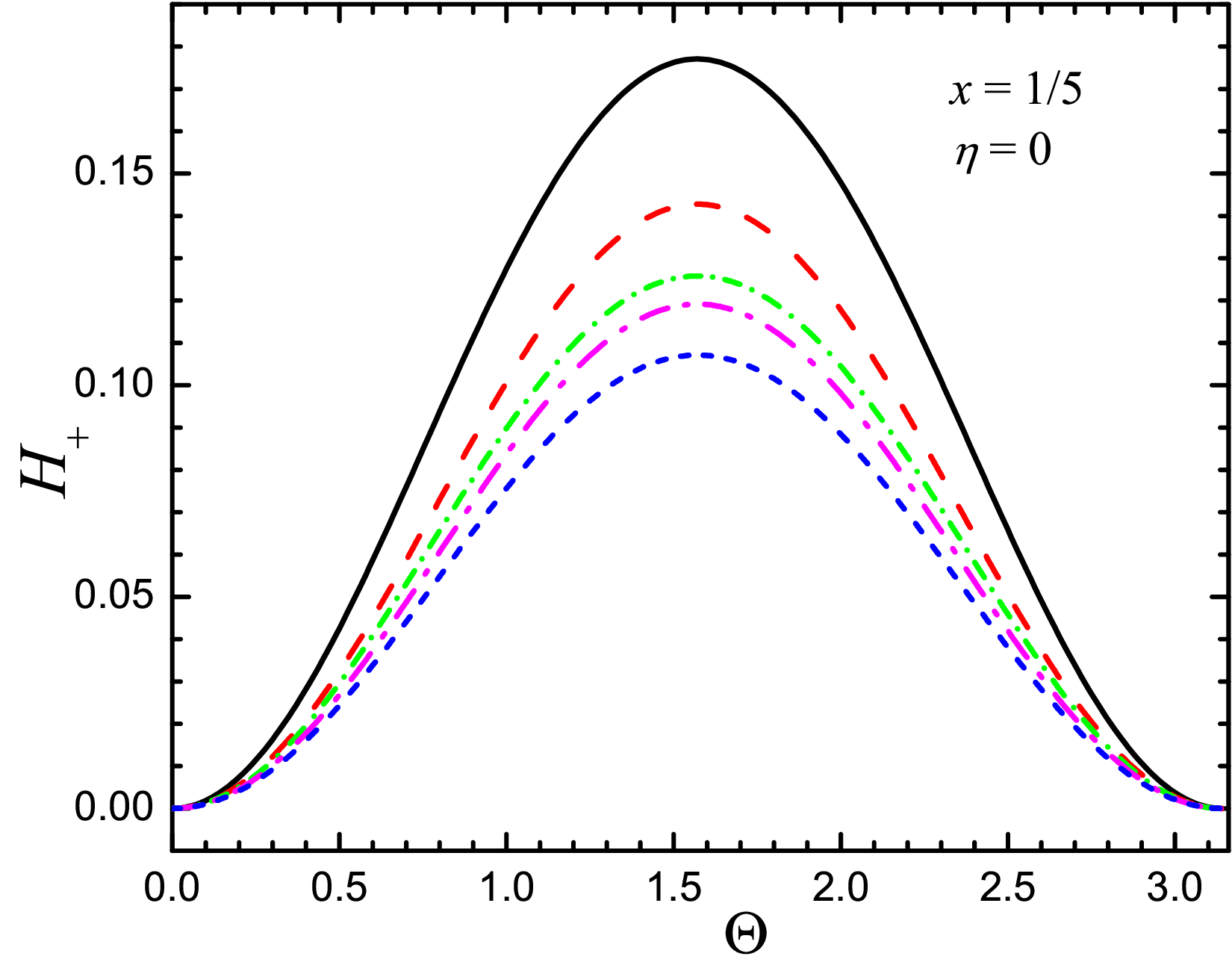} &
\includegraphics[angle=0, width=0.475\textwidth]{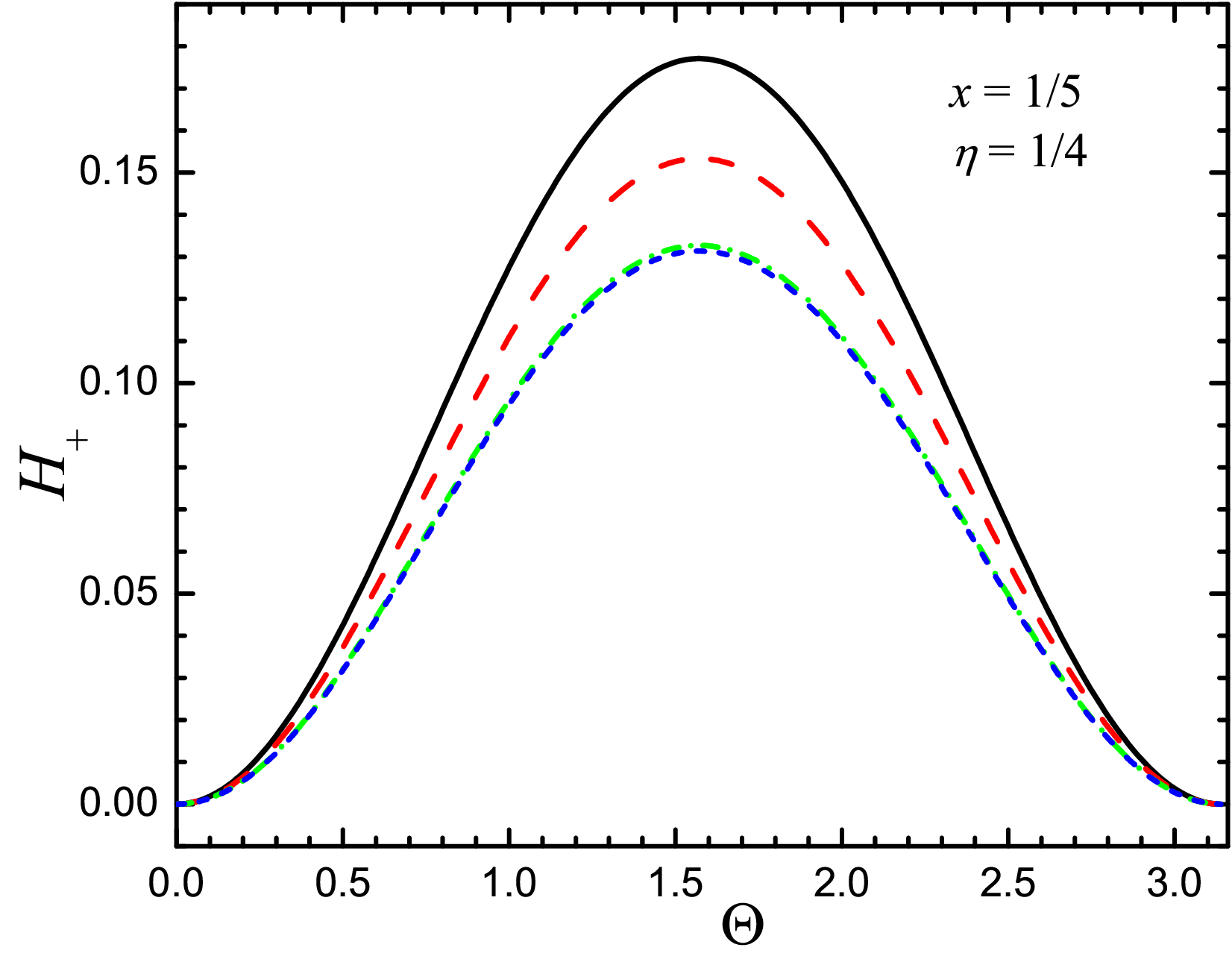} \\
\includegraphics[angle=0, width=0.48\textwidth]{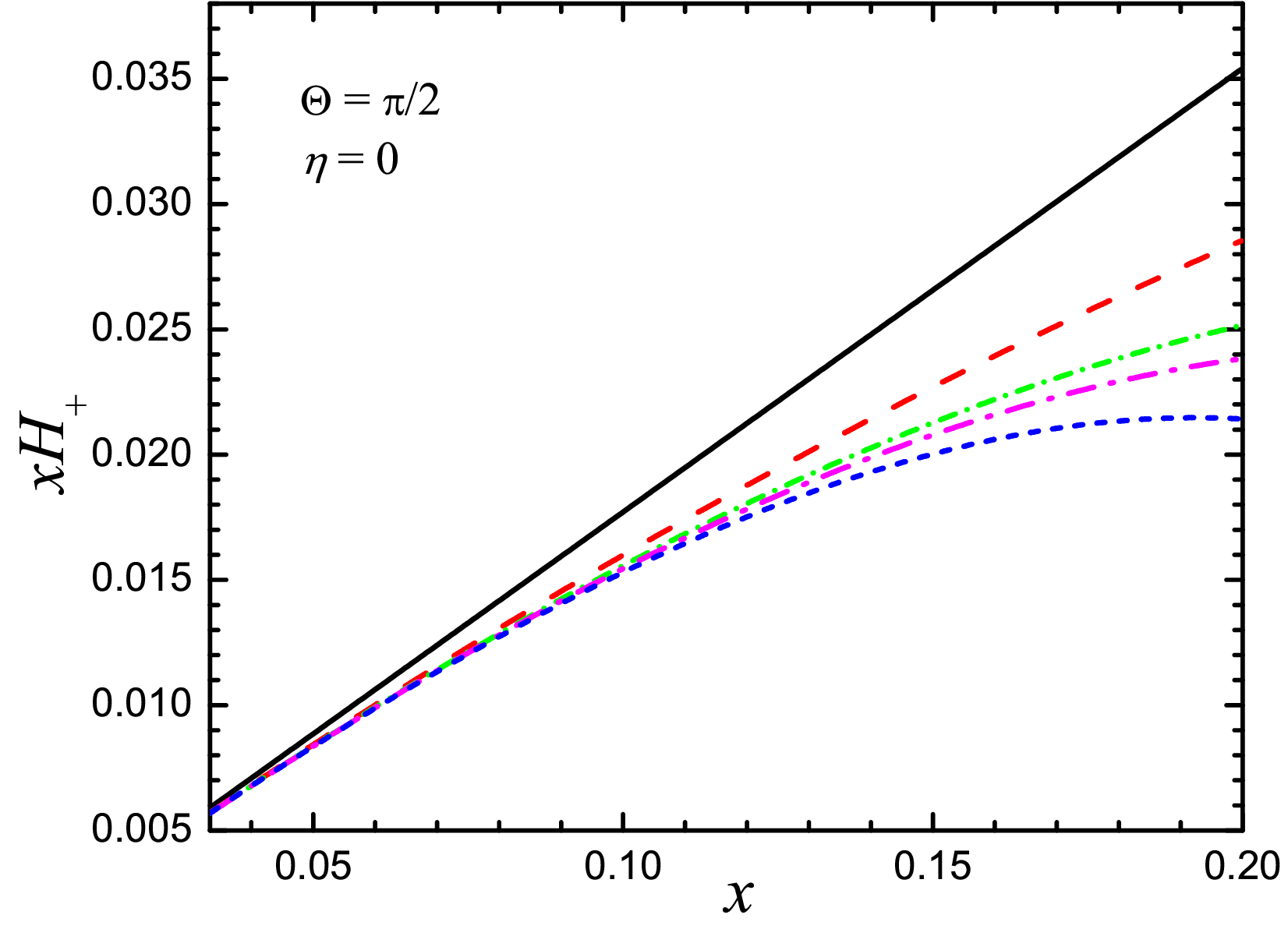} &
\includegraphics[angle=0, width=0.48\textwidth]{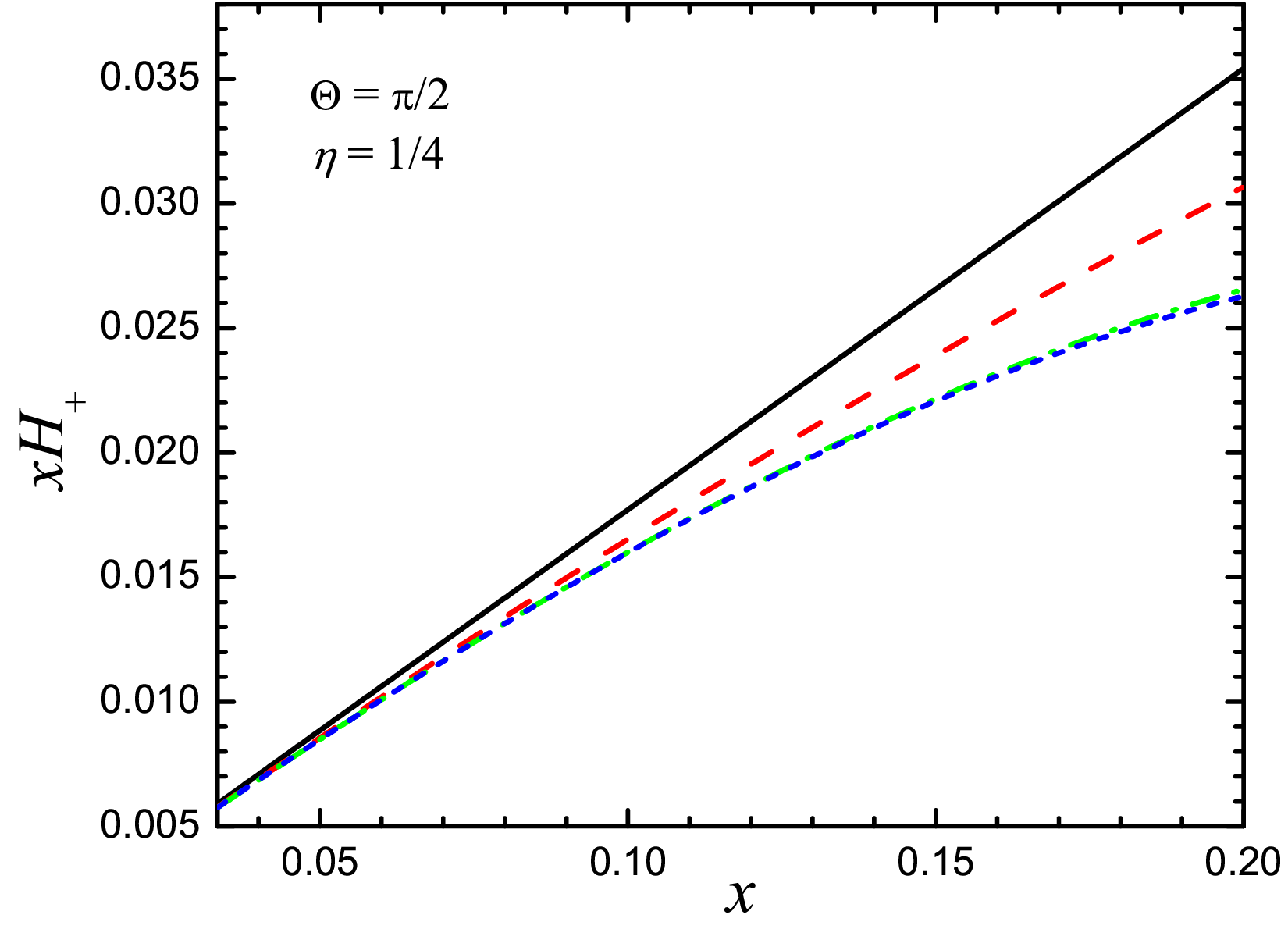}
\end{array}
$
\caption{\label{fig:H-theta}(color online). Dependence of the post-Newtonian (PN) corrections to the Christodoulou memory on binary inclination and orbital separation. The plots in the top row show the memory contribution to $H_+$ [Eq.~\eqref{eq:Hplus}] at each cumulative PN order as a function of $\Theta$ (the polar angle to the observer; $\Theta=0$ points along the binary's orbital angular momentum) for $x=1/5$ and $\eta=0$ (left) and $\eta=0.25$ (right). The different curves represent terms up to the following cumulative PN orders: solid (black) 0PN; long-dashed (red) 1PN; short-dash-dotted (green) 2PN; long-dash-dotted (magenta) 2.5PN; short-dashed (blue) 3PN.  The bottom row plots $xH_{+}$ [see Eqs.~\eqref{eq:hplusfactor} and \eqref{eq:Hplus}] and shows the memory's dependence on the PN parameter $x$ (which equals $M/r$ at Newtonian order, where $r$ is the orbital separation in harmonic coordinates). The labeling scheme is the same as in the top row. The PN corrections do not qualitatively change the angular dependence, but tend to decrease the magnitude of the memory. Since the 2.5PN correction vanishes for $\eta=0.25$, the 2.5PN curve is identical to the 2PN one and is not displayed in the right column plots. For $\eta=0.25$ the 3PN curve is nearly coincident with the 2PN curve.}
\end{figure*}
where $c_{\Theta}=\cos\Theta$ and $s_{\Theta} = \sin\Theta$.
Choosing $\alpha=-1$ and $\Theta=\iota$, these expressions can be directly combined with Eqs.~(8.9) of Ref.~\cite{blanchet3pnwaveform} to yield the total $+$ polarization. For reference, we also recall the leading-order, nonmemory contributions to the polarizations:
\bs
\label{eq:hquad}
\begin{align}
\label{eq:hplusquad}
h_{+}^{(0)} &=  -2 \alpha \frac{\eta M}{R} x (1+c^2_{\Theta}) \cos2(\varphi - \Phi), \\
\label{eq:hcrossquad}
h_{\times}^{(0)} &=  -4 \alpha \frac{\eta M}{R} x c_{\Theta} \sin2(\varphi - \Phi),
\end{align}
\es
where $\Theta=\iota$ and $\Phi=\pi/2$ in the conventions of Ref.~\cite{blanchet3pnwaveform}.

We note the following features of the memory waveform: (i) As pointed out previously by Ref.~\cite{blanchet3pnwaveform}, there is no memory term at 0.5PN order; we see here that the 1.5PN contribution to the memory also vanishes. (ii) The 2.5PN term vanishes for equal-mass binaries. Even in the $\eta \rightarrow 0$ limit, the 2.5PN term has the smallest magnitude of all the nonvanishing PN terms computed here. (iii) Aside from the nonhereditary DC term discussed below, all of the hereditary memory terms only affect the $+$ polarization. This arises from the choice of our polarization triad, the fact that the $U_{l0}^{\rm (mem)}$ are all real, and the absence of hereditary-memory contributions in the radiative-current multipoles. A rotation of the polarization triad would cause nonlinear memory contributions to both polarizations. (iv) As expected for $m=0$ modes in a planar system, the memory contribution to the polarizations is independent of the angle $\Phi$. (v) Lastly, note that the memory pieces of the waveform are entirely free of any arbitrary constants (such as $r_0$) that arise in the MPM formalism (Sec.~\ref{sec:MPM}).

Figure \ref{fig:H-theta} indicates several more features of the PN corrections to the memory waveform: modulo some factors, these plots display the memory contribution to the $+$ waveform polarization [Eqs.~\eqref{eq:hplusfactor}--\eqref{eq:Hplusmem}] from 0PN to 3PN orders (e.g., the 3PN curves contain all PN terms up to and including the 3PN terms). In the top row we see that the angular dependence is qualitatively similar at each PN order, with the memory's amplitude peaking at $\Theta=\pi/2$. The bottom plots show how the PN corrections to the memory depend on the PN parameter $x$, with $x$ ranging from $1/30$ ($r \approx 30 M$) to $1/5$ (approximately corresponding to the last stable orbit at harmonic coordinate radius $r\approx 5M$).\footnote{Note that $x=(M/r) [1 + O(2)]$, where the PN correction terms can be inferred from Eq.~(7.21) of Ref.~\cite{blanchet-faye-3pn-PRD2001}.}
While the 0PN piece of the memory increases linearly with $x$, the higher PN correction terms tend to decrease the memory: for example, for $\eta=0.25$ and $x=1/5$, the 3PN memory is smaller than the 0PN memory by a factor of $\approx 0.74$. Although the memory contribution to $h_+$ vanishes in the $\eta \rightarrow 0$ limit, the plots have factored out the leading-order $\eta$ dependence and allow us to compare the equal-mass and extreme-mass-ratio limits. Notice that as the PN order is increased, the memory waveforms seem to be converging rapidly in the equal-mass limit, but converge more slowly for small mass-ratios. This is consistent with the behavior of the oscillatory pieces of the waveform and the GW luminosity.
\section{\label{sec:discussion}Discussion}
Now that we have completed our derivation of the higher-order PN corrections to the memory waveform, we can address several other memory-related issues. The first two subsections below discuss additional nonoscillatory contributions to the waveform. In the last two subsections we address the challenges associated with computing the nonlinear memory in numerical relativity (NR) simulations and observing memory in gravitational-wave (GW) detectors.

Throughout this paper and elsewhere in the literature, memory effects are often referred to as DC (for ``direct-current'') effects in the waveform. This terminology should be taken to mean simply that the memory is a \emph{nonoscillatory} modulation of the waveform, in contrast to the AC or \emph{oscillatory} waveform modulations. In particular we note that DC waveform corrections are generally not constant offsets in the polarizations but can vary with time. For bound binaries this variation usually proceeds on a slow (radiation-reaction) time scale rather than an orbital time scale. For example, the GW modes for quasicircular binaries can be expanded in a Fourier series of the form
\be
\label{eq:Hlmnmodes}
h_{lm}(t) = \sum_{n=-\infty}^{+\infty} {\mathcal H}_{lm}^{\{n\}}(t) e^{-in\Omega(t) t},
\ee
where $\Omega \equiv 1/t \int \omega(t) dt$ and ${\mathcal H}_{lm}^{\{n\}}$ is some mode amplitude function.
In the absence of radiation-reaction $\Omega$ and ${\mathcal H}_{lm}^{\{n\}}$ are constant, but they are slowly evolving functions of time when radiation-reaction is included. In such a sum the DC terms are those corresponding to $n=0$. We note that this Fourier index $n$ is generally distinct from the azimuthal angular index $m$, but, for the conventional choice in which the orbit lies in the $x$-$y$ plane,  $n=m$, and we can remove the sum in Eq.~\eqref{eq:Hlmnmodes}.
\subsection{\label{sec:crosmem}The nonlinear, nonhereditary DC term}
In addition to the nonlinear, Christodoulou memory which has been the primary focus of this investigation, a new type of nonlinear, zero-frequency term has recently been discovered by Arun et al.~\cite{arun25PNamp}. Unlike the hereditary memory discussed here, this term is \emph{nonhereditary} and has its origin in the (mass quadrupole)$\times$(mass quadrupole) and (current dipole)$\times$(mass quadrupole) terms that arise in the 1.5PN corrections to the radiative current-octupole moment [Eq.~(5.6b) of Ref.~\cite{blanchet3pnwaveform}]:
\begin{multline}
\label{eq:Vijk}
{\mathcal V}_{ijk}(T_R) = {\mathcal S}_{ijk}^{(3)}(T_R) \\ + \frac{G}{c^3} \bigg\{ 2{\mathcal M} \int_{-\infty}^{T_R} \left[ \ln\left(\frac{T_R-\tau}{2\tau_0}\right) + \frac{5}{3}\right] {\mathcal S}_{ijk}^{(5)}(\tau) d\tau  \\
+ \frac{1}{10} \epsilon_{ab<i}{\mathcal M}_{j\underline{a}}^{(5)}{\mathcal M}_{k>b} \! - \frac{1}{2} \epsilon_{ab<i}{\mathcal M}_{j\underline{a}}^{(4)}{\mathcal M}_{k>b}^{(1)} \! - 2 {\mathcal S}_{<i}{\mathcal M}_{jk>}^{(4)} \! \bigg\} \\ + O(5),
\end{multline}
where
\be
{\mathcal S}_i = {\mathcal J}_i = \eta M \epsilon_{i a b} x_a v_b + O(c^{-2}),
\ee
and the last three terms in Eq.~\eqref{eq:Vijk} give rise to the nonhereditary DC effect.

We can gain further insight into this effect by expanding Eq.~\eqref{eq:Vijk} on the basis of STF spherical harmonics. Using Eq.~\eqref{eq:Vlmdef}  to compute $V_{3m}$ from ${\mathcal V}_{ijk}$ and Eq.~\eqref{eq:Ustfdef} to express ${\mathcal M}_{ij}$ in terms of $M_{3m}$, the radiative current-octupole moment can be written as
\be
\label{eq:V3m}
V_{3m} = V_{3m}^{\rm (A)} + V_{3m}^{\rm (B)},
\ee
where
\be
\label{eq:V3mA}
V_{3m}^{\rm (A)} = S_{3m}^{(3)} + \frac{2G {\mathcal M}}{c^3} \int_{-\infty}^{T_R} \left[ \ln\left(\frac{T_R-\tau}{2\tau_0}\right) + \frac{5}{3}\right] {S}_{3m}^{(5)}(\tau) d\tau,
\ee
\begin{multline}
\label{eq:V30}
V_{30}^{\rm (B)} = \frac{i}{3360} \sqrt{\frac{105}{\pi}}\frac{G}{c^3} \left\{ \left[ M_{2-2}^{(5)} M_{22} - M_{22}^{(5)} M_{2-2} \right] \right.
\\ \left.
+ 5 \left[ M_{22}^{(4)} M_{2-2}^{(1)} - M_{2-2}^{(4)} M_{22}^{(1)} \right] - 32 i \sqrt{15\pi} J_z M_{20}^{(4)} \right\},
\end{multline}
\pagebreak
\begin{multline}
\label{eq:V32}
V_{3\pm2}^{\rm (B)} = \pm \frac{i}{672} \sqrt{\frac{21}{\pi}}\frac{G}{c^3} \left\{ \left[ M_{20}^{(5)} M_{2\pm2} - M_{2\pm2}^{(5)} M_{20} \right] \right.
\\ \left.
+ 5 \left[ M_{2\pm2}^{(4)} M_{20}^{(1)} - M_{20}^{(4)} M_{2\pm2}^{(1)} \right] \mp 32 i \sqrt{5\pi/3} J_z M_{2\pm2}^{(4)} \right\},
\end{multline}
where we have used ${\mathcal S}_i = J_z \delta^z_i$ and $M_{2\pm1}=0$ (for orbits in the $x$-$y$ plane). Since for quasicircular and planar orbits the canonical-mass moments are proportional to $M_{lm} \approx I_{lm} \propto e^{-im\varphi}$, one can easily see that the sinusoidal dependence on $\varphi$ cancels in $V_{30}^{\rm (B)}$ above.

Unlike the Christodoulou memory the nonlinear, nonhereditary DC effect originating from $V_{30}^{\rm (B)}$ modifies the $\times$ polarization waveform at the 2.5PN order. In the notation of Sec.~\ref{sec:results}, this term provides the following contribution to the $h_{lm}$ modes [Eq.~(9.4g) of Ref.~\cite{blanchet3pnwaveform}],
\be
\label{eq:H30}
\hat{H}_{30} = \alpha \frac{2}{5} i \sqrt{\frac{6}{7}} x^{5/2} \eta ,
\ee
and to the $\times$ polarization [Eq.~(5.10) of Ref.~\cite{arun25PNamp}],
\be
\label{eq:Hcrossmem}
H_{\times}^{(2.5, {\rm mem})} = -\alpha \frac{6}{5} s_{\Theta}^2 c_{\Theta} \eta .
\ee

It is not clear if there is a simple physical explanation for this nonlinear, nonhereditary DC term. Because of its high PN order, it is likely to be of much less observational significance than the Christodoulou memory. It is also not clear if this term leads to a ``true'' memory in a GW detector, i.e., a displacement in the detector that persists after the GW has passed \cite{bala-luc-privcomm}. It is possible that this nonhereditary, DC waveform correction grows during the inspiral phase, but then decays to zero during the merger and ringdown, leaving no net memory.  Nonetheless, it is interesting that this type of nonlinearity [the canonical-moment coupling in Eq.~\eqref{eq:Vijk}] can give rise to a DC effect in the waveform. It is possible that other multipole interactions of this type could contribute additional nonhereditary DC terms at higher PN orders.
\subsection{\label{sec:linmem}Linear DC effects in bound binaries}
Linear memory is generally considered to arise from permanent differences between late and early times in the derivatives of the source moments $\Delta {\mathcal I}_L^{(l)}$ and $\Delta {\mathcal J}_L^{(l)}$ [or equivalently $\Delta I_{lm}^{(l)}$ and $\Delta J_{lm}^{(l)}$]. While unbound orbits derive their memory from changes in the derivatives of the source moments, bound orbits are generally thought to have vanishing linear memory. Here we argue that bound orbits that undergo gravitational radiation-reaction---or any other nonperiodic, secular change---can also display nonoscillatory (DC) waveform components resulting from long-timescale changes in the derivatives of the source-multipole moments.

For example, consider the source mass-multipole moments written in the form
\be
\label{eq:Ilmform}
I_{lm} \propto \eta M r(t)^l e^{-im\varphi(t)} [ 1+ O(2)] .
\ee
The $m\neq 0$ modes have time derivatives proportional to $I_{lm}^{(l)} \propto e^{-im\varphi(t)}$. For unbound orbits the time derivatives $I_{lm}^{(l)}$ that enter the waveform can display linear memory from differences in the phase angle $\varphi$.\footnote{For example, by parameterizing the orbit equations in terms of the true anomaly, one can show that at leading-PN order an extreme hyperbolic orbit with eccentricity $e_0 \gg 1$ and impact parameter $b$ has linear memory
\begin{align*} \Delta I_{2 \pm 2}^{(2)} \approx  &-\sqrt{\frac{8\pi}{5}} \frac{\eta M^2}{b} [5 (e^{\mp i\varphi^{(+)}} - e^{\mp i\varphi^{(-)}}) \\
&+ (e^{\mp 3i\varphi^{(+)}} - e^{\mp 3i\varphi^{(-)}})],
\end{align*}
where the argument of periastron lies on the $+x$ axis. Here the linear memory arises from different incoming and outgoing phase angles $\varphi^{(\pm)} = \pm \arccos(-1/e_0)$. See also Eqs.~(15) of Ref.~\cite{wiseman-will-memory} which are equivalent to this result.}
But for bound orbits, these modes yield purely oscillatory contributions to the polarizations.

To see a linear DC effect for bound orbits one has to examine the $m=0$ modes. For quasicircular orbits the time derivatives of the $m\neq 0$ moments are dominated by the derivatives $d^n/dt^n(e^{-i m \varphi}) \propto \omega^n e^{-i m \varphi}$, with the derivatives of $r(t)$ contributing small, oscillatory corrections at 2.5PN and higher orders.  But for the $m=0$ modes, the derivatives of $r(t)$ dominate. For example, for $l=2, m=0$, the second derivative of the source mass-quadrupole moment is
\be
\label{eq:d2I20}
I_{20}^{(2)} \propto \eta^3 M x^6 [1+O(2)].
\ee
This term arises from the noncircularity of the orbit induced by radiation-reaction. Analogous formulas hold for higher $l$ values and for the source-current moments.
Terms of this form vanish in the distant past, are slowly growing, and nonoscillatory. They contribute DC terms to the waveform at 5PN and higher orders (and are thus negligible for most sources).  Furthermore, these DC effects in bound binaries need not only result from gravitational radiation-reaction. Consider, for example, a compact body orbiting inside a dense accretion disk that surrounds a massive black hole. Viscous forces inside the disk will also cause an inward radial velocity $\dot{r}$, producing a slowly changing, nonoscillatory contribution to  $I_{l0}^{(l)}$ and $J_{l0}^{(l)}$, and a resulting DC contribution to the waveform. One can imagine other gravitational and nongravitational forces that might produce a similar DC effect.

As with the nonlinear, nonhereditary DC term discussed in Sec.~\ref{sec:crosmem}, it is not clear if these linear DC effects give rise to a permanent memory. It is possible that these DC contributions to the waveform will vanish at late times.
\subsection{\label{sec:NRmemory}Memory in numerical simulations}
While post-Newtonian calculations can only reliably determine the memory during the inspiral phase, in a quasicircular binary most of the memory will accumulate during the merger and ringdown when GW emission is greatest \cite{favata-memory-saturation} (see also the discussion in Sec.~\ref{sec:timeintegrals} above). Numerical relativity (NR) simulations would thus seem to be an obvious approach for calculating the full nonlinear memory from coalescing compact binaries. However the accurate computation of the GW memory from NR simulations faces several challenges.
\begin{figure*}[t]
$
\begin{array}{cc}
\includegraphics[angle=0, width=0.48\textwidth]{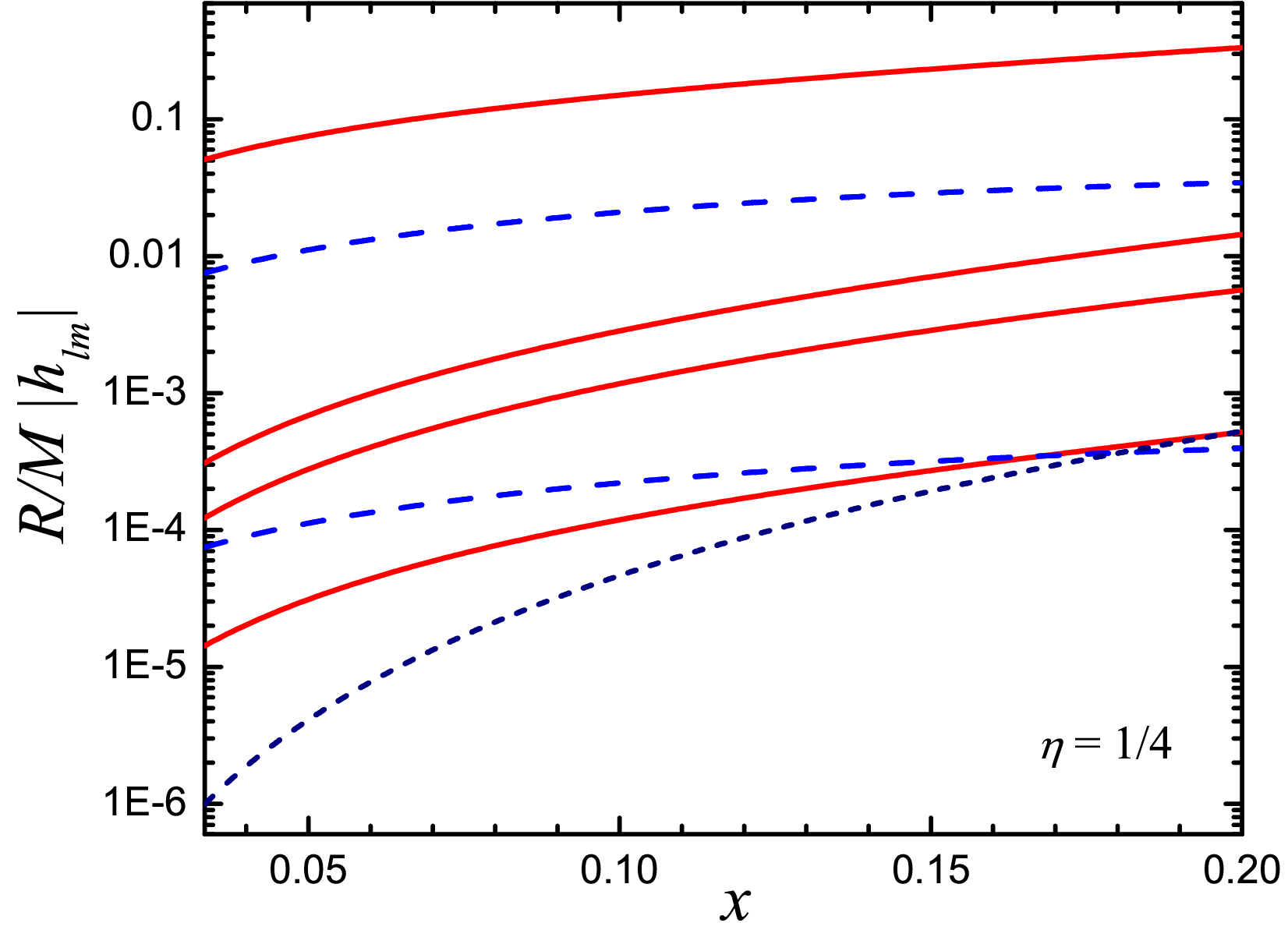} &
\includegraphics[angle=0, width=0.49\textwidth]{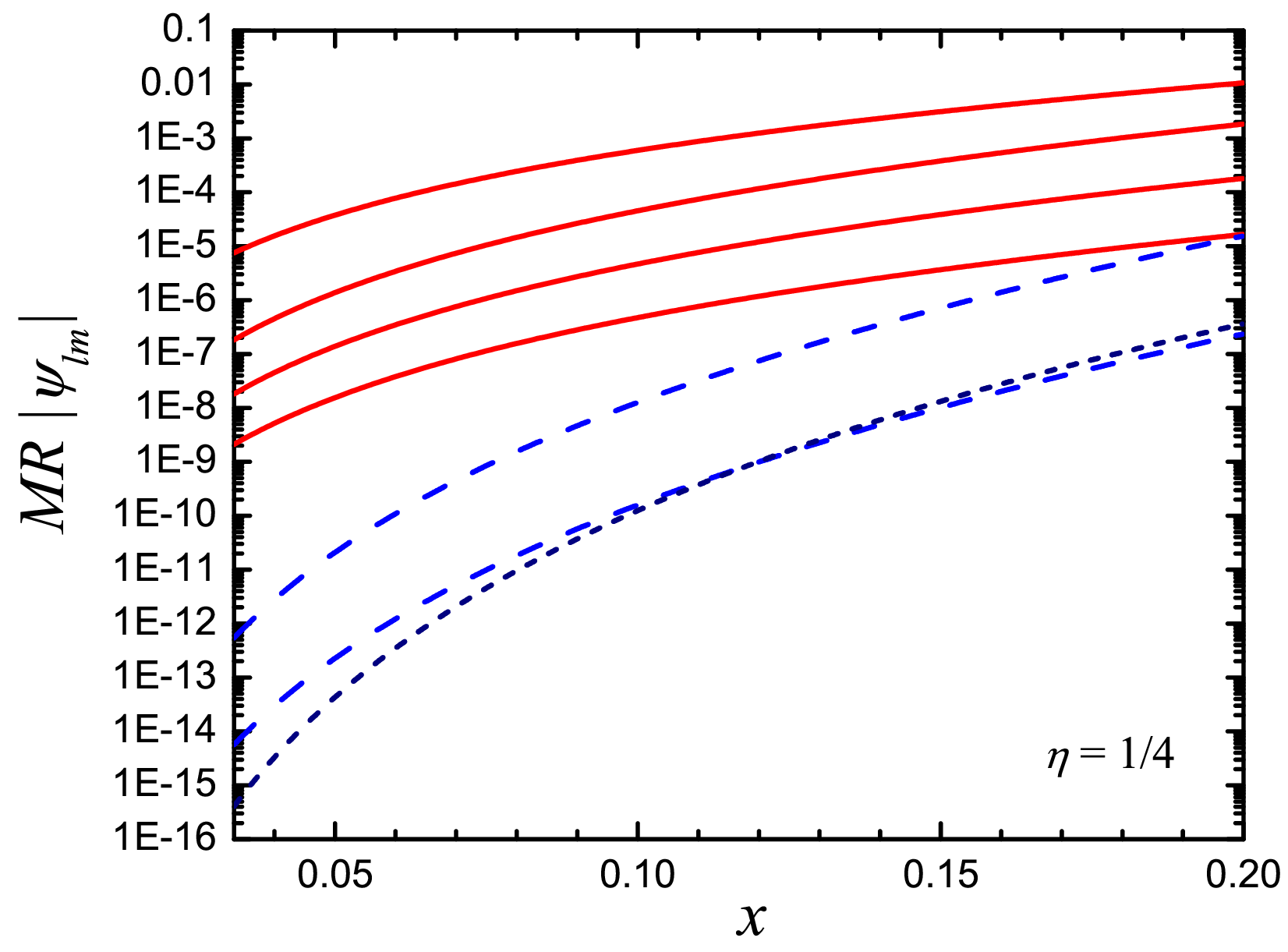}
\end{array}
$
\caption{\label{fig:x-hlm}(color online). Comparison of memory and nonmemory modes. The left plot shows the absolute value of some of the largest $h_{lm}$ modes---as well as the largest $m=0$ modes---as a function of the post-Newtonian (PN) parameter $x$. The $m\neq 0$ modes are the solid (red) curves. From top to bottom they are $h_{22}$, $h_{44}$, $h_{32}$, and $h_{42}$. These are computed from Eqs~(9.3)-(9.4) of Ref.~\cite{blanchet3pnwaveform}. The remaining curves are the $m=0$ nonoscillatory modes. The long-dashed (blue) curves are the $h_{20}$ (top) and $h_{40}$ modes [Eqs.~\eqref{eq:Hl0-mem}] expanded to 3PN order. The short-dashed (navy) curve at the bottom is the $h_{30}$ mode [the nonlinear, nonhereditary DC term given by Eq.~\eqref{eq:H30}]. The right plot is similar except it shows the absolute value of the corresponding $\psi_{lm}$ modes (see equations in Appendix \ref{app:psilm}). Note how the relative values of the oscillatory ($m\neq0$) and the memory ($m=0$) modes change in the two plots. Although the memory is relatively large in the metric-perturbation modes, the memory modes are significantly suppressed relative to the other curvature-perturbation modes.
Both plots are for equal-mass binaries ($\eta=1/4$) and span the range $x=1/30 - 1/5$. Recall that larger values of $x$ correspond to smaller orbital separations and later times. Note that the modes with odd $m$ vanish for equal-mass binaries. Several other numerically smaller modes would also appear on these plots, but are suppressed for clarity.}
\end{figure*}

Restricting our discussion to nonspinning, quasicircular binaries orbiting in the $x$-$y$ plane, numerical relativity simulations can most accurately calculate the dominant $l=m=2$ mode of the waveform. However no memory effect is present in this mode. The nonlinear memory is only present in the $m=0$ modes which have not been examined (to my knowledge) in any of the published papers on NR simulations of quasicircular, inspiralling binaries. These $m=0$ modes tend to be much smaller in magnitude and depend more sensitively than other modes on the initial conditions of the simulations. To see why this is so, we begin by considering the two ways in which gravitational radiation is typically calculated in NR simulations. The most widely used approach is to calculate the curvature scalar $\Psi_4$ and decompose its value at large $R$ into a sum over spin-weighted spherical-harmonic modes:\footnote{Note that different conventions exist in the literature for the relation between $\Psi_4$, $h_{+}$, and $h_{\times}$.}
\be
\label{eq:Psi4}
\Psi_4 = \ddot{h}_{+} - i \ddot{h}_{\times} = \sum_{l=2}^{\infty} \sum_{m=-l}^{l} \psi_{lm}(t,R) {}_{-2}Y_{lm}(\Theta,\Phi) .
\ee
Comparison with Eq.~\eqref{eq:hdecompose} shows that the modes of $\Psi_4$ can be directly related to the modes of the metric perturbation:
\be
\label{eq:psilm}
\psi_{lm} = \ddot{h}_{lm} .
\ee
Some NR simulations \cite{koppitz-etal-PRL2007,pollney-etal-spinorbitrecoil,nagar-rezzolla-metricpert,nagar-rezzolla-metricpert-erratum} also compute quantities which are more directly related to the $h_{lm}$ metric perturbation modes [see Eq.~(30) of Ref.~\cite{pollney-etal-spinorbitrecoil} for the exact relation].

We can use PN waveforms to gain insight into the relative magnitudes of the $h_{lm}$ and  $\psi_{lm}$ modes. The $h_{lm}$ modes have been calculated to 3PN order in Refs.~\cite{blanchet3pnwaveform,kidder08}. The $m=0$ pieces of $h_{lm}$ are computed to 3PN order in Sec.~\ref{sec:results} above. From these $h_{lm}$ modes it is straightforward to compute the $\psi_{lm}$ modes. This is detailed in Appendix \ref{app:psilm}. Using these results, we plot in Fig.~\ref{fig:x-hlm} the absolute values of $h_{lm}$ and $\psi_{lm}$ for some selected modes as a function of the PN parameter $x$.\footnote{Similar plots from NR simulations that compare the evolution of different $(l,m\neq0)$ modes during the merger and ringdown can be found in Fig.~3 of Ref.~\cite{schnittman-multipolarrecoil}, Fig.~11 of Ref.~\cite{baker-etal-PRD2008-gravlradcharacteristics}, Fig.~2 of Ref.~\cite{berti-etal-multipolarnonspinning}, and Fig.~24 of Ref.~\cite{buonanno-cook-pretorius}.} Both plots show some of the largest oscillatory modes for $\eta=1/4$---the $(2,2)$, $(4,4)$, $(3,2)$, and $(4,2)$ modes---as well as selected $m=0$ ``memory'' modes: $(2,0)$, $(3,0)$, and $(4,0)$. Notice that when the $h_{lm}$ modes are compared, the $h_{20}$ mode has the second largest magnitude after the $h_{22}$ mode. Note also the PN scaling of these $h_{lm}$ modes:
\bs
\label{eq:hlmscaling}
\begin{align}
&|h_{22}| \sim |h_{20}| \sim |h_{40}| \sim O(\eta x), \\
&|h_{44}| \sim |h_{32}| \sim |h_{42}| \sim O[\eta x^2 (1-3\eta)], \\
&|h_{30}| \sim O(\eta^2 x^{7/2}).
\end{align}
\es
Sill, even though the modes corresponding to the leading-order Christodoulou memory---$h_{20}$ and $h_{40}$---enter at the same PN order as $h_{22}$, they are significantly smaller in magnitude: $|h_{22}| \approx 9.7 |h_{20}| \approx 850 |h_{40}|$ at $x\approx 1/5$. This several orders-of-magnitude difference between the dominant $(2,2)$ mode and the leading-order memory modes means that the memory effect will be difficult to resolve even for simulations that directly compute the modes of the metric perturbation.

As the right plot of Fig.~\ref{fig:x-hlm} shows, the situation is significantly worse for simulations that extract the gravitational-wave content from the $\psi_{lm}$ modes. In this case the largest memory modes are many orders of magnitude smaller than the largest oscillatory modes. For comparison, the PN scaling of the $\psi_{lm}$ modes is
\bs
\label{eq:psilmscaling}
\begin{align}
&|\psi_{22}| \sim O(\eta x^4), \\
&|\psi_{44}| \sim |\psi_{32}| \sim |\psi_{42}| \sim O[\eta x^5 (1-3\eta)], \\
&|\psi_{20}| \sim |\psi_{40}| \sim O(\eta^3 x^{9} ), \\
&|\psi_{30}| \sim O(\eta^4 x^{23/2} ).
\end{align}
\es
In terms of the $\psi_{lm}$ modes, the leading-order Christodoulou-memory modes are smaller than the $\psi_{22}$ mode by \emph{five} PN orders. The relative mode magnitudes are $|\psi_{22}| \approx 690 |\psi_{20}| \approx 4.5 \times 10^4 |\psi_{40}|$ at $x \approx 1/5$. Significant improvements in the accuracy of NR simulations will be required to resolve these memory modes. Comparison with Fig.~2 of Ref.~\cite{berti-etal-multipolarnonspinning} suggests that current simulations can resolve $\psi_{lm}$ modes that are about two orders of magnitude smaller than the dominant $\psi_{22}$ mode.

Numerical relativity simulations face further difficulties in calculating the memory. For those simulations that compute the $\psi_{lm}$ curvature perturbation modes, two time integrations are required to construct the $h_{lm}$ metric perturbation modes. As discussed in detail in Ref.~\cite{berti-etal-multipolarnonspinning}, choosing these integration constants to be zero causes an artificial memory and a slope in the $h_{lm}$ modes (see their Figs.~4 and 5). This artificial memory is completely nonphysical and appears for all $(l,m)$ modes. It arises from the finite size of the initial separation and extraction radius. The initial burst of ``junk'' radiation could also contribute to this artificial memory. (These effects are likely responsible for the memory seen in the ``Lazarus'' simulations of Ref.~\cite{lazarus-PRD02}.)  Reference \cite{berti-etal-multipolarnonspinning} suggests choosing one integration constant so that the $h_{lm}$ modes have zero slope, and choosing the other constant so that the $h_{lm}$ modes have zero offset at late times. This procedure should produce physically accurate metric waveforms for the $m\neq 0$ modes (which do not possess any physical memory). However, this procedure will completely cancel any physical memory effect present in the $m=0$ modes. Since the $\psi_{l0}$ modes are so small anyway, this will not be an important issue until NR simulations become more accurate. When that time arrives, one approach to computing those integration constants is to simply match the $h_{lm}$ modes computed from the simulations with their corresponding PN expansions [Eqs.~\eqref{eq:Hl0-mem}] at large orbital separations.

Simulations that can compute the $h_{lm}$ modes directly do not have the problem of choosing two integration constants. They implicitly assume that the initial value of the asymptotic metric is zero at the start of the simulation. However, in addition to the problem of the $h_{40}$ and higher-order memory modes being small in magnitude (and thus hard to resolve), these simulations (as well as those that compute $\psi_{lm}$) must also deal with the memory's sensitivity to the past history of the source. To see how this sensitivity comes about, let us examine the leading-(Newtonian)-order piece of the largest memory mode $h_{20}$:
\be
\label{eq:h20NR}
h^{\rm NR}_{20}(T_R) = \frac{1}{\sqrt{2} R} \int_{T_0}^{T_R} U_{20}^{(\rm mem)(1)}(t) \,dt ,
\ee
where $U_{20}^{(\rm mem)(1)}$ is given by the leading-order piece of Eq.~\eqref{eq:U20mem1}, and $T_0$ is the starting time of the simulation. For a realistic, quasicircular binary that inspirals from a large initial separation, $T_0\rightarrow -\infty$. Using the Newtonian-order piece of Eq.~\eqref{eq:xdot35pn}, we can reexpress Eq.~\eqref{eq:h20NR} as an integral from $x_0(T_0) \approx M/r_0(T_0)$ to $x(T_R) \approx M/r(T_R)$, yielding
\be
h^{\rm NR}_{20}(T_R) = \frac{4}{7} \sqrt{\frac{5\pi}{6}} \frac{\eta M}{R} \left(\frac{M}{r}-\frac{M}{r_0} \right) .
\ee
Since this reduces to the memory from a realistic binary when $r_0 \rightarrow \infty$ (or at least when $r\gg M$), we can easily see that NR simulations of binaries at a finite initial separation $r_0$ would \emph{underestimate} the size of the memory. The resulting fractional error in the memory when the binary is at a smaller separation $r$ is
\be
\frac{|\delta h^{\rm NR}_{20}|}{h_{20}} \approx \frac{r}{r_0}.
\ee
For a simulation that starts with an initial separation of $10M$, the error in the memory at a harmonic coordinate separation of $5M$ is $50\%$. Thus, very large initial separations ($\gtrsim 50M$) are required to compute the memory accurately. (See also Sec.~4.2 of Ref.~\cite{arun25PNamp} for a related discussion.) As mentioned above, this issue could be circumvented by using PN expressions for the memory as initial conditions at separations where PN theory is accurate.

In addition to the problems mentioned above, other sources of error---such as gauge effects or a finite extraction radius---can contaminate the numerical waveforms and possibly swamp a small memory signal \cite{larry-privcomm}.

While directly extracting the memory from the $h_{lm}$ or $\psi_{lm}$ modes will be difficult, NR could be used to compute the memory indirectly. The effective-one-body (EOB) formalism (see Damour\cite{EOB-damour-lecnotes} for references) provides a means to compute the $h_{lm}$ modes semianalytically, calibrating their values to NR simulations. Reference \cite{favata-memory-saturation} has used an EOB calculation of the $h_{22}$ mode, combined with leading-order expressions for the energy flux, to compute the memory for the merger and ringdown phases of binary black-hole coalescence. Another approach is to use a hybrid PN/NR method to compute the memory. This would involve taking the numerical $m\neq 0$ $h_{lm}$ modes (or the time integral of the $\psi_{lm}$ modes) computed in an NR simulation and substituting them directly into Eq.~\eqref{eq:dUlmmem}. The resulting angular integrals could be evaluated analytically as was done in Sec.~\ref{sec:angularint}. The final time integral must be performed numerically to compute the $h_{l0}^{(\rm mem)}$ modes, with the PN expressions for the memory in Eqs.~\eqref{eq:Hl0-mem} used to compute the integration constant. This hybrid PN/NR procedure is currently being implemented \cite{favata-PNNR-memory} with the accurate merger waveforms described in Ref.~\cite{scheel-merger}.
\subsection{\label{sec:memdetec}Detecting memory with gravitational-wave interferometers}
While the gravitational-wave memory is an intrinsically interesting effect of general relativity, its scientific relevance relies on its potential for detection. The memory from a GW that passed through a region of space in the distant past (long before the start of the observation) is itself undetectable. In this case, the memory near a detector is simply a small, constant shift in the Minkowski metric, which is unobservable. For example, choosing a TT-coordinate system whose z axis is along the direction of the wave's propagation, the metric at late times is
\be
ds^2 = -dt^2 + (1+h_{+})dx^2 + (1-h_{+})dy^2 + 2 h_{\times} dx dy + dz^2,
\ee
where $h_{+}$ and $h_{\times}$ are constant in space and time near the detector. This metric has vanishing connection coefficients ($\Gamma^{\alpha}_{\beta \gamma} =0)$ and can be transformed to a new coordinate system via\footnote{If we require the coordinate transformation to be real-valued than we must impose the conditions $h_+\geq 0$ and $h_+^2 + h_{\times}^2 < 1$.}
\bs
\label{eq:coordtranshmem}
\begin{align}
T &=t ,\\
Z &=z ,\\
U &= x + \frac{y}{1+h_{+}} \left( h_{\times} - h_{+}^{1/2} \sqrt{1-h_{+}^2 - h_{\times}^2} \right) , \\
V &= x h_{+}^{1/2} + \frac{ y \left( h_{\times} h_{+}^{1/2} + \sqrt{1-h_{+}^2 - h_{\times}^2} \right)}{1+h_+} ,
\end{align}
\es
which has the flat metric
\[ ds^2 = -dT^2 + dU^2 + dV^2 + dZ^2. \]
What is actually observable is the buildup of the memory over some observation time as the GWs pass through the detector.

The detectability of the memory was first studied by Thorne and Braginsky \cite{braginskii-thorne,kipmemory} for unmodeled burst sources. Different types of detectors respond differently to the memory. For a memory with a characteristic rise time $\tau$, a resonant mass detector will have no significant response to the memory unless its resonant frequency is $\ll 1/\tau$ \cite{braginskii-thorne}. A typical rise time for the Christodoulou memory from a merging black-hole binary is $\tau \sim (50 - 100) M$ \cite{favata-memory-saturation}. Resonant mass detectors typically operate at frequencies $> 700\text{ Hz}$---far too high for them to be sensitive to the GW memory. An ideal laser interferometric detector whose test masses are completely unconstrained would experience a permanent deformation after the GW has passed. In practice, a LIGO-type interferometer has servomechanisms that keep the test mass positions fixed. So while LIGO is sensitive to the buildup of the memory, its design prevents it from physically ``storing'' the memory indefinitely. In contrast a detector like LISA is designed to be truly freely-falling and could (in principle) be permanently ``deformed'' by the passage of a GW with memory. However, we emphasize that this late-time deformation is not in itself directly observable except in comparison with some earlier nondeformed state.

Properly computing the memory signal that a detector could measure requires a detailed knowledge of the memory waveform throughout the inspiral and coalescence. This is particularly important for the memory (as opposed to the oscillatory pieces of the GW) because most of the signal amplitude accumulates during the merger and ringdown phases. Kennefick \cite{kennefick-memory} has computed the signal-to-noise ratio (SNR) for the memory by truncating the Newtonian-order memory formula [Eq.\eqref{eq:Hplus0pn}] at some critical radius $r_k/M > (64 \pi \eta)^{2/5} \approx 4.8 \text{ for } \eta=1/4$ [see his Eq.~(21)] and assuming the memory remains constant at that value for later times. His approach has several important limitations: (i) the Newtonian approximation is used in the strong field region where it is no longer valid; (ii) the final saturation value of the memory is set by a cutoff separation which is---to some extent---determined arbitrarily; and (iii) the rise and abrupt cutoff of the memory signal does not necessarily mimic its true evolution. The details of how the memory rises and asymptotes to its final saturation value can significantly affect the Fourier transform of the memory signal and the resulting value for the SNR [cf.~Fig.~2 of Ref.~\cite{favata-memory-saturation}].

Rather than model the evolution of the memory signal, the memory's SNR can be crudely estimated using the zero-frequency limit (ZFL) approximation \cite{smarr-zfl,bontz-price,wagoner-lowfreq,turner-neutrinomemory}. This is essentially the approach followed by Thorne \cite{kipmemory}, who examined the Christodoulou memory from unmodeled GW bursts from merging BHs.\footnote{Note that Thorne's \cite{kipmemory} normalization of the SNR should be corrected by multiplying his Eq.~(A1) by $\sqrt{2}$ (see footnote 61 of Ref.~\cite{flanagan-hughesI}).} However the ZFL is inadequate for computing the memory's SNR for some sources. As is also the case with Kennefick's analysis, the ZFL approach accurately computes the Fourier transform of the memory signal only in the regime where $f\ll 1/\tau \sim 1/(70 M)$ \cite{favata-memory-saturation}. For two $10^6 M_{\odot}$ BHs, $1/\tau \approx 0.00145 \text{Hz}$; for two $10 M_{\odot}$ BHs, $1/\tau \approx 145 \text{Hz}$. These frequencies lie near the peak sensitivity of space-based \cite{lisaweb} and terrestrial \cite{ligoweb} interferometric detectors, making the ZFL a poor tool for estimating the memory's SNR for these sources. Kennefick's \cite{kennefick-memory} estimate of the Fourier transform improves upon Thorne's ZFL approach at low frequencies by incorporating a model for the memory's rise during the early inspiral. But Kennefick's model also has large errors at frequencies $\sim 1/\tau$ where much of the SNR is accumulated. Both approaches can potentially make order-of-magnitude errors in the SNR. An improved estimate of the memory's detectability needs to incorporate a model for the memory's evolution during the merger and ringdown, as well as the inspiral. Such a model is developed in Ref.~\cite{favata-memory-saturation} and compared with the approaches of Thorne \cite{kipmemory} and Kennefick \cite{kennefick-memory}.
\section{\label{sec:conclusion}Conclusions}
The nonlinear (Christodoulou) memory is an interesting and potentially observable effect. Previous work has computed only the leading-order memory contributions to the gravitational-wave polarizations \cite{wiseman-will-memory,kennefick-memory,arun25PNamp,blanchet3pnwaveform}. Here the post-Newtonian corrections to the memory for quasicircular, inspiralling binaries have been computed. The main results are the equations listed in Sec.~\ref{sec:results} which gives the memory's contribution to the waveform's spin-weighted spherical-harmonic modes and the $+$ polarization. These results are illustrated graphically in Fig.~\ref{fig:H-theta} and show that the post-Newtonian corrections to the memory are important in the late inspiral. The calculations presented here complete the memory to 3PN order. Other nonoscillatory effects on the waveform were also discussed. In particular, a nonlinear, nonhereditary DC (nonoscillatory) term is present in the $\times$ polarization at 2.5PN order \cite{arun25PNamp}, while linear DC terms can enter the waveform at 5PN and higher orders.

While most studies have focused on the memory in quasicircular orbits, the memory should also be calculated for other types of orbits. For hyperbolic orbits with small deflection angles, the nonlinear memory has been calculated in Ref.~\cite{wiseman-will-memory}. In future work the memory for inspiralling eccentric binaries \cite{favata-eccentricmemory} will also be computed. This is especially important because circularized binaries have growing eccentricity in the past, and the nonlinear memory is sensitive to the binary's past history.

One of the main conclusions of this paper is that it will be difficult (but not impossible) to directly extract the memory from numerical relativity waveforms. Nonetheless, numerical relativity will be critical in providing a full understanding of the memory. These simulations will eventually better determine the size of the memory in the late inspiral, and they will also be able to compute the evolution and saturation value of the memory during the merger and ringdown phases of coalescence \cite{favata-memory-saturation}. The formulas presented here could be compared with the inspiral phase of those future simulations. They could also be used to set the memory's initial value at the start of a simulation.

Detecting the memory in binary black-hole coalescences could serve as a test of general relativity. However, the nonoscillatory nature of the memory makes its detection difficult. Detections will only be likely for sources whose primary waves will have high signal-to-noise ratios, such as supermassive black-hole binaries in the LISA band. This paper argues that previous signal-to-noise ratio estimates of the memory may contain large errors. This issue is investigated further in Ref.~\cite{favata-memory-saturation}.

Aside from the formal mathematical descriptions in Refs.~\cite{christodoulou-mem,frauendiener-memnote}, the memory has only been explored in detail with post-Newtonian theory. It would be interesting to also explore the memory using second-order black-hole perturbation theory. Recent work along these lines has demonstrated a memory effect in the second-order quasinormal modes of the Schwarzschild spacetime \cite{nakano-ioka-2ndorderQNM}. It would also be interesting to see how the nonlinear memory manifests itself in alternative theories of gravity.
\begin{acknowledgments}
I gratefully acknowledge: Luc Blanchet, Bala Iyer, and Larry Kidder for providing several comments and suggestions on an earlier version of this manuscript; Emanuele Berti, Christian Ott, Mark Scheel, Ulrich Sperhake, and other members of the Caltech relativity group for their comments and helpful conversations during my visit there; John Baker, Mark Hannam, Sascha Husa, Luciano Rezzolla, and other attendees of the Seventh International LISA Symposium for helpful discussions; and Andrew Gould for a helpful conversation during his visit to the KITP. I especially thank the anonymous referees for their very detailed reports which improved this paper. This research was supported in part by the National Science Foundation under
Grant No. PHY05-51164.
\end{acknowledgments}
\appendix
\section{\label{app:angularint}Angular integral of the triple product of spin-weighted spherical harmonics}
In this appendix we briefly derive a general formula for the angular integral of a product of three spin-weighted spherical-harmonic functions. Integrals of this form are needed when we evaluate the nonlinear memory contribution to the radiative mass-multipole moments $U_{lm}$ [see Eq.~\eqref{eq:dUlmmem}]. We begin by rewriting Eq.~\eqref{eq:dlms} as
\be
d^{l_j}_{m_j s_j}(\Theta) = \sum_{k_j=k_{i(j)}}^{k_{f(j)}} g_j(k_j) \left(\sin\frac{\Theta}{2}\right)^{p_j} \left(\cos\frac{\Theta}{2}\right)^{2l_j - p_j} \,,
\ee
where
\begin{widetext}
\be
g_j(k_j) = \frac{(-1)^{k_j} \left[ (l_j+m_j)!(l_j-m_j)!(l_j+s_j)!(l_j-s_j)!\right]^{1/2}}{k_j! (l_j+m_j - k_j)! (l_j-s_j-k_j)! (s_j-m_j+k_j)!} \,,
\ee
$p_j = 2k_j + s_j - m_j$, and $k_{i(j)}$ and $k_{f(j)}$ are defined analogously to the limits in Eq.~\eqref{eq:dlms}. The index $j=1,2,3$ serves only to label the three harmonics.
The integrals we wish to evaluate can then be written as:
\begin{align}
\label{eq:int3Y}
G^{s_1 s_2 s_3}_{l_1 l_2 l_2 m_1 m_2 m_3} &\equiv \int {}_{-s_1}Y^{l_1 m_1}(\Theta, \Phi) \, {}_{-s_2}Y^{l_2 m_2}(\Theta, \Phi) \, {}_{-s_3}Y^{l_3 m_3}(\Theta, \Phi) \, d\Omega 
\nonumber \\ 
&=  (-1)^{s_1 + s_2 + s_3} \frac{\left[(2l_1+1)(2l_2+1)(2l_3+1)\right]^{1/2}}{(4\pi)^{3/2}}  \int_0^{2\pi} e^{i (m_1+m_2+m_3)\Phi} \, d\Phi \, \int_0^{\pi} d^{l_1}_{m_1 s_1} d^{l_2}_{m_2 s_2} d^{l_3}_{m_3 s_3} \sin\Theta \, d\Theta \,.
\end{align}
The $\Phi$ integral is simply
\be
\int_0^{2\pi} e^{i (m_1+m_2+m_3)\Phi} \, d\Phi = 2\pi \delta^{-m_1}_{m_2+m_3} \;.
\ee
The $\Theta$ integral can be written as
\be
\int_0^{\pi} d^{l_1}_{m_1 s_1} d^{l_2}_{m_2 s_2} d^{l_3}_{m_3 s_3} \sin\Theta \, d\Theta  = 2 \sum_{k_1,k_2,k_3} g_1(k_1) g_2(k_2) g_3(k_3) \int_0^{\pi} \left(\sin\frac{\Theta}{2}\right)^{2a-1} \left(\cos\frac{\Theta}{2}\right)^{2b-1} \, d\Theta \,,
\ee
\end{widetext}
where
$a=1+(p_1+p_2+p_3)/2$ and $b=1+l_1+l_2+l_3-(p_1+p_2+p_3)/2$. The remaining $\Theta$ integral is expressible in terms of the Beta or Gamma functions and can be found in standard tables \cite{spiegel}:
\be
\int_0^{\pi} \! \left(\sin\frac{\Theta}{2}\right)^{2a-1} \! \left(\cos\frac{\Theta}{2}\right)^{2b-1} \! d\Theta = \frac{\Gamma(a) \Gamma(b)}{\Gamma(a+b)} = B(a,b).
\ee

This solution is easily extended to arbitrary products of the spin-weighted spherical harmonics. Integrals involving the complex conjugates of one or more of the ${}_{-s}Y^{lm}$ in the integrand can be computed by using ${}_{s}Y^{lm\ast} = (-1)^{s+m} {}_{-s}Y^{l\, -m}$. While computer algebra programs such as Maple or Mathematica can symbolically evaluate the angular integrals in Eq.~\eqref{eq:int3Y} directly, it is significantly faster for those programs to evaluate the result when expressed as the above sum over Gamma functions. This is helpful when many such integrals need to be evaluated for large $l$ values.

We also note that the integral \eqref{eq:int3Y} is proportional to the product of two $3$-$j$ symbols or two Clebsch-Gordon coefficients. This follows from the fact that the spin-weighted spherical harmonics are related to the Wigner matrices, which have a well-known triple-product integral [cf.~ Eq.(4.6.2) of Ref.~\cite{edmonds}, Eq.~(4.62) of Ref.~\cite{rose}, Eq.~(110.3) of Ref.~\cite{landaulifshitzQM}, or Eq.~(1.42) of Ref.~\cite{3jbook}. However, one must use extreme care when comparing notation, definitions, and phase conventions among different authors. The above formulas provide an explicit result in terms of the Gamma function, which is uniformly defined.
\section{\label{app:psilm}Calculation of the $\dot{h}_{lm}$ and $\psi_{lm}$ modes }
In this appendix we outline the computation of the explicit PN expansions (for quasicircular orbits) of the $\psi_{lm}$ modes defined in Eq.~\eqref{eq:Psi4}. These expressions are used to justify the results of Sec.~\ref{sec:NRmemory} and in generating the curves in Fig.~\ref{fig:x-hlm}. The $\psi_{lm}$ modes are directly related to the metric perturbation modes via $\psi_{lm} = \ddot{h}_{lm}$. The $h_{lm}$ modes are written as
\be
h_{lm} = 8 \sqrt{\frac{\pi}{5}} \frac{\eta M}{R} x \hat{H}_{lm}(x) e^{-im\psi} ,
\ee
where the $\hat{H}_{lm}$ are given up to 3PN order by Eqs.~(9.4) of Ref.~\cite{blanchet3pnwaveform} and Eqs.~\eqref{eq:Hl0-mem} above. The phase variable $\psi$ is related to the orbital phase $\varphi$ via [Eq.~(8.8) of Ref.~\cite{blanchet3pnwaveform}]
\be
\label{eq:psiphase}
\psi = \varphi - 3x^{3/2} \left[ 1-\frac{\eta}{2} x\right] \ln\left(\frac{x}{x_0}\right) ,
\ee
where $\ln x_0=\frac{11}{18} - \frac{2}{3}\gamma_E - \frac{4}{3}\ln 2 + \frac{2}{3}\ln\left(\frac{M}{r_0}\right)$ is related to the arbitrary constant $r_0$ appearing in the coordinate transformation in Eq.~\eqref{eq:coordtrans}. The orbital phase can be expressed explicitly as a function of time, or more conveniently as a function of $x$, to 3.5PN order [Eq.~(5) of Ref.~\cite{blanchet35PNphaseerratum}]:
\begin{widetext}
\begin{multline}
\label{eq:phi-phase}
\varphi(x) = -\frac{1}{32 \eta x^{5/2}} \left\{1 + x \left( \frac{3715}{1008} + \frac{55}{12} \eta \right) - 10\pi x^{3/2} + x^{2} \left( \frac{15\,293\,365}{1\,016\,064} + \frac{27\,145}{1008}\eta + \frac{3085}{144}\eta^2 \right)
\right. \\
+ \pi x^{5/2}\ln\left(\frac{x}{x_0}\right) \left( \frac{38\,645}{1344} - \frac{65}{16}\eta \right)
+ x^{3} \left[ \frac{12\,348\,611\,926\,451}{18\,776\,862\,720} - \frac{160}{3}\pi^2 -\frac{856}{21} (2\gamma_E + \ln16x)
\right. \\ \left. \left.
+ \left( -\frac{15\,737\,765\,635}{12\,192\,768} + \frac{2255}{48}\pi^2  \right) \! \eta
+ \frac{76\,055}{6912}\eta^2 - \frac{127\,825}{5184}\eta^3 \right] + \pi x^{7/2} \left( \frac{77\,096\,675}{2\,032\,128} + \frac{378\,515}{12\,096} \eta - \frac{74\,045}{6048} \eta^2 \right) \! +O(8) \!\! \right\} \!\! ,
\end{multline}
where we have substituted the values $\lambda = -1987/3080$ and $\theta=-11\,831/9240$ for the ambiguity parameters \cite{blanchet-damour-farese-dimreg,blanchet-iyer-hadreg,blanchetdamour3PNprl}. For reference, we also list the full $l=m=2$ mode [Eq.~(9.4a) of Ref.~\cite{blanchet3pnwaveform}]:
\begin{multline}
\label{eq:Hhat22}
\hat{H}_{22} = -\alpha \left\{ 1 + x \left(-\frac{107}{42} + \frac{55}{42} \eta\right) + 2\pi x^{3/2} + x^2 \left( - \frac{2173}{1512} - \frac{1069}{216} \eta + \frac{2047}{1512} \eta^2 \right) + x^{5/2} \left[ -\frac{107}{21}\pi + \left( \frac{34}{21}\pi - 24 i \right) \eta \right] \right.
\\ \left.
+ x^{3} \left[ \frac{27\,027\,409}{646\,800} + \frac{2\pi^2}{3} + \frac{428}{105} \left( i \pi - 2\gamma_E - \ln 16x \right) + \left( - \frac{278\,185}{33\,264} + \frac{41}{96} \pi^2 \right) \eta - \frac{20\,261}{2772} \eta^2 + \frac{114\,635}{99\,792} \eta^3 \right] +O(7) \right\} ,
\end{multline}
where $\alpha$ is $+1$ for the polarization conventions in Kidder \cite{kidder08} and $-1$ for the conventions in Blanchet et al.~\cite{blanchet3pnwaveform}.

To compute the time derivatives of the metric perturbation modes, we express $h_{lm}$ and $\psi$ entirely in terms of $x$ using the above formulas. We then use Eq.~\eqref{eq:xdot35pn} to express the time derivatives in terms of $x$ derivatives: $\dot{f}\equiv df/dt = \dot{x} f'$, where $f'\equiv df/dx$ for an arbitrary function $f[x(t)]$. A second-time derivative is then given by $\ddot{f} = \dot{x}(f'' \dot{x} + f' \dot{x}')$, where $\dot{x}' \equiv \frac{d}{dx}[\text{right-hand side of Eq.~}\eqref{eq:xdot35pn}]$. Defining $\tilde{H}_{lm} = x \hat{H}_{lm}$, the first-time derivatives of the $h_{lm}$ modes are easily seen to be
\begin{align}
\label{eq:dothlm}
\dot{h}_{lm} &= 8 \sqrt{\frac{\pi}{5}} \frac{\eta M}{R} \dot{x} e^{-im \psi} \left( \tilde{H}'_{lm} - i m \psi' \tilde{H}_{lm}  \right), \nonumber \\
&= 8 \sqrt{\frac{\pi}{5}} \frac{\eta M}{R} \dot{x} e^{-im \psi} \left[ \hat{H}_{lm}( 1 - i m \psi' x ) + x \hat{H}'_{lm} \right] , \nonumber \\
&\equiv -16 i \sqrt{\frac{\pi}{5}} \frac{\eta}{R} x^{5/2} \hat{\dot{h}}_{lm} e^{-im\psi}
\end{align}
These expressions are used in evaluating the energy flux [Eq.~\eqref{eq:dEdtdOmega-hlm}] and the memory contribution to the radiative-mass multipoles [Eq.~\eqref{eq:dUlmmem}]. For reference we list only the value for the $l=m=2$ mode:
\begin{multline}
\label{eq:hdothat22}
\hat{\dot{h}}_{22} = -\alpha \left\{ 1 + x \left(-\frac{107}{42} + \frac{55}{42} \eta\right) + 2\pi x^{3/2} + x^2 \left( - \frac{2173}{1512} - \frac{1069}{216} \eta + \frac{2047}{1512} \eta^2 \right) + x^{5/2} \left[ -\frac{107}{21}\pi + \left( \frac{34}{21}\pi - \frac{88}{5} i \right) \eta \right] \right.
\\ \left.
+ x^{3} \left[ \frac{27\,027\,409}{646\,800} + \frac{2\pi^2}{3} + \frac{428}{105} \left( i \pi - 2\gamma_E - \ln 16x \right) + \left( - \frac{278\,185}{33\,264} + \frac{41}{96} \pi^2 \right) \eta - \frac{20\,261}{2772} \eta^2 + \frac{114\,635}{99\,792} \eta^3 \right] +O(7) \right\}.
\end{multline}
The values of the $\dot{h}_{l0}$ modes can be inferred from Eqs.~\eqref{eq:hlm} and \eqref{eq:dUlmPNexpand} above.

Taking another time derivative lets us write the $\psi_{lm}$ modes as
\begin{align}
\label{eq:psilmderiv}
\psi_{lm} &=\ddot{h}_{lm} = 8 \sqrt{\frac{\pi}{5}} \frac{\eta M}{R} \dot{x} e^{-im \psi} \left\{ \dot{x}' \left[ (-im \psi') \tilde{H}_{lm} + \tilde{H}_{lm}' \right] + \dot{x} \left[ \tilde{H}_{lm}'' - 2im \psi' \tilde{H}_{lm}' - m(i \psi'' + m{\psi'}^2 ) \tilde{H}_{lm} \right] \right\} , \nonumber \\
&\equiv -32 \sqrt{\frac{\pi}{5}} \frac{\eta}{MR} x^4 \hat{\psi}_{lm} e^{-i m \psi} .
\end{align}
We list here the $l=m=2$ and the $m=0$ modes:\footnote{Note that the expression for $\psi_{22}$ agrees with the 2.5PN order result in Eqs.~(3.7a) and (A1a) of Ref.~\cite{berti-etal-multipolarnonspinning} when their parameter $\varpi=-24$. Note also that the 3PN order expressions for $\hat{H}_{22}$, $\hat{\dot{h}}_{22}$, and $\hat{\psi}_{22}$ are almost identical except for the imaginary part of the 2.5PN order term.}
\bs
\label{eq:psihatlm}
\begin{multline}
\label{eq:psihat22}
\hat{\psi}_{22} = -\alpha \left\{ 1 + x \left(-\frac{107}{42} + \frac{55}{42} \eta\right) + 2\pi x^{3/2} + x^2 \left( - \frac{2173}{1512} - \frac{1069}{216} \eta + \frac{2047}{1512} \eta^2 \right) + x^{5/2} \left[ -\frac{107}{21}\pi + \left( \frac{34}{21}\pi - \frac{8}{5} i \right) \eta \right] \right.
\\ \left.
+ x^{3} \left[ \frac{27\,027\,409}{646\,800} + \frac{2\pi^2}{3} + \frac{428}{105} \left( i \pi - 2\gamma_E - \ln 16x \right) + \left( - \frac{278\,185}{33\,264} + \frac{41}{96} \pi^2 \right) \eta - \frac{20\,261}{2772} \eta^2 + \frac{114\,635}{99\,792} \eta^3 \right] +O(7) \right\} ,
\end{multline}
\begin{multline}
\label{eq:psihat20}
\hat{\psi}_{20} =  -\alpha \frac{512}{7\sqrt{6}} \eta^2 x^5 \left\{ 1 + x \left( -\frac{1531}{210} - \frac{27}{10}\eta \right) + \frac{46}{5} \pi x^{3/2} + x^2 \left( \frac{99\,697\,607}{7\,983\,360} + \frac{1\,296\,181}{73\,920} \eta - \frac{63}{110} \eta^2 \right)
\right. \\
+ \pi x^{5/2} \left( - \frac{5323}{84} -\frac{203}{5} \eta \right)  + x^3 \left[ \frac{581\,338\,215\,151}{2\,223\,936\,000} + \frac{104 \pi^2}{3} -\frac{11\,128}{525} (2 \gamma_E + \ln16x)
\right. \\ \left. \left.
+ \left( -\frac{410\,304\,892\,447}{1\,245\,404\,160} + \frac{861\pi^2}{80}  \right) \eta + \frac{475\,729\,817}{8\,648\,640} \eta^2 + \frac{600\,931}{45\,760} \eta^3  \right]    +  O(7) \right\},
\end{multline}
\be
\label{eq:psihat30}
\hat{\psi}_{30} = -\alpha \frac{64\,512}{25\sqrt{42}} i \eta^3 x^{15/2} [1+O(2)] ,
\ee
\begin{multline}
\label{eq:psihat40}
\hat{\psi}_{40} = -\alpha \frac{128}{315 \sqrt{2}} \eta^2 x^5 \left\{ 1 + x \left( -\frac{720\,109}{36\,960} + \frac{4913}{88}\eta \right) + \frac{46}{5}\pi x^{3/2} + x^2 \left( \frac{28\,800\,120\,359}{207\,567\,360} - \frac{744\,931\,807}{1\,441\,440} \eta - \frac{3\,054\,527}{13\,728} \eta^2 \right)
\right. \\
+ \pi x^{5/2} \left( - \frac{8\,388\,377}{36\,960} +\frac{28\,063}{40} \eta \right)  + x^3 \left[ - \frac{7\,998\,029\,861}{1\,482\,624\,000} + \frac{104 \pi^2}{3} -\frac{11\,128}{525} (2 \gamma_E + \ln16x)
\right. \\ \left. \left.
+ \left( \frac{604\,409\,577\,763}{1\,245\,404\,160} + \frac{861\pi^2}{80}  \right) \eta + \frac{76\,345\,025\,027}{34\,594\,560} \eta^2 + \frac{434\,249}{1248} \eta^3  \right] +  O(7) \right\},
\end{multline}
\begin{multline}
\label{eq:psihat60}
\hat{\psi}_{60} = \alpha \frac{1678}{1155\sqrt{273}} \eta^2 x^6 \left[ 1 - \frac{3612}{839} \eta + x \left( - \frac{220\,594\,721}{12\,685\,680} + \frac{428\,789}{5034} \eta - \frac{62\,685}{839} \eta^2 \right) + \pi x^{3/2} \left( \frac{10\,281}{839} - \frac{43\,428}{839} \eta \right)
\right. \\ \left.
+ x^2 \left( \frac{46\,290\,704\,197}{369\,696\,960} - \frac{425\,708\,869\,429}{646\,969\,680} \eta + \frac{274\,872\,827}{342\,312} \eta^2 + \frac{28\,016\,681}{85\,578} \eta^3 \right) + O(5) \right],
\end{multline}
\begin{multline}
\label{eq:psihat80}
\hat{\psi}_{80} = -\alpha \frac{529\,207}{1\,737\,450 \sqrt{119}} \eta^2 x^7 \left[ 1- \frac{452\,070}{75\,601} \eta + \frac{733\,320}{75\,601} \eta^2
\right. \\ \left.
+ x \left( -\frac{60\,519\,713}{3\,628\,848} + \frac{26\,190\,406\,133}{241\,318\,392} \eta - \frac{2\,034\,568\,425}{10\,054\,933} \eta^2 + \frac{116\,967\,870}{1\,436\,419} \eta^3 \right)
 + O(3) \right],
\end{multline}
\be
\label{eq:psihat100}
\hat{\psi}_{10\,0} = \alpha \frac{525\,221}{4\,922\,775 \sqrt{154}} \eta^2 x^8 \left[ 1 - \frac{79\,841\,784}{9\,979\,199} \eta + \frac{198\,570\,240}{9\,979\,199} \eta^2 - \frac{172\,307\,520}{9979\,199} \eta^3 + O(2) \right].
\ee
\es
The remaining $\psi_{lm}$ modes for $m \neq 0$ can be easily computed by applying Eq.~\eqref{eq:psilmderiv} to Eqs.~(9.4) of Ref.~\cite{blanchet3pnwaveform} (see also Appendix A of Ref.~\cite{berti-etal-multipolarnonspinning}). These modes (including some not listed here) are used in the construction of Fig.~\ref{fig:x-hlm}.
\end{widetext}
\bibliography{PNmemArXivV2}

\begin{thebibliography}{115}
\expandafter\ifx\csname natexlab\endcsname\relax\def\natexlab#1{#1}\fi
\expandafter\ifx\csname bibnamefont\endcsname\relax
  \def\bibnamefont#1{#1}\fi
\expandafter\ifx\csname bibfnamefont\endcsname\relax
  \def\bibfnamefont#1{#1}\fi
\expandafter\ifx\csname citenamefont\endcsname\relax
  \def\citenamefont#1{#1}\fi
\expandafter\ifx\csname url\endcsname\relax
  \def\url#1{\texttt{#1}}\fi
\expandafter\ifx\csname urlprefix\endcsname\relax\def\urlprefix{URL }\fi
\providecommand{\bibinfo}[2]{#2}
\providecommand{\eprint}[2][]{\url{#2}}

\bibitem[{\citenamefont{{Zel'Dovich} and
  {Polnarev}}(1974)}]{zeldovich-polnarev}
\bibinfo{author}{\bibfnamefont{Y.~B.} \bibnamefont{{Zel'Dovich}}}
  \bibnamefont{and} \bibinfo{author}{\bibfnamefont{A.~G.}
  \bibnamefont{{Polnarev}}}, \bibinfo{journal}{Astron.~Zh.}
  \textbf{\bibinfo{volume}{51}}, \bibinfo{pages}{30} (\bibinfo{year}{1974}),
  \bibinfo{note}{[Sov.~Astron.~{\bf 18}, 17 (1974)]}.

\bibitem[{\citenamefont{{Braginsky} and
  {Grishchuk}}(1985)}]{braginskii-grishchuk}
\bibinfo{author}{\bibfnamefont{V.~B.} \bibnamefont{{Braginsky}}}
  \bibnamefont{and} \bibinfo{author}{\bibfnamefont{L.~P.}
  \bibnamefont{{Grishchuk}}}, \bibinfo{journal}{Zh.~Eksp.~Teor.~Fiz.}
  \textbf{\bibinfo{volume}{89}}, \bibinfo{pages}{744} (\bibinfo{year}{1985}),
  \bibinfo{note}{[Sov.~Phys.~JETP {\bf 62}, 427 (1985)]}.

\bibitem[{\citenamefont{{Braginsky} and {Thorne}}(1987)}]{braginskii-thorne}
\bibinfo{author}{\bibfnamefont{V.~B.} \bibnamefont{{Braginsky}}}
  \bibnamefont{and} \bibinfo{author}{\bibfnamefont{K.~S.}
  \bibnamefont{{Thorne}}}, \bibinfo{journal}{Nature (London)}
  \textbf{\bibinfo{volume}{327}}, \bibinfo{pages}{123} (\bibinfo{year}{1987}).

\bibitem[{\citenamefont{{Turner}}(1977)}]{turner-unbound}
\bibinfo{author}{\bibfnamefont{M.}~\bibnamefont{{Turner}}},
  \bibinfo{journal}{Astrophys.~J.} \textbf{\bibinfo{volume}{216}},
  \bibinfo{pages}{610} (\bibinfo{year}{1977}).

\bibitem[{\citenamefont{{Turner} and {Will}}(1978)}]{turner-will}
\bibinfo{author}{\bibfnamefont{M.}~\bibnamefont{{Turner}}} \bibnamefont{and}
  \bibinfo{author}{\bibfnamefont{C.~M.} \bibnamefont{{Will}}},
  \bibinfo{journal}{Astrophys.~J.} \textbf{\bibinfo{volume}{220}},
  \bibinfo{pages}{1107} (\bibinfo{year}{1978}).

\bibitem[{\citenamefont{{Kovacs} and {Thorne}}(1978)}]{kovacs-thorne-IV}
\bibinfo{author}{\bibfnamefont{S.~J.} \bibnamefont{{Kovacs}},
  \bibfnamefont{Jr.}} \bibnamefont{and} \bibinfo{author}{\bibfnamefont{K.~S.}
  \bibnamefont{{Thorne}}}, \bibinfo{journal}{Astrophys.~J.}
  \textbf{\bibinfo{volume}{224}}, \bibinfo{pages}{62} (\bibinfo{year}{1978}).

\bibitem[{\citenamefont{{Burrows} and {Hayes}}(1996)}]{burrows-hayesPRL96}
\bibinfo{author}{\bibfnamefont{A.}~\bibnamefont{{Burrows}}} \bibnamefont{and}
  \bibinfo{author}{\bibfnamefont{J.}~\bibnamefont{{Hayes}}},
  \bibinfo{journal}{Phys.~Rev.~Lett.} \textbf{\bibinfo{volume}{76}},
  \bibinfo{pages}{352} (\bibinfo{year}{1996}), \eprint{arXiv:astro-ph/9511106}.

\bibitem[{\citenamefont{{Buonanno} et~al.}(2005)\citenamefont{{Buonanno},
  {Sigl}, {Raffelt}, {Janka}, and {M{\"u}ller}}}]{buonanno-stochasticBGmemory}
\bibinfo{author}{\bibfnamefont{A.}~\bibnamefont{{Buonanno}}},
  \bibinfo{author}{\bibfnamefont{G.}~\bibnamefont{{Sigl}}},
  \bibinfo{author}{\bibfnamefont{G.~G.} \bibnamefont{{Raffelt}}},
  \bibinfo{author}{\bibfnamefont{H.-T.} \bibnamefont{{Janka}}},
  \bibnamefont{and}
  \bibinfo{author}{\bibfnamefont{E.}~\bibnamefont{{M{\"u}ller}}},
  \bibinfo{journal}{Phys.~Rev.~D} \textbf{\bibinfo{volume}{72}},
  \bibinfo{pages}{084001} (\bibinfo{year}{2005}),
  \eprint{arXiv:astro-ph/0412277}.

\bibitem[{\citenamefont{{M{\"u}ller} et~al.}(2004)\citenamefont{{M{\"u}ller},
  {Rampp}, {Buras}, {Janka}, and {Shoemaker}}}]{muller-GWcorecollapse}
\bibinfo{author}{\bibfnamefont{E.}~\bibnamefont{{M{\"u}ller}}},
  \bibinfo{author}{\bibfnamefont{M.}~\bibnamefont{{Rampp}}},
  \bibinfo{author}{\bibfnamefont{R.}~\bibnamefont{{Buras}}},
  \bibinfo{author}{\bibfnamefont{H.-T.} \bibnamefont{{Janka}}},
  \bibnamefont{and} \bibinfo{author}{\bibfnamefont{D.~H.}
  \bibnamefont{{Shoemaker}}}, \bibinfo{journal}{Astrophys.~J.}
  \textbf{\bibinfo{volume}{603}}, \bibinfo{pages}{221} (\bibinfo{year}{2004}),
  \eprint{arXiv:astro-ph/0309833}.

\bibitem[{\citenamefont{{Kotake} et~al.}(2006)\citenamefont{{Kotake}, {Sato},
  and {Takahashi}}}]{kotake-sato-SNmemory}
\bibinfo{author}{\bibfnamefont{K.}~\bibnamefont{{Kotake}}},
  \bibinfo{author}{\bibfnamefont{K.}~\bibnamefont{{Sato}}}, \bibnamefont{and}
  \bibinfo{author}{\bibfnamefont{K.}~\bibnamefont{{Takahashi}}},
  \bibinfo{journal}{Rep.~Prog.~Phys.} \textbf{\bibinfo{volume}{69}},
  \bibinfo{pages}{971} (\bibinfo{year}{2006}), \eprint{arXiv:astro-ph/0509456}.

\bibitem[{\citenamefont{{Davies} et~al.}(2002)\citenamefont{{Davies}, {King},
  {Rosswog}, and {Wynn}}}]{davies-king-SNkick-memory}
\bibinfo{author}{\bibfnamefont{M.~B.} \bibnamefont{{Davies}}},
  \bibinfo{author}{\bibfnamefont{A.}~\bibnamefont{{King}}},
  \bibinfo{author}{\bibfnamefont{S.}~\bibnamefont{{Rosswog}}},
  \bibnamefont{and} \bibinfo{author}{\bibfnamefont{G.}~\bibnamefont{{Wynn}}},
  \bibinfo{journal}{Astrophys.~J.~Lett.} \textbf{\bibinfo{volume}{579}},
  \bibinfo{pages}{L63} (\bibinfo{year}{2002}), \eprint{arXiv:astro-ph/0204358}.

\bibitem[{\citenamefont{{Mosquera Cuesta} and {Bonilla
  Quintero}}(2008)}]{cuesta-pulsarkickmem}
\bibinfo{author}{\bibfnamefont{H.~J.} \bibnamefont{{Mosquera Cuesta}}}
  \bibnamefont{and} \bibinfo{author}{\bibfnamefont{C.~A.} \bibnamefont{{Bonilla
  Quintero}}}, \bibinfo{journal}{Journal of Cosmology and Astro-Particle
  Physics} \textbf{\bibinfo{volume}{11}}, \bibinfo{pages}{6}
  (\bibinfo{year}{2008}), \eprint{arXiv:0711.3046v2 [astro-ph]}.

\bibitem[{\citenamefont{{Kotake} et~al.}(2009)\citenamefont{{Kotake},
  {Iwakami}, {Ohnishi}, and {Yamada}}}]{kotake-etal-sasiApJL09}
\bibinfo{author}{\bibfnamefont{K.}~\bibnamefont{{Kotake}}},
  \bibinfo{author}{\bibfnamefont{W.}~\bibnamefont{{Iwakami}}},
  \bibinfo{author}{\bibfnamefont{N.}~\bibnamefont{{Ohnishi}}},
  \bibnamefont{and} \bibinfo{author}{\bibfnamefont{S.}~\bibnamefont{{Yamada}}},
  \bibinfo{journal}{Astrophys.~J.~Lett.} \textbf{\bibinfo{volume}{697}},
  \bibinfo{pages}{L133} (\bibinfo{year}{2009}), \eprint{arXiv:0904.4300
  [astro-ph.HE]}.

\bibitem[{\citenamefont{{Turner}}(1978)}]{turner-neutrinomemory}
\bibinfo{author}{\bibfnamefont{M.~S.} \bibnamefont{{Turner}}},
  \bibinfo{journal}{Nature (London)} \textbf{\bibinfo{volume}{274}},
  \bibinfo{pages}{565} (\bibinfo{year}{1978}).

\bibitem[{\citenamefont{{Epstein}}(1978)}]{epstein-neutrinomemory}
\bibinfo{author}{\bibfnamefont{R.}~\bibnamefont{{Epstein}}},
  \bibinfo{journal}{Astrophys.~J.} \textbf{\bibinfo{volume}{223}},
  \bibinfo{pages}{1037} (\bibinfo{year}{1978}).

\bibitem[{\citenamefont{{Loveridge}}(2004)}]{loveridge-pulsarkickmemory}
\bibinfo{author}{\bibfnamefont{L.~C.} \bibnamefont{{Loveridge}}},
  \bibinfo{journal}{Phys.~Rev.~D} \textbf{\bibinfo{volume}{69}},
  \bibinfo{pages}{024008} (\bibinfo{year}{2004}),
  \eprint{arXiv:astro-ph/0309362}.

\bibitem[{\citenamefont{{Ott}}(2009)}]{ott-corecollapsereview}
\bibinfo{author}{\bibfnamefont{C.~D.} \bibnamefont{{Ott}}},
  \bibinfo{journal}{Classical and Quantum Gravity}
  \textbf{\bibinfo{volume}{26}}, \bibinfo{pages}{063001}
  (\bibinfo{year}{2009}), \eprint{arXiv:0809.0695 [astro-ph]}.

\bibitem[{\citenamefont{{Sago} et~al.}(2004)\citenamefont{{Sago}, {Ioka},
  {Nakamura}, and {Yamazaki}}}]{sago-GRBmemory}
\bibinfo{author}{\bibfnamefont{N.}~\bibnamefont{{Sago}}},
  \bibinfo{author}{\bibfnamefont{K.}~\bibnamefont{{Ioka}}},
  \bibinfo{author}{\bibfnamefont{T.}~\bibnamefont{{Nakamura}}},
  \bibnamefont{and}
  \bibinfo{author}{\bibfnamefont{R.}~\bibnamefont{{Yamazaki}}},
  \bibinfo{journal}{Phys.~Rev.~D} \textbf{\bibinfo{volume}{70}},
  \bibinfo{pages}{104012} (\bibinfo{year}{2004}), \eprint{arXiv:gr-qc/0405067}.

\bibitem[{\citenamefont{{Hiramatsu} et~al.}(2005)\citenamefont{{Hiramatsu},
  {Kotake}, {Kudoh}, and {Taruya}}}]{hiramatsu-kotake-GRBmemory}
\bibinfo{author}{\bibfnamefont{T.}~\bibnamefont{{Hiramatsu}}},
  \bibinfo{author}{\bibfnamefont{K.}~\bibnamefont{{Kotake}}},
  \bibinfo{author}{\bibfnamefont{H.}~\bibnamefont{{Kudoh}}}, \bibnamefont{and}
  \bibinfo{author}{\bibfnamefont{A.}~\bibnamefont{{Taruya}}},
  \bibinfo{journal}{Mon.~Not.~R.~Astron.~Soc.} \textbf{\bibinfo{volume}{364}},
  \bibinfo{pages}{1063} (\bibinfo{year}{2005}),
  \eprint{arXiv:astro-ph/0509787}.

\bibitem[{\citenamefont{{Segalis} and {Ori}}(2001)}]{segalis-ori-GRBmemory}
\bibinfo{author}{\bibfnamefont{E.~B.} \bibnamefont{{Segalis}}}
  \bibnamefont{and} \bibinfo{author}{\bibfnamefont{A.}~\bibnamefont{{Ori}}},
  \bibinfo{journal}{Phys.~Rev.~D} \textbf{\bibinfo{volume}{64}},
  \bibinfo{pages}{064018} (\bibinfo{year}{2001}), \eprint{arXiv:gr-qc/0101117}.

\bibitem[{\citenamefont{{Favata}}(2009{\natexlab{a}})}]{favata-lisa7confproc}
\bibinfo{author}{\bibfnamefont{M.}~\bibnamefont{{Favata}}},
  \bibinfo{journal}{J.~Phys.: Conf.~Ser.} p. \bibinfo{pages}{012043}
  (\bibinfo{year}{2009}{\natexlab{a}}), \eprint{arXiv:0811.3451 [astro-ph]}.

\bibitem[{\citenamefont{{Thorne}}(1992)}]{kipmemory}
\bibinfo{author}{\bibfnamefont{K.~S.} \bibnamefont{{Thorne}}},
  \bibinfo{journal}{Phys.~Rev.~D} \textbf{\bibinfo{volume}{45}},
  \bibinfo{pages}{520} (\bibinfo{year}{1992}).

\bibitem[{\citenamefont{{Payne}}(1983)}]{payne-zfl}
\bibinfo{author}{\bibfnamefont{P.~N.} \bibnamefont{{Payne}}},
  \bibinfo{journal}{Phys.~Rev.~D} \textbf{\bibinfo{volume}{28}},
  \bibinfo{pages}{1894} (\bibinfo{year}{1983}).

\bibitem[{\citenamefont{{Blanchet} and {Damour}}(1990)}]{blanchet-thesis}
\bibinfo{author}{\bibfnamefont{L.}~\bibnamefont{{Blanchet}}} \bibnamefont{and}
  \bibinfo{author}{\bibfnamefont{T.}~\bibnamefont{{Damour}}},
  \emph{\bibinfo{title}{Luc Blanchet, Th\`{e}se d'Habilitation}}
  (\bibinfo{address}{Universit\'{e} Pierre et Marie Curie, Paris},
  \bibinfo{year}{1990}), chap. \bibinfo{chapter}{{Tail effects in the
  generation of gravitational waves}}, pp. \bibinfo{pages}{195--227}.

\bibitem[{\citenamefont{{Blanchet} and
  {Damour}}(1992)}]{blanchet-damour-hereditary}
\bibinfo{author}{\bibfnamefont{L.}~\bibnamefont{{Blanchet}}} \bibnamefont{and}
  \bibinfo{author}{\bibfnamefont{T.}~\bibnamefont{{Damour}}},
  \bibinfo{journal}{Phys.~Rev.~D} \textbf{\bibinfo{volume}{46}},
  \bibinfo{pages}{4304} (\bibinfo{year}{1992}).

\bibitem[{\citenamefont{{Christodoulou}}(1991)}]{christodoulou-mem}
\bibinfo{author}{\bibfnamefont{D.}~\bibnamefont{{Christodoulou}}},
  \bibinfo{journal}{Phys.~Rev.~Lett.} \textbf{\bibinfo{volume}{67}},
  \bibinfo{pages}{1486} (\bibinfo{year}{1991}).

\bibitem[{\citenamefont{Thorne}(1980)}]{kiprmp}
\bibinfo{author}{\bibfnamefont{K.~S.} \bibnamefont{Thorne}},
  \bibinfo{journal}{Rev. Mod. Phys.} \textbf{\bibinfo{volume}{52}},
  \bibinfo{pages}{299} (\bibinfo{year}{1980}).

\bibitem[{\citenamefont{Blanchet}(2006)}]{blanchetLRR}
\bibinfo{author}{\bibfnamefont{L.}~\bibnamefont{Blanchet}},
  \bibinfo{journal}{Living Rev. Relativity} \textbf{\bibinfo{volume}{9}},
  \bibinfo{pages}{4} (\bibinfo{year}{2006}), \bibinfo{note}{[Online Article]:
  cited [30 Sept.~2008]}, \eprint{gr-qc/0202016},
  \urlprefix\url{http://www.livingreviews.org/lrr-2006-4}.

\bibitem[{\citenamefont{Will and Wiseman}(1996)}]{will-wiseman-2pn}
\bibinfo{author}{\bibfnamefont{C.~M.} \bibnamefont{Will}} \bibnamefont{and}
  \bibinfo{author}{\bibfnamefont{A.~G.} \bibnamefont{Wiseman}},
  \bibinfo{journal}{Phys. Rev. D} \textbf{\bibinfo{volume}{54}},
  \bibinfo{pages}{4813} (\bibinfo{year}{1996}), \eprint{arXiv:gr-qc/9608012v1}.

\bibitem[{\citenamefont{{Isaacson}}(1968)}]{isaacson}
\bibinfo{author}{\bibfnamefont{R.~A.} \bibnamefont{{Isaacson}}},
  \bibinfo{journal}{Phys.~Rev.} \textbf{\bibinfo{volume}{166}},
  \bibinfo{pages}{1272} (\bibinfo{year}{1968}).

\bibitem[{\citenamefont{Misner et~al.}(1973)\citenamefont{Misner, Thorne, and
  Wheeler}}]{mtw}
\bibinfo{author}{\bibfnamefont{C.~W.} \bibnamefont{Misner}},
  \bibinfo{author}{\bibfnamefont{K.~S.} \bibnamefont{Thorne}},
  \bibnamefont{and} \bibinfo{author}{\bibfnamefont{J.~A.}
  \bibnamefont{Wheeler}}, \emph{\bibinfo{title}{Gravitation}}
  (\bibinfo{publisher}{Freeman}, \bibinfo{address}{San Francisco},
  \bibinfo{year}{1973}).

\bibitem[{\citenamefont{{Wiseman} and {Will}}(1991)}]{wiseman-will-memory}
\bibinfo{author}{\bibfnamefont{A.~G.} \bibnamefont{{Wiseman}}}
  \bibnamefont{and} \bibinfo{author}{\bibfnamefont{C.~M.}
  \bibnamefont{{Will}}}, \bibinfo{journal}{Phys.~Rev.~D}
  \textbf{\bibinfo{volume}{44}}, \bibinfo{pages}{R2945} (\bibinfo{year}{1991}).

\bibitem[{\citenamefont{{Blanchet}}(1997)}]{blanchet-marck-lasota}
\bibinfo{author}{\bibfnamefont{L.}~\bibnamefont{{Blanchet}}}, in
  \emph{\bibinfo{booktitle}{Relativistic Gravitation and Gravitational
  Radiation}}, edited by \bibinfo{editor}{\bibfnamefont{J.-A.}
  \bibnamefont{{Marck}}} \bibnamefont{and}
  \bibinfo{editor}{\bibfnamefont{J.-P.} \bibnamefont{{Lasota}}}
  (\bibinfo{publisher}{Cambridge University Press},
  \bibinfo{address}{Cambridge}, \bibinfo{year}{1997}), p.~\bibinfo{pages}{33},
  \eprint{arXiv:gr-qc/9607025}.

\bibitem[{\citenamefont{{Wiseman}}(1993)}]{wiseman-tails}
\bibinfo{author}{\bibfnamefont{A.~G.} \bibnamefont{{Wiseman}}},
  \bibinfo{journal}{Phys.~Rev.~D} \textbf{\bibinfo{volume}{48}},
  \bibinfo{pages}{4757} (\bibinfo{year}{1993}).

\bibitem[{\citenamefont{{Arun} et~al.}(2004)\citenamefont{{Arun}, {Blanchet},
  {Iyer}, and {Qusailah}}}]{arun25PNamp}
\bibinfo{author}{\bibfnamefont{K.~G.} \bibnamefont{{Arun}}},
  \bibinfo{author}{\bibfnamefont{L.}~\bibnamefont{{Blanchet}}},
  \bibinfo{author}{\bibfnamefont{B.~R.} \bibnamefont{{Iyer}}},
  \bibnamefont{and} \bibinfo{author}{\bibfnamefont{M.~S.~S.}
  \bibnamefont{{Qusailah}}}, \bibinfo{journal}{Class.~Quantum Grav.}
  \textbf{\bibinfo{volume}{21}}, \bibinfo{pages}{3771} (\bibinfo{year}{2004}),
  \bibinfo{note}{{\bf 22}, 3115 (2005)}, \eprint{arXiv:gr-qc/0404085v4}.

\bibitem[{\citenamefont{{Favata}}(2009{\natexlab{b}})}]{favata-memory-saturati%
on}
\bibinfo{author}{\bibfnamefont{M.}~\bibnamefont{{Favata}}},
  \bibinfo{journal}{Astrophys.~J.~Lett.} p. \bibinfo{pages}{159}
  (\bibinfo{year}{2009}{\natexlab{b}}), \eprint{arXiv:0902.3660 [astro-ph.SR]}.

\bibitem[{\citenamefont{{Wagoner} and {Will}}(1976)}]{wagoner-will}
\bibinfo{author}{\bibfnamefont{R.~V.} \bibnamefont{{Wagoner}}}
  \bibnamefont{and} \bibinfo{author}{\bibfnamefont{C.~M.}
  \bibnamefont{{Will}}}, \bibinfo{journal}{Astrophys.~J.}
  \textbf{\bibinfo{volume}{210}}, \bibinfo{pages}{764} (\bibinfo{year}{1976}),
  \bibinfo{note}{{\bf 215}, 984 (1977)}.

\bibitem[{\citenamefont{{Blanchet} et~al.}(1996)\citenamefont{{Blanchet},
  {Iyer}, {Will}, and {Wiseman}}}]{blanchet-iyer-will-wiseman-CQG-2PNwaveform}
\bibinfo{author}{\bibfnamefont{L.}~\bibnamefont{{Blanchet}}},
  \bibinfo{author}{\bibfnamefont{B.~R.} \bibnamefont{{Iyer}}},
  \bibinfo{author}{\bibfnamefont{C.~M.} \bibnamefont{{Will}}},
  \bibnamefont{and} \bibinfo{author}{\bibfnamefont{A.~G.}
  \bibnamefont{{Wiseman}}}, \bibinfo{journal}{Class.~Quantum Grav.}
  \textbf{\bibinfo{volume}{13}}, \bibinfo{pages}{575} (\bibinfo{year}{1996}),
  \eprint{arXiv:gr-qc/9602024}.

\bibitem[{\citenamefont{{Kidder} et~al.}(2007)\citenamefont{{Kidder},
  {Blanchet}, and {Iyer}}}]{kidder-blanchet-iyer-25PNwaveform}
\bibinfo{author}{\bibfnamefont{L.~E.} \bibnamefont{{Kidder}}},
  \bibinfo{author}{\bibfnamefont{L.}~\bibnamefont{{Blanchet}}},
  \bibnamefont{and} \bibinfo{author}{\bibfnamefont{B.~R.}
  \bibnamefont{{Iyer}}}, \bibinfo{journal}{Class.~Quantum Grav.}
  \textbf{\bibinfo{volume}{24}}, \bibinfo{pages}{5307} (\bibinfo{year}{2007}),
  \eprint{arXiv:0706.0726}.

\bibitem[{\citenamefont{{Blanchet} et~al.}(2008)\citenamefont{{Blanchet},
  {Faye}, {Iyer}, and {Sinha}}}]{blanchet3pnwaveform}
\bibinfo{author}{\bibfnamefont{L.}~\bibnamefont{{Blanchet}}},
  \bibinfo{author}{\bibfnamefont{G.}~\bibnamefont{{Faye}}},
  \bibinfo{author}{\bibfnamefont{B.~R.} \bibnamefont{{Iyer}}},
  \bibnamefont{and} \bibinfo{author}{\bibfnamefont{S.}~\bibnamefont{{Sinha}}},
  \bibinfo{journal}{Class.~Quantum Grav.} \textbf{\bibinfo{volume}{25}},
  \bibinfo{pages}{165003} (\bibinfo{year}{2008}), \eprint{arXiv:0802.1249v2
  [gr-qc]}.

\bibitem[{lig()}]{ligoweb}
\emph{\bibinfo{title}{\textsc{LIGO}}},
  \bibinfo{howpublished}{\url{http://www.ligo.caltech.edu}}.

\bibitem[{geo()}]{geoweb}
\emph{\bibinfo{title}{\textsc{GEO600}}},
  \bibinfo{howpublished}{\url{http://www.geo600.uni-hannover.de}}.

\bibitem[{vir()}]{virgoweb}
\emph{\bibinfo{title}{\textsc{VIRGO}}},
  \bibinfo{howpublished}{\url{http://www.virgo.infn.it}}.

\bibitem[{tam()}]{tamaweb}
\emph{\bibinfo{title}{\textsc{TAMA300}}},
  \bibinfo{howpublished}{\url{http://tamago.mtk.nao.ac.jp}}.

\bibitem[{\citenamefont{{Pretorius}}(2005)}]{pretorius-PRL2005}
\bibinfo{author}{\bibfnamefont{F.}~\bibnamefont{{Pretorius}}},
  \bibinfo{journal}{Phys.~Rev.~Lett.} \textbf{\bibinfo{volume}{95}},
  \bibinfo{pages}{121101} (\bibinfo{year}{2005}), \eprint{arXiv:gr-qc/0507014}.

\bibitem[{\citenamefont{{Pretorius}}(2006)}]{pretorius-CQG2006}
\bibinfo{author}{\bibfnamefont{F.}~\bibnamefont{{Pretorius}}},
  \bibinfo{journal}{Class.~Quantum Grav.} \textbf{\bibinfo{volume}{23}},
  \bibinfo{pages}{S529} (\bibinfo{year}{2006}), \eprint{arXiv:gr-qc/0602115}.

\bibitem[{\citenamefont{Pretorius}(2009)}]{pretorius-BBHreview}
\bibinfo{author}{\bibfnamefont{F.}~\bibnamefont{Pretorius}}, in
  \emph{\bibinfo{booktitle}{Physics of Relativistic Objects in Compact
  Binaries: from Birth to Coalescence}}, edited by
  \bibinfo{editor}{\bibfnamefont{M.}~\bibnamefont{Colpi}},
  \bibinfo{editor}{\bibfnamefont{P.}~\bibnamefont{Casella}},
  \bibinfo{editor}{\bibfnamefont{V.}~\bibnamefont{Gorini}},
  \bibinfo{editor}{\bibfnamefont{U.}~\bibnamefont{Moschella}},
  \bibnamefont{and} \bibinfo{editor}{\bibfnamefont{A.}~\bibnamefont{Possenti}}
  (\bibinfo{publisher}{Springer Verlag, Canopus Publishing Limited},
  \bibinfo{year}{2009}), \eprint{arXiv:0710.1338v1 [gr-qc]}.

\bibitem[{\citenamefont{{Baker}
  et~al.}(2006{\natexlab{a}})\citenamefont{{Baker}, {Centrella}, {Choi},
  {Koppitz}, and {van Meter}}}]{baker-etal-PRL2006}
\bibinfo{author}{\bibfnamefont{J.~G.} \bibnamefont{{Baker}}},
  \bibinfo{author}{\bibfnamefont{J.}~\bibnamefont{{Centrella}}},
  \bibinfo{author}{\bibfnamefont{D.-I.} \bibnamefont{{Choi}}},
  \bibinfo{author}{\bibfnamefont{M.}~\bibnamefont{{Koppitz}}},
  \bibnamefont{and} \bibinfo{author}{\bibfnamefont{J.}~\bibnamefont{{van
  Meter}}}, \bibinfo{journal}{Phys.~Rev.~Lett.} \textbf{\bibinfo{volume}{96}},
  \bibinfo{pages}{111102} (\bibinfo{year}{2006}{\natexlab{a}}),
  \eprint{arXiv:gr-qc/0511103}.

\bibitem[{\citenamefont{{Baker}
  et~al.}(2006{\natexlab{b}})\citenamefont{{Baker}, {Centrella}, {Choi},
  {Koppitz}, and {van Meter}}}]{baker-etal-PRD2006}
\bibinfo{author}{\bibfnamefont{J.~G.} \bibnamefont{{Baker}}},
  \bibinfo{author}{\bibfnamefont{J.}~\bibnamefont{{Centrella}}},
  \bibinfo{author}{\bibfnamefont{D.-I.} \bibnamefont{{Choi}}},
  \bibinfo{author}{\bibfnamefont{M.}~\bibnamefont{{Koppitz}}},
  \bibnamefont{and} \bibinfo{author}{\bibfnamefont{J.}~\bibnamefont{{van
  Meter}}}, \bibinfo{journal}{Phys.~Rev.~D} \textbf{\bibinfo{volume}{73}},
  \bibinfo{pages}{104002} (\bibinfo{year}{2006}{\natexlab{b}}),
  \eprint{arXiv:gr-qc/0602026}.

\bibitem[{\citenamefont{{Campanelli}
  et~al.}(2006{\natexlab{a}})\citenamefont{{Campanelli}, {Lousto},
  {Marronetti}, and {Zlochower}}}]{campanelli-etal-PRL2006}
\bibinfo{author}{\bibfnamefont{M.}~\bibnamefont{{Campanelli}}},
  \bibinfo{author}{\bibfnamefont{C.~O.} \bibnamefont{{Lousto}}},
  \bibinfo{author}{\bibfnamefont{P.}~\bibnamefont{{Marronetti}}},
  \bibnamefont{and}
  \bibinfo{author}{\bibfnamefont{Y.}~\bibnamefont{{Zlochower}}},
  \bibinfo{journal}{Phys.~Rev.~Lett.} \textbf{\bibinfo{volume}{96}},
  \bibinfo{pages}{111101} (\bibinfo{year}{2006}{\natexlab{a}}),
  \eprint{arXiv:gr-qc/0511048}.

\bibitem[{\citenamefont{{Campanelli}
  et~al.}(2006{\natexlab{b}})\citenamefont{{Campanelli}, {Lousto}, and
  {Zlochower}}}]{campanelli-etal-PRD2006}
\bibinfo{author}{\bibfnamefont{M.}~\bibnamefont{{Campanelli}}},
  \bibinfo{author}{\bibfnamefont{C.~O.} \bibnamefont{{Lousto}}},
  \bibnamefont{and}
  \bibinfo{author}{\bibfnamefont{Y.}~\bibnamefont{{Zlochower}}},
  \bibinfo{journal}{Phys.~Rev.~D} \textbf{\bibinfo{volume}{73}},
  \bibinfo{pages}{061501(R)} (\bibinfo{year}{2006}{\natexlab{b}}),
  \eprint{arXiv:gr-qc/0601091}.

\bibitem[{\citenamefont{{Herrmann} et~al.}(2007)\citenamefont{{Herrmann},
  {Hinder}, {Shoemaker}, and {Laguna}}}]{herrmann-etal-CQG2007}
\bibinfo{author}{\bibfnamefont{F.}~\bibnamefont{{Herrmann}}},
  \bibinfo{author}{\bibfnamefont{I.}~\bibnamefont{{Hinder}}},
  \bibinfo{author}{\bibfnamefont{D.}~\bibnamefont{{Shoemaker}}},
  \bibnamefont{and} \bibinfo{author}{\bibfnamefont{P.}~\bibnamefont{{Laguna}}},
  \bibinfo{journal}{Class.~Quantum Grav.} \textbf{\bibinfo{volume}{24}},
  \bibinfo{pages}{S33} (\bibinfo{year}{2007}), \eprint{arXiv:gr-qc/0601026}.

\bibitem[{\citenamefont{{Sperhake}}(2007)}]{sperhake-PRD2007}
\bibinfo{author}{\bibfnamefont{U.}~\bibnamefont{{Sperhake}}},
  \bibinfo{journal}{Phys.~Rev.~D} \textbf{\bibinfo{volume}{76}},
  \bibinfo{pages}{104015} (\bibinfo{year}{2007}), \eprint{arXiv:gr-qc/0606079}.

\bibitem[{\citenamefont{{Scheel} et~al.}(2006)\citenamefont{{Scheel},
  {Pfeiffer}, {Lindblom}, {Kidder}, {Rinne}, and
  {Teukolsky}}}]{scheel-etal-PRD2006}
\bibinfo{author}{\bibfnamefont{M.~A.} \bibnamefont{{Scheel}}},
  \bibinfo{author}{\bibfnamefont{H.~P.} \bibnamefont{{Pfeiffer}}},
  \bibinfo{author}{\bibfnamefont{L.}~\bibnamefont{{Lindblom}}},
  \bibinfo{author}{\bibfnamefont{L.~E.} \bibnamefont{{Kidder}}},
  \bibinfo{author}{\bibfnamefont{O.}~\bibnamefont{{Rinne}}}, \bibnamefont{and}
  \bibinfo{author}{\bibfnamefont{S.~A.} \bibnamefont{{Teukolsky}}},
  \bibinfo{journal}{Phys.~Rev.~D} \textbf{\bibinfo{volume}{74}},
  \bibinfo{pages}{104006} (\bibinfo{year}{2006}), \eprint{arXiv:gr-qc/0607056}.

\bibitem[{\citenamefont{{Koppitz} et~al.}(2007)\citenamefont{{Koppitz},
  {Pollney}, {Reisswig}, {Rezzolla}, {Thornburg}, {Diener}, and
  {Schnetter}}}]{koppitz-etal-PRL2007}
\bibinfo{author}{\bibfnamefont{M.}~\bibnamefont{{Koppitz}}},
  \bibinfo{author}{\bibfnamefont{D.}~\bibnamefont{{Pollney}}},
  \bibinfo{author}{\bibfnamefont{C.}~\bibnamefont{{Reisswig}}},
  \bibinfo{author}{\bibfnamefont{L.}~\bibnamefont{{Rezzolla}}},
  \bibinfo{author}{\bibfnamefont{J.}~\bibnamefont{{Thornburg}}},
  \bibinfo{author}{\bibfnamefont{P.}~\bibnamefont{{Diener}}}, \bibnamefont{and}
  \bibinfo{author}{\bibfnamefont{E.}~\bibnamefont{{Schnetter}}},
  \bibinfo{journal}{Phys.~Rev.~Lett.} \textbf{\bibinfo{volume}{99}},
  \bibinfo{pages}{041102} (\bibinfo{year}{2007}), \eprint{arXiv:gr-qc/0701163}.

\bibitem[{\citenamefont{{Pollney} et~al.}(2007)\citenamefont{{Pollney},
  {Reisswig}, {Rezzolla}, {Szil{\'a}gyi}, {Ansorg}, {Deris}, {Diener},
  {Dorband}, {Koppitz}, {Nagar} et~al.}}]{pollney-etal-spinorbitrecoil}
\bibinfo{author}{\bibfnamefont{D.}~\bibnamefont{{Pollney}}},
  \bibinfo{author}{\bibfnamefont{C.}~\bibnamefont{{Reisswig}}},
  \bibinfo{author}{\bibfnamefont{L.}~\bibnamefont{{Rezzolla}}},
  \bibinfo{author}{\bibfnamefont{B.}~\bibnamefont{{Szil{\'a}gyi}}},
  \bibinfo{author}{\bibfnamefont{M.}~\bibnamefont{{Ansorg}}},
  \bibinfo{author}{\bibfnamefont{B.}~\bibnamefont{{Deris}}},
  \bibinfo{author}{\bibfnamefont{P.}~\bibnamefont{{Diener}}},
  \bibinfo{author}{\bibfnamefont{E.~N.} \bibnamefont{{Dorband}}},
  \bibinfo{author}{\bibfnamefont{M.}~\bibnamefont{{Koppitz}}},
  \bibinfo{author}{\bibfnamefont{A.}~\bibnamefont{{Nagar}}},
  \bibnamefont{et~al.}, \bibinfo{journal}{Phys.~Rev.~D}
  \textbf{\bibinfo{volume}{76}}, \bibinfo{pages}{124002}
  (\bibinfo{year}{2007}), \eprint{arXiv:0707.2559v1 [gr-qc]}.

\bibitem[{\citenamefont{{Scheel} et~al.}(2009)\citenamefont{{Scheel}, {Boyle},
  {Chu}, {Kidder}, {Matthews}, and {Pfeiffer}}}]{scheel-merger}
\bibinfo{author}{\bibfnamefont{M.~A.} \bibnamefont{{Scheel}}},
  \bibinfo{author}{\bibfnamefont{M.}~\bibnamefont{{Boyle}}},
  \bibinfo{author}{\bibfnamefont{T.}~\bibnamefont{{Chu}}},
  \bibinfo{author}{\bibfnamefont{L.~E.} \bibnamefont{{Kidder}}},
  \bibinfo{author}{\bibfnamefont{K.~D.} \bibnamefont{{Matthews}}},
  \bibnamefont{and} \bibinfo{author}{\bibfnamefont{H.~P.}
  \bibnamefont{{Pfeiffer}}}, \bibinfo{journal}{Phys.~Rev.~D}
  \textbf{\bibinfo{volume}{79}}, \bibinfo{pages}{024003}
  (\bibinfo{year}{2009}), \eprint{arXiv:0810.1767v2 [gr-qc]}.

\bibitem[{\citenamefont{{Hannam} et~al.}(2009)\citenamefont{{Hannam}, {Husa},
  {Baker}, {Boyle}, {Br{\"u}gmann}, {Chu}, {Dorband}, {Herrmann}, {Hinder},
  {Kelly} et~al.}}]{samurai09}
\bibinfo{author}{\bibfnamefont{M.}~\bibnamefont{{Hannam}}},
  \bibinfo{author}{\bibfnamefont{S.}~\bibnamefont{{Husa}}},
  \bibinfo{author}{\bibfnamefont{J.~G.} \bibnamefont{{Baker}}},
  \bibinfo{author}{\bibfnamefont{M.}~\bibnamefont{{Boyle}}},
  \bibinfo{author}{\bibfnamefont{B.}~\bibnamefont{{Br{\"u}gmann}}},
  \bibinfo{author}{\bibfnamefont{T.}~\bibnamefont{{Chu}}},
  \bibinfo{author}{\bibfnamefont{N.}~\bibnamefont{{Dorband}}},
  \bibinfo{author}{\bibfnamefont{F.}~\bibnamefont{{Herrmann}}},
  \bibinfo{author}{\bibfnamefont{I.}~\bibnamefont{{Hinder}}},
  \bibinfo{author}{\bibfnamefont{B.~J.} \bibnamefont{{Kelly}}},
  \bibnamefont{et~al.}, \bibinfo{journal}{Phys.~Rev.~D}
  \textbf{\bibinfo{volume}{79}}, \bibinfo{pages}{084025}
  (\bibinfo{year}{2009}), \eprint{arXiv:0901.2437 [gr-qc]}.

\bibitem[{\citenamefont{Aylott et~al.}(2009)}]{ninja09}
\bibinfo{author}{\bibfnamefont{B.}~\bibnamefont{Aylott}} \bibnamefont{et~al.}
  (\bibinfo{year}{2009}), \eprint{arXiv:0901.4399 [gr-qc]}.

\bibitem[{\citenamefont{{Buonanno}
  et~al.}(2007{\natexlab{a}})\citenamefont{{Buonanno}, {Cook}, and
  {Pretorius}}}]{buonanno-cook-pretorius}
\bibinfo{author}{\bibfnamefont{A.}~\bibnamefont{{Buonanno}}},
  \bibinfo{author}{\bibfnamefont{G.~B.} \bibnamefont{{Cook}}},
  \bibnamefont{and}
  \bibinfo{author}{\bibfnamefont{F.}~\bibnamefont{{Pretorius}}},
  \bibinfo{journal}{Phys.~Rev.~D} \textbf{\bibinfo{volume}{75}},
  \bibinfo{pages}{124018} (\bibinfo{year}{2007}{\natexlab{a}}),
  \eprint{arXiv:gr-qc/0610122}.

\bibitem[{\citenamefont{{Boyle} et~al.}(2007)\citenamefont{{Boyle}, {Brown},
  {Kidder}, {Mrou{\'e}}, {Pfeiffer}, {Scheel}, {Cook}, and
  {Teukolsky}}}]{boyle-etal-PRD2007}
\bibinfo{author}{\bibfnamefont{M.}~\bibnamefont{{Boyle}}},
  \bibinfo{author}{\bibfnamefont{D.~A.} \bibnamefont{{Brown}}},
  \bibinfo{author}{\bibfnamefont{L.~E.} \bibnamefont{{Kidder}}},
  \bibinfo{author}{\bibfnamefont{A.~H.} \bibnamefont{{Mrou{\'e}}}},
  \bibinfo{author}{\bibfnamefont{H.~P.} \bibnamefont{{Pfeiffer}}},
  \bibinfo{author}{\bibfnamefont{M.~A.} \bibnamefont{{Scheel}}},
  \bibinfo{author}{\bibfnamefont{G.~B.} \bibnamefont{{Cook}}},
  \bibnamefont{and} \bibinfo{author}{\bibfnamefont{S.~A.}
  \bibnamefont{{Teukolsky}}}, \bibinfo{journal}{Phys.~Rev.~D}
  \textbf{\bibinfo{volume}{76}}, \bibinfo{pages}{124038}
  (\bibinfo{year}{2007}), \eprint{arXiv:0710.0158v2 [gr-qc]}.

\bibitem[{\citenamefont{{Mrou{\'e}} et~al.}(2008)\citenamefont{{Mrou{\'e}},
  {Kidder}, and {Teukolsky}}}]{mroue-kidder-saul-PRD2008}
\bibinfo{author}{\bibfnamefont{A.~H.} \bibnamefont{{Mrou{\'e}}}},
  \bibinfo{author}{\bibfnamefont{L.~E.} \bibnamefont{{Kidder}}},
  \bibnamefont{and} \bibinfo{author}{\bibfnamefont{S.~A.}
  \bibnamefont{{Teukolsky}}}, \bibinfo{journal}{Phys.~Rev.~D}
  \textbf{\bibinfo{volume}{78}}, \bibinfo{pages}{044004}
  (\bibinfo{year}{2008}), \eprint{arXiv:0805.2390v3 [gr-qc]}.

\bibitem[{\citenamefont{{Boyle} et~al.}(2008)\citenamefont{{Boyle}, {Buonanno},
  {Kidder}, {Mrou{\'e}}, {Pan}, {Pfeiffer}, and
  {Scheel}}}]{boyle-etal-Efluxcomparison}
\bibinfo{author}{\bibfnamefont{M.}~\bibnamefont{{Boyle}}},
  \bibinfo{author}{\bibfnamefont{A.}~\bibnamefont{{Buonanno}}},
  \bibinfo{author}{\bibfnamefont{L.~E.} \bibnamefont{{Kidder}}},
  \bibinfo{author}{\bibfnamefont{A.~H.} \bibnamefont{{Mrou{\'e}}}},
  \bibinfo{author}{\bibfnamefont{Y.}~\bibnamefont{{Pan}}},
  \bibinfo{author}{\bibfnamefont{H.~P.} \bibnamefont{{Pfeiffer}}},
  \bibnamefont{and} \bibinfo{author}{\bibfnamefont{M.~A.}
  \bibnamefont{{Scheel}}}, \bibinfo{journal}{Phys.~Rev.~D}
  \textbf{\bibinfo{volume}{78}}, \bibinfo{pages}{104020}
  (\bibinfo{year}{2008}), \eprint{arXiv:0804.4184v2 [gr-qc]}.

\bibitem[{\citenamefont{{Baker}
  et~al.}(2007{\natexlab{a}})\citenamefont{{Baker}, {van Meter}, {McWilliams},
  {Centrella}, and {Kelly}}}]{baker-etal-PRL2007-NRPN}
\bibinfo{author}{\bibfnamefont{J.~G.} \bibnamefont{{Baker}}},
  \bibinfo{author}{\bibfnamefont{J.~R.} \bibnamefont{{van Meter}}},
  \bibinfo{author}{\bibfnamefont{S.~T.} \bibnamefont{{McWilliams}}},
  \bibinfo{author}{\bibfnamefont{J.}~\bibnamefont{{Centrella}}},
  \bibnamefont{and} \bibinfo{author}{\bibfnamefont{B.~J.}
  \bibnamefont{{Kelly}}}, \bibinfo{journal}{Phys.~Rev.~Lett.}
  \textbf{\bibinfo{volume}{99}}, \bibinfo{pages}{181101}
  (\bibinfo{year}{2007}{\natexlab{a}}), \eprint{arXiv:gr-qc/0612024}.

\bibitem[{\citenamefont{{Baker}
  et~al.}(2007{\natexlab{b}})\citenamefont{{Baker}, {McWilliams}, {van Meter},
  {Centrella}, {Choi}, {Kelly}, and {Koppitz}}}]{baker-etal-PRD2007-NRPN}
\bibinfo{author}{\bibfnamefont{J.~G.} \bibnamefont{{Baker}}},
  \bibinfo{author}{\bibfnamefont{S.~T.} \bibnamefont{{McWilliams}}},
  \bibinfo{author}{\bibfnamefont{J.~R.} \bibnamefont{{van Meter}}},
  \bibinfo{author}{\bibfnamefont{J.}~\bibnamefont{{Centrella}}},
  \bibinfo{author}{\bibfnamefont{D.-I.} \bibnamefont{{Choi}}},
  \bibinfo{author}{\bibfnamefont{B.~J.} \bibnamefont{{Kelly}}},
  \bibnamefont{and}
  \bibinfo{author}{\bibfnamefont{M.}~\bibnamefont{{Koppitz}}},
  \bibinfo{journal}{Phys.~Rev.~D} \textbf{\bibinfo{volume}{75}},
  \bibinfo{pages}{124024} (\bibinfo{year}{2007}{\natexlab{b}}),
  \eprint{arXiv:gr-qc/0612117}.

\bibitem[{\citenamefont{{Hinder} et~al.}(2008)\citenamefont{{Hinder},
  {Herrmann}, {Laguna}, and {Shoemaker}}}]{hinder-etal-eccentricPN-NR}
\bibinfo{author}{\bibfnamefont{I.}~\bibnamefont{{Hinder}}},
  \bibinfo{author}{\bibfnamefont{F.}~\bibnamefont{{Herrmann}}},
  \bibinfo{author}{\bibfnamefont{P.}~\bibnamefont{{Laguna}}}, \bibnamefont{and}
  \bibinfo{author}{\bibfnamefont{D.}~\bibnamefont{{Shoemaker}}}
  (\bibinfo{year}{2008}), \eprint{arXiv:0806.1037v1 [gr-qc]}.

\bibitem[{\citenamefont{{Campanelli} et~al.}(2009)\citenamefont{{Campanelli},
  {Lousto}, {Nakano}, and
  {Zlochower}}}]{campanelli-etal-spinning-NR-PN-compare}
\bibinfo{author}{\bibfnamefont{M.}~\bibnamefont{{Campanelli}}},
  \bibinfo{author}{\bibfnamefont{C.~O.} \bibnamefont{{Lousto}}},
  \bibinfo{author}{\bibfnamefont{H.}~\bibnamefont{{Nakano}}}, \bibnamefont{and}
  \bibinfo{author}{\bibfnamefont{Y.}~\bibnamefont{{Zlochower}}},
  \bibinfo{journal}{Phys.~Rev.~D} \textbf{\bibinfo{volume}{79}},
  \bibinfo{pages}{084010} (\bibinfo{year}{2009}), \eprint{arXiv:0808.0713v2
  [gr-qc]}.

\bibitem[{\citenamefont{{Hannam}
  et~al.}(2008{\natexlab{a}})\citenamefont{{Hannam}, {Husa}, {Gonz{\'a}lez},
  {Sperhake}, and {Br{\"u}gmann}}}]{hannam-etal-PN-NR-meet}
\bibinfo{author}{\bibfnamefont{M.}~\bibnamefont{{Hannam}}},
  \bibinfo{author}{\bibfnamefont{S.}~\bibnamefont{{Husa}}},
  \bibinfo{author}{\bibfnamefont{J.~A.} \bibnamefont{{Gonz{\'a}lez}}},
  \bibinfo{author}{\bibfnamefont{U.}~\bibnamefont{{Sperhake}}},
  \bibnamefont{and}
  \bibinfo{author}{\bibfnamefont{B.}~\bibnamefont{{Br{\"u}gmann}}},
  \bibinfo{journal}{Phys.~Rev.~D} \textbf{\bibinfo{volume}{77}},
  \bibinfo{pages}{044020} (\bibinfo{year}{2008}{\natexlab{a}}),
  \eprint{arXiv:0706.1305v2 [gr-qc]}.

\bibitem[{\citenamefont{{Berti} et~al.}(2007)\citenamefont{{Berti}, {Cardoso},
  {Gonzalez}, {Sperhake}, {Hannam}, {Husa}, and
  {Br{\"u}gmann}}}]{berti-etal-multipolarnonspinning}
\bibinfo{author}{\bibfnamefont{E.}~\bibnamefont{{Berti}}},
  \bibinfo{author}{\bibfnamefont{V.}~\bibnamefont{{Cardoso}}},
  \bibinfo{author}{\bibfnamefont{J.~A.} \bibnamefont{{Gonzalez}}},
  \bibinfo{author}{\bibfnamefont{U.}~\bibnamefont{{Sperhake}}},
  \bibinfo{author}{\bibfnamefont{M.}~\bibnamefont{{Hannam}}},
  \bibinfo{author}{\bibfnamefont{S.}~\bibnamefont{{Husa}}}, \bibnamefont{and}
  \bibinfo{author}{\bibfnamefont{B.}~\bibnamefont{{Br{\"u}gmann}}},
  \bibinfo{journal}{Phys.~Rev.~D} \textbf{\bibinfo{volume}{76}},
  \bibinfo{pages}{064034} (\bibinfo{year}{2007}), \eprint{arXiv:gr-qc/0703053}.

\bibitem[{\citenamefont{{Berti} et~al.}(2008)\citenamefont{{Berti}, {Cardoso},
  {Gonz{\'a}lez}, {Sperhake}, and
  {Br{\"u}gmann}}}]{berti-etal-mulitpolarspinning}
\bibinfo{author}{\bibfnamefont{E.}~\bibnamefont{{Berti}}},
  \bibinfo{author}{\bibfnamefont{V.}~\bibnamefont{{Cardoso}}},
  \bibinfo{author}{\bibfnamefont{J.~A.} \bibnamefont{{Gonz{\'a}lez}}},
  \bibinfo{author}{\bibfnamefont{U.}~\bibnamefont{{Sperhake}}},
  \bibnamefont{and}
  \bibinfo{author}{\bibfnamefont{B.}~\bibnamefont{{Br{\"u}gmann}}},
  \bibinfo{journal}{Class.~Quantum Grav.} \textbf{\bibinfo{volume}{25}},
  \bibinfo{pages}{114035} (\bibinfo{year}{2008}), \eprint{arXiv:0711.1097v2
  [gr-qc]}.

\bibitem[{\citenamefont{{Hannam}
  et~al.}(2008{\natexlab{b}})\citenamefont{{Hannam}, {Husa}, {Br{\"u}gmann},
  and {Gopakumar}}}]{hannam-gopa-NRPN-spin}
\bibinfo{author}{\bibfnamefont{M.}~\bibnamefont{{Hannam}}},
  \bibinfo{author}{\bibfnamefont{S.}~\bibnamefont{{Husa}}},
  \bibinfo{author}{\bibfnamefont{B.}~\bibnamefont{{Br{\"u}gmann}}},
  \bibnamefont{and}
  \bibinfo{author}{\bibfnamefont{A.}~\bibnamefont{{Gopakumar}}},
  \bibinfo{journal}{\prd} \textbf{\bibinfo{volume}{78}},
  \bibinfo{pages}{104007} (\bibinfo{year}{2008}{\natexlab{b}}),
  \eprint{0712.3787}.

\bibitem[{\citenamefont{{Damour}
  et~al.}(2008{\natexlab{a}})\citenamefont{{Damour}, {Nagar}, {Hannam}, {Husa},
  and {Brugmann}}}]{damour-nagar-jena}
\bibinfo{author}{\bibfnamefont{T.}~\bibnamefont{{Damour}}},
  \bibinfo{author}{\bibfnamefont{A.}~\bibnamefont{{Nagar}}},
  \bibinfo{author}{\bibfnamefont{M.}~\bibnamefont{{Hannam}}},
  \bibinfo{author}{\bibfnamefont{S.}~\bibnamefont{{Husa}}}, \bibnamefont{and}
  \bibinfo{author}{\bibfnamefont{B.}~\bibnamefont{{Brugmann}}},
  \bibinfo{journal}{Phys.~Rev.~D} \textbf{\bibinfo{volume}{78}},
  \bibinfo{pages}{044039} (\bibinfo{year}{2008}{\natexlab{a}}),
  \eprint{arXiv:0803.3162v2 [gr-qc]}.

\bibitem[{\citenamefont{{Damour}
  et~al.}(2008{\natexlab{b}})\citenamefont{{Damour}, {Nagar}, {Dorband},
  {Pollney}, and {Rezzolla}}}]{damour-nagar-AEI}
\bibinfo{author}{\bibfnamefont{T.}~\bibnamefont{{Damour}}},
  \bibinfo{author}{\bibfnamefont{A.}~\bibnamefont{{Nagar}}},
  \bibinfo{author}{\bibfnamefont{E.~N.} \bibnamefont{{Dorband}}},
  \bibinfo{author}{\bibfnamefont{D.}~\bibnamefont{{Pollney}}},
  \bibnamefont{and}
  \bibinfo{author}{\bibfnamefont{L.}~\bibnamefont{{Rezzolla}}},
  \bibinfo{journal}{Phys.~Rev.~D} \textbf{\bibinfo{volume}{77}},
  \bibinfo{pages}{084017} (\bibinfo{year}{2008}{\natexlab{b}}),
  \eprint{arXiv:0712.3003v2 [gr-qc]}.

\bibitem[{\citenamefont{{Damour} and
  {Nagar}}(2008)}]{damour-nagar-caltechcornell}
\bibinfo{author}{\bibfnamefont{T.}~\bibnamefont{{Damour}}} \bibnamefont{and}
  \bibinfo{author}{\bibfnamefont{A.}~\bibnamefont{{Nagar}}},
  \bibinfo{journal}{Phys.~Rev.~D} \textbf{\bibinfo{volume}{77}},
  \bibinfo{pages}{024043} (\bibinfo{year}{2008}), \eprint{arXiv:0711.2628v2
  [gr-qc]}.

\bibitem[{\citenamefont{{Damour} and {Nagar}}(2009)}]{damour-nagar-PRD09}
\bibinfo{author}{\bibfnamefont{T.}~\bibnamefont{{Damour}}} \bibnamefont{and}
  \bibinfo{author}{\bibfnamefont{A.}~\bibnamefont{{Nagar}}},
  \bibinfo{journal}{Phys.~Rev.~D} \textbf{\bibinfo{volume}{79}},
  \bibinfo{pages}{081503(R)} (\bibinfo{year}{2009}), \eprint{arXiv:0902.0136
  [gr-qc]}.

\bibitem[{\citenamefont{{Gopakumar} et~al.}(2008)\citenamefont{{Gopakumar},
  {Hannam}, {Husa}, and {Br{\"u}gmann}}}]{gopa-jena-eccentricPNNR}
\bibinfo{author}{\bibfnamefont{A.}~\bibnamefont{{Gopakumar}}},
  \bibinfo{author}{\bibfnamefont{M.}~\bibnamefont{{Hannam}}},
  \bibinfo{author}{\bibfnamefont{S.}~\bibnamefont{{Husa}}}, \bibnamefont{and}
  \bibinfo{author}{\bibfnamefont{B.}~\bibnamefont{{Br{\"u}gmann}}},
  \bibinfo{journal}{\prd} \textbf{\bibinfo{volume}{78}},
  \bibinfo{pages}{064026} (\bibinfo{year}{2008}), \eprint{0712.3737}.

\bibitem[{\citenamefont{{Pan} et~al.}(2008)\citenamefont{{Pan}, {Buonanno},
  {Baker}, {Centrella}, {Kelly}, {McWilliams}, {Pretorius}, and {van
  Meter}}}]{pan-buonanno-baker-etal-NRPN}
\bibinfo{author}{\bibfnamefont{Y.}~\bibnamefont{{Pan}}},
  \bibinfo{author}{\bibfnamefont{A.}~\bibnamefont{{Buonanno}}},
  \bibinfo{author}{\bibfnamefont{J.~G.} \bibnamefont{{Baker}}},
  \bibinfo{author}{\bibfnamefont{J.}~\bibnamefont{{Centrella}}},
  \bibinfo{author}{\bibfnamefont{B.~J.} \bibnamefont{{Kelly}}},
  \bibinfo{author}{\bibfnamefont{S.~T.} \bibnamefont{{McWilliams}}},
  \bibinfo{author}{\bibfnamefont{F.}~\bibnamefont{{Pretorius}}},
  \bibnamefont{and} \bibinfo{author}{\bibfnamefont{J.~R.} \bibnamefont{{van
  Meter}}}, \bibinfo{journal}{Phys.~Rev.~D} \textbf{\bibinfo{volume}{77}},
  \bibinfo{pages}{024014} (\bibinfo{year}{2008}), \eprint{arXiv:0704.1964v2
  [gr-qc]}.

\bibitem[{\citenamefont{{Buonanno}
  et~al.}(2007{\natexlab{b}})\citenamefont{{Buonanno}, {Pan}, {Baker},
  {Centrella}, {Kelly}, {McWilliams}, and {van
  Meter}}}]{buonanno-pan-baker-etal-nonspinningEOB}
\bibinfo{author}{\bibfnamefont{A.}~\bibnamefont{{Buonanno}}},
  \bibinfo{author}{\bibfnamefont{Y.}~\bibnamefont{{Pan}}},
  \bibinfo{author}{\bibfnamefont{J.~G.} \bibnamefont{{Baker}}},
  \bibinfo{author}{\bibfnamefont{J.}~\bibnamefont{{Centrella}}},
  \bibinfo{author}{\bibfnamefont{B.~J.} \bibnamefont{{Kelly}}},
  \bibinfo{author}{\bibfnamefont{S.~T.} \bibnamefont{{McWilliams}}},
  \bibnamefont{and} \bibinfo{author}{\bibfnamefont{J.~R.} \bibnamefont{{van
  Meter}}}, \bibinfo{journal}{Phys.~Rev.~D} \textbf{\bibinfo{volume}{76}},
  \bibinfo{pages}{104049} (\bibinfo{year}{2007}{\natexlab{b}}),
  \eprint{arXiv:0706.3732v3 [gr-qc]}.

\bibitem[{\citenamefont{Buonanno et~al.}(2009)\citenamefont{Buonanno, Pan,
  Pfeiffer, Scheel, Buchman, and Kidder}}]{buonanno-caltechEOB09}
\bibinfo{author}{\bibfnamefont{A.}~\bibnamefont{Buonanno}},
  \bibinfo{author}{\bibfnamefont{Y.}~\bibnamefont{Pan}},
  \bibinfo{author}{\bibfnamefont{H.~P.} \bibnamefont{Pfeiffer}},
  \bibinfo{author}{\bibfnamefont{M.~A.} \bibnamefont{Scheel}},
  \bibinfo{author}{\bibfnamefont{L.~T.} \bibnamefont{Buchman}},
  \bibnamefont{and} \bibinfo{author}{\bibfnamefont{L.~E.}
  \bibnamefont{Kidder}}, \bibinfo{journal}{Phys.~Rev.~D}
  \textbf{\bibinfo{volume}{79}}, \bibinfo{pages}{124028}
  (\bibinfo{year}{2009}), \eprint{arXiv:0902.0790 [gr-qc]}.

\bibitem[{\citenamefont{{Kennefick}}(1994)}]{kennefick-memory}
\bibinfo{author}{\bibfnamefont{D.}~\bibnamefont{{Kennefick}}},
  \bibinfo{journal}{Phys.~Rev.~D} \textbf{\bibinfo{volume}{50}},
  \bibinfo{pages}{3587} (\bibinfo{year}{1994}).

\bibitem[{lis()}]{lisaweb}
\emph{\bibinfo{title}{\textsc{LISA}}},
  \bibinfo{howpublished}{\url{http://lisa.jpl.nasa.gov}}.

\bibitem[{\citenamefont{{Blanchet} et~al.}(2002)\citenamefont{{Blanchet},
  {Faye}, {Iyer}, and {Joguet}}}]{blanchet35PNphase}
\bibinfo{author}{\bibfnamefont{L.}~\bibnamefont{{Blanchet}}},
  \bibinfo{author}{\bibfnamefont{G.}~\bibnamefont{{Faye}}},
  \bibinfo{author}{\bibfnamefont{B.~R.} \bibnamefont{{Iyer}}},
  \bibnamefont{and} \bibinfo{author}{\bibfnamefont{B.}~\bibnamefont{{Joguet}}},
  \bibinfo{journal}{Phys.~Rev.~D} \textbf{\bibinfo{volume}{65}},
  \bibinfo{pages}{061501(R)} (\bibinfo{year}{2002}),
  \eprint{arXiv:gr-qc/0105099v3}.

\bibitem[{\citenamefont{{Blanchet} et~al.}(2005)\citenamefont{{Blanchet},
  {Faye}, {Iyer}, and {Joguet}}}]{blanchet35PNphaseerratum}
\bibinfo{author}{\bibfnamefont{L.}~\bibnamefont{{Blanchet}}},
  \bibinfo{author}{\bibfnamefont{G.}~\bibnamefont{{Faye}}},
  \bibinfo{author}{\bibfnamefont{B.~R.} \bibnamefont{{Iyer}}},
  \bibnamefont{and} \bibinfo{author}{\bibfnamefont{B.}~\bibnamefont{{Joguet}}},
  \bibinfo{journal}{Phys.~Rev.~D} \textbf{\bibinfo{volume}{71}},
  \bibinfo{pages}{129902(E)} (\bibinfo{year}{2005}),
  \eprint{arXiv:gr-qc/0105099v3}.

\bibitem[{\citenamefont{{Kidder}}(2008)}]{kidder08}
\bibinfo{author}{\bibfnamefont{L.~E.} \bibnamefont{{Kidder}}},
  \bibinfo{journal}{Phys.~Rev.~D} \textbf{\bibinfo{volume}{77}},
  \bibinfo{pages}{044016} (\bibinfo{year}{2008}), \eprint{arXiv:0710.0614v1
  [gr-qc]}.

\bibitem[{\citenamefont{Iyer}(1993)}]{iyermultipolenotes}
\bibinfo{author}{\bibfnamefont{B.~R.} \bibnamefont{Iyer}}, in
  \emph{\bibinfo{booktitle}{Quantum Gravity, Gravitational Radiation and Large
  Scale Structure in the Universe}}, edited by
  \bibinfo{editor}{\bibfnamefont{B.~R.} \bibnamefont{Iyer}},
  \bibinfo{editor}{\bibfnamefont{S.~V.} \bibnamefont{Dhurandhar}},
  \bibnamefont{and} \bibinfo{editor}{\bibfnamefont{K.~B.} \bibnamefont{Joseph}}
  (\bibinfo{publisher}{Inter University Centre for Astronomy and Astrophysics},
  \bibinfo{address}{Pune, India}, \bibinfo{year}{1993}).

\bibitem[{\citenamefont{{Pati} and {Will}}(2000)}]{pati-will-DIRE1}
\bibinfo{author}{\bibfnamefont{M.~E.} \bibnamefont{{Pati}}} \bibnamefont{and}
  \bibinfo{author}{\bibfnamefont{C.~M.} \bibnamefont{{Will}}},
  \bibinfo{journal}{Phys.~Rev.~D} \textbf{\bibinfo{volume}{62}},
  \bibinfo{pages}{124015} (\bibinfo{year}{2000}), \eprint{arXiv:gr-qc/0007087}.

\bibitem[{\citenamefont{{Goldberger} and
  {Rothstein}}(2006)}]{goldberger-rothstein-PRD2006}
\bibinfo{author}{\bibfnamefont{W.~D.} \bibnamefont{{Goldberger}}}
  \bibnamefont{and} \bibinfo{author}{\bibfnamefont{I.~Z.}
  \bibnamefont{{Rothstein}}}, \bibinfo{journal}{Phys.~Rev.~D}
  \textbf{\bibinfo{volume}{73}}, \bibinfo{pages}{104029}
  (\bibinfo{year}{2006}), \eprint{arXiv:hep-th/0409156v2}.

\bibitem[{\citenamefont{{Blanchet}}(2005)}]{blanchet-tailsoftails-erratrum}
\bibinfo{author}{\bibfnamefont{L.}~\bibnamefont{{Blanchet}}},
  \bibinfo{journal}{Class.~Quantum Grav.} \textbf{\bibinfo{volume}{22}},
  \bibinfo{pages}{3381} (\bibinfo{year}{2005}), \eprint{arXiv:gr-qc/9710038v2}.

\bibitem[{\citenamefont{{Blanchet}}(1998{\natexlab{a}})}]{blanchet-quadquad}
\bibinfo{author}{\bibfnamefont{L.}~\bibnamefont{{Blanchet}}},
  \bibinfo{journal}{Class.~Quantum Grav.} \textbf{\bibinfo{volume}{15}},
  \bibinfo{pages}{89} (\bibinfo{year}{1998}{\natexlab{a}}),
  \eprint{arXiv:gr-qc/9710037}.

\bibitem[{\citenamefont{{Blanchet}}(1998{\natexlab{b}})}]{blanchet-tailsoftail%
s}
\bibinfo{author}{\bibfnamefont{L.}~\bibnamefont{{Blanchet}}},
  \bibinfo{journal}{Class.~Quantum Grav.} \textbf{\bibinfo{volume}{15}},
  \bibinfo{pages}{113} (\bibinfo{year}{1998}{\natexlab{b}}),
  \eprint{arXiv:gr-qc/9710038v2}.

\bibitem[{\citenamefont{{Favata}}(2009{\natexlab{c}})}]{favata-eccentricmemory}
\bibinfo{author}{\bibfnamefont{M.}~\bibnamefont{{Favata}}}
  (\bibinfo{year}{2009}{\natexlab{c}}), \bibinfo{note}{(in preparation)}.

\bibitem[{\citenamefont{{Blanchet}
  et~al.}(2004{\natexlab{a}})\citenamefont{{Blanchet}, {Damour}, and
  {Esposito-Far{\`e}se}}}]{blanchet-damour-farese-dimreg}
\bibinfo{author}{\bibfnamefont{L.}~\bibnamefont{{Blanchet}}},
  \bibinfo{author}{\bibfnamefont{T.}~\bibnamefont{{Damour}}}, \bibnamefont{and}
  \bibinfo{author}{\bibfnamefont{G.}~\bibnamefont{{Esposito-Far{\`e}se}}},
  \bibinfo{journal}{Phys.~Rev.~D} \textbf{\bibinfo{volume}{69}},
  \bibinfo{pages}{124007} (\bibinfo{year}{2004}{\natexlab{a}}),
  \eprint{arXiv:gr-qc/0311052}.

\bibitem[{\citenamefont{{Blanchet} and {Iyer}}(2005)}]{blanchet-iyer-hadreg}
\bibinfo{author}{\bibfnamefont{L.}~\bibnamefont{{Blanchet}}} \bibnamefont{and}
  \bibinfo{author}{\bibfnamefont{B.~R.} \bibnamefont{{Iyer}}},
  \bibinfo{journal}{Phys.~Rev.~D} \textbf{\bibinfo{volume}{71}},
  \bibinfo{pages}{024004} (\bibinfo{year}{2005}), \eprint{arXiv:gr-qc/0409094}.

\bibitem[{\citenamefont{{Blanchet}
  et~al.}(2004{\natexlab{b}})\citenamefont{{Blanchet}, {Damour},
  {Esposito-Far{\`e}se}, and {Iyer}}}]{blanchetdamour3PNprl}
\bibinfo{author}{\bibfnamefont{L.}~\bibnamefont{{Blanchet}}},
  \bibinfo{author}{\bibfnamefont{T.}~\bibnamefont{{Damour}}},
  \bibinfo{author}{\bibfnamefont{G.}~\bibnamefont{{Esposito-Far{\`e}se}}},
  \bibnamefont{and} \bibinfo{author}{\bibfnamefont{B.~R.}
  \bibnamefont{{Iyer}}}, \bibinfo{journal}{Phys.~Rev.~Lett.}
  \textbf{\bibinfo{volume}{93}}, \bibinfo{pages}{091101}
  (\bibinfo{year}{2004}{\natexlab{b}}), \eprint{arXiv:gr-qc/0406012v1}.

\bibitem[{\citenamefont{{Blanchet} and
  {Faye}}(2001)}]{blanchet-faye-3pn-PRD2001}
\bibinfo{author}{\bibfnamefont{L.}~\bibnamefont{{Blanchet}}} \bibnamefont{and}
  \bibinfo{author}{\bibfnamefont{G.}~\bibnamefont{{Faye}}},
  \bibinfo{journal}{\prd} \textbf{\bibinfo{volume}{63}},
  \bibinfo{pages}{062005} (\bibinfo{year}{2001}), \eprint{arXiv:gr-qc/0007051}.

\bibitem[{bal()}]{bala-luc-privcomm}
\bibinfo{note}{Bala Iyer and Luc Blanchet, (private communication)}.

\bibitem[{\citenamefont{{Nagar} and
  {Rezzolla}}(2005)}]{nagar-rezzolla-metricpert}
\bibinfo{author}{\bibfnamefont{A.}~\bibnamefont{{Nagar}}} \bibnamefont{and}
  \bibinfo{author}{\bibfnamefont{L.}~\bibnamefont{{Rezzolla}}},
  \bibinfo{journal}{Class.~Quantum Grav.} \textbf{\bibinfo{volume}{22}},
  \bibinfo{pages}{R167} (\bibinfo{year}{2005}), \eprint{arXiv:gr-qc/0502064v4}.

\bibitem[{\citenamefont{{Nagar} and
  {Rezzolla}}(2006)}]{nagar-rezzolla-metricpert-erratum}
\bibinfo{author}{\bibfnamefont{A.}~\bibnamefont{{Nagar}}} \bibnamefont{and}
  \bibinfo{author}{\bibfnamefont{L.}~\bibnamefont{{Rezzolla}}},
  \bibinfo{journal}{Class.~Quantum Grav.} \textbf{\bibinfo{volume}{23}},
  \bibinfo{pages}{4297} (\bibinfo{year}{2006}), \eprint{arXiv:gr-qc/0502064}.

\bibitem[{\citenamefont{{Schnittman} et~al.}(2008)\citenamefont{{Schnittman},
  {Buonanno}, {van Meter}, {Baker}, {Boggs}, {Centrella}, {Kelly}, and
  {McWilliams}}}]{schnittman-multipolarrecoil}
\bibinfo{author}{\bibfnamefont{J.~D.} \bibnamefont{{Schnittman}}},
  \bibinfo{author}{\bibfnamefont{A.}~\bibnamefont{{Buonanno}}},
  \bibinfo{author}{\bibfnamefont{J.~R.} \bibnamefont{{van Meter}}},
  \bibinfo{author}{\bibfnamefont{J.~G.} \bibnamefont{{Baker}}},
  \bibinfo{author}{\bibfnamefont{W.~D.} \bibnamefont{{Boggs}}},
  \bibinfo{author}{\bibfnamefont{J.}~\bibnamefont{{Centrella}}},
  \bibinfo{author}{\bibfnamefont{B.~J.} \bibnamefont{{Kelly}}},
  \bibnamefont{and} \bibinfo{author}{\bibfnamefont{S.~T.}
  \bibnamefont{{McWilliams}}}, \bibinfo{journal}{Phys.~Rev.~D}
  \textbf{\bibinfo{volume}{77}}, \bibinfo{pages}{044031}
  (\bibinfo{year}{2008}), \eprint{arXiv:0707.0301}.

\bibitem[{\citenamefont{{Baker} et~al.}(2008)\citenamefont{{Baker}, {Boggs},
  {Centrella}, {Kelly}, {McWilliams}, and {van
  Meter}}}]{baker-etal-PRD2008-gravlradcharacteristics}
\bibinfo{author}{\bibfnamefont{J.~G.} \bibnamefont{{Baker}}},
  \bibinfo{author}{\bibfnamefont{W.~D.} \bibnamefont{{Boggs}}},
  \bibinfo{author}{\bibfnamefont{J.}~\bibnamefont{{Centrella}}},
  \bibinfo{author}{\bibfnamefont{B.~J.} \bibnamefont{{Kelly}}},
  \bibinfo{author}{\bibfnamefont{S.~T.} \bibnamefont{{McWilliams}}},
  \bibnamefont{and} \bibinfo{author}{\bibfnamefont{J.~R.} \bibnamefont{{van
  Meter}}}, \bibinfo{journal}{Phys.~Rev.~D} \textbf{\bibinfo{volume}{78}},
  \bibinfo{pages}{044046} (\bibinfo{year}{2008}), \eprint{arXiv:0805.1428v2
  [gr-qc]}.

\bibitem[{\citenamefont{{Baker} et~al.}(2002)\citenamefont{{Baker},
  {Campanelli}, {Lousto}, and {Takahashi}}}]{lazarus-PRD02}
\bibinfo{author}{\bibfnamefont{J.}~\bibnamefont{{Baker}}},
  \bibinfo{author}{\bibfnamefont{M.}~\bibnamefont{{Campanelli}}},
  \bibinfo{author}{\bibfnamefont{C.~O.} \bibnamefont{{Lousto}}},
  \bibnamefont{and}
  \bibinfo{author}{\bibfnamefont{R.}~\bibnamefont{{Takahashi}}},
  \bibinfo{journal}{\prd} \textbf{\bibinfo{volume}{65}},
  \bibinfo{pages}{124012} (\bibinfo{year}{2002}),
  \eprint{arXiv:astro-ph/0202469}.

\bibitem[{lar()}]{larry-privcomm}
\bibinfo{note}{Larry Kidder, (private communication)}.

\bibitem[{\citenamefont{{Damour}}(2008)}]{EOB-damour-lecnotes}
\bibinfo{author}{\bibfnamefont{T.}~\bibnamefont{{Damour}}},
  \bibinfo{journal}{Int.~J.~Mod.~Phys.~A} \textbf{\bibinfo{volume}{23}},
  \bibinfo{pages}{1130} (\bibinfo{year}{2008}), \eprint{arXiv:0802.4047v1
  [gr-qc]}.

\bibitem[{\citenamefont{{Favata}}(2009{\natexlab{d}})}]{favata-PNNR-memory}
\bibinfo{author}{\bibfnamefont{M.}~\bibnamefont{{Favata}}}
  (\bibinfo{year}{2009}{\natexlab{d}}), \bibinfo{note}{(in preparation)}.

\bibitem[{\citenamefont{{Smarr}}(1977)}]{smarr-zfl}
\bibinfo{author}{\bibfnamefont{L.}~\bibnamefont{{Smarr}}},
  \bibinfo{journal}{Phys.~Rev.~D} \textbf{\bibinfo{volume}{15}},
  \bibinfo{pages}{2069} (\bibinfo{year}{1977}).

\bibitem[{\citenamefont{{Bontz} and {Price}}(1979)}]{bontz-price}
\bibinfo{author}{\bibfnamefont{R.~J.} \bibnamefont{{Bontz}}} \bibnamefont{and}
  \bibinfo{author}{\bibfnamefont{R.~H.} \bibnamefont{{Price}}},
  \bibinfo{journal}{Astrophys.~J.} \textbf{\bibinfo{volume}{228}},
  \bibinfo{pages}{560} (\bibinfo{year}{1979}).

\bibitem[{\citenamefont{{Wagoner}}(1979)}]{wagoner-lowfreq}
\bibinfo{author}{\bibfnamefont{R.~V.} \bibnamefont{{Wagoner}}},
  \bibinfo{journal}{Phys.~Rev.~D} \textbf{\bibinfo{volume}{19}},
  \bibinfo{pages}{2897} (\bibinfo{year}{1979}).

\bibitem[{\citenamefont{{Flanagan} and {Hughes}}(1998)}]{flanagan-hughesI}
\bibinfo{author}{\bibfnamefont{{\'E}.~{\'E}.} \bibnamefont{{Flanagan}}}
  \bibnamefont{and} \bibinfo{author}{\bibfnamefont{S.~A.}
  \bibnamefont{{Hughes}}}, \bibinfo{journal}{Phys.~Rev.~D}
  \textbf{\bibinfo{volume}{57}}, \bibinfo{pages}{4535} (\bibinfo{year}{1998}),
  \eprint{arXiv:gr-qc/9701039}.

\bibitem[{\citenamefont{{Frauendiener}}(1992)}]{frauendiener-memnote}
\bibinfo{author}{\bibfnamefont{J.}~\bibnamefont{{Frauendiener}}},
  \bibinfo{journal}{Class.~Quantum Grav.} \textbf{\bibinfo{volume}{9}},
  \bibinfo{pages}{1639} (\bibinfo{year}{1992}).

\bibitem[{\citenamefont{{Nakano} and {Ioka}}(2007)}]{nakano-ioka-2ndorderQNM}
\bibinfo{author}{\bibfnamefont{H.}~\bibnamefont{{Nakano}}} \bibnamefont{and}
  \bibinfo{author}{\bibfnamefont{K.}~\bibnamefont{{Ioka}}},
  \bibinfo{journal}{Phys.~Rev.~D} \textbf{\bibinfo{volume}{76}},
  \bibinfo{pages}{084007} (\bibinfo{year}{2007}), \eprint{arXiv:0708.0450v1
  [gr-qc]}.

\bibitem[{\citenamefont{Spiegel and Liu}(1999)}]{spiegel}
\bibinfo{author}{\bibfnamefont{M.~R.} \bibnamefont{Spiegel}} \bibnamefont{and}
  \bibinfo{author}{\bibfnamefont{J.}~\bibnamefont{Liu}},
  \emph{\bibinfo{title}{Schaum's Mathematical Handbook of Formulas and Tables}}
  (\bibinfo{publisher}{McGraw-Hill}, \bibinfo{address}{New York},
  \bibinfo{year}{1999}), \bibinfo{edition}{2nd} ed., \bibinfo{note}{(see
  integral 18.32)}.

\bibitem[{\citenamefont{{Edmonds}}(1960)}]{edmonds}
\bibinfo{author}{\bibfnamefont{A.~R.} \bibnamefont{{Edmonds}}},
  \emph{\bibinfo{title}{{Angular Momentum in Quantum Mechanics}}}
  (\bibinfo{publisher}{Princeton University Press},
  \bibinfo{address}{Princeton}, \bibinfo{year}{1960}), \bibinfo{edition}{2nd}
  ed.

\bibitem[{\citenamefont{{Rose}}(1957)}]{rose}
\bibinfo{author}{\bibfnamefont{M.~E.} \bibnamefont{{Rose}}},
  \emph{\bibinfo{title}{{Elementary Theory of Angular Momentum}}}
  (\bibinfo{publisher}{Wiley}, \bibinfo{address}{New York},
  \bibinfo{year}{1957}).

\bibitem[{\citenamefont{{Landau} and {Lifshitz}}(1977)}]{landaulifshitzQM}
\bibinfo{author}{\bibfnamefont{L.~D.} \bibnamefont{{Landau}}} \bibnamefont{and}
  \bibinfo{author}{\bibfnamefont{E.~M.} \bibnamefont{{Lifshitz}}},
  \emph{\bibinfo{title}{{Quantum mechanics}}} (\bibinfo{publisher}{Pergamon
  Press}, \bibinfo{address}{Oxford}, \bibinfo{year}{1977}),
  \bibinfo{edition}{3rd} ed.

\bibitem[{\citenamefont{{Rotenberg} et~al.}(1959)\citenamefont{{Rotenberg},
  {Bivins}, {Metropolis}, and {Wooten}}}]{3jbook}
\bibinfo{author}{\bibfnamefont{M.}~\bibnamefont{{Rotenberg}}},
  \bibinfo{author}{\bibfnamefont{R.}~\bibnamefont{{Bivins}}},
  \bibinfo{author}{\bibfnamefont{N.}~\bibnamefont{{Metropolis}}},
  \bibnamefont{and} \bibinfo{author}{\bibfnamefont{J.~K.}
  \bibnamefont{{Wooten}}}, \emph{\bibinfo{title}{{The 3-j and 6-j symbols}}}
  (\bibinfo{publisher}{The Technology Press (MIT)},
  \bibinfo{address}{Cambridge, MA}, \bibinfo{year}{1959}),
  \bibinfo{edition}{2nd} ed.

\end{thebibliography}
\end{document}